\documentclass[a4wide]{article}          
\usepackage{graphicx}
\usepackage{amsmath}
\usepackage{amssymb}
\usepackage{xspace}

\usepackage{array}

\usepackage[ruled]{algorithm2e}

\usepackage{color}
\usepackage{xcolor}

\newcommand{\props}{P}
\newcommand{\model}{\mathcal{M}}

\newcommand{\closure}{\mathcal{C}}

\newcommand{\dist}[1]{\mathcal{D}^{#1}}

\newcommand{\reals}{\mathbb{R}}
\newcommand{\nats}{\mathbb{N}}

\newcommand{\arel}{\mathcal{R}}

\newcommand{\topochecker}{{\ttfamily{topochecker}}\xspace}
\newcommand{\SLCS}{SLCS\xspace}
\newcommand{\SLCSMI}{ImgQL\xspace}

\newcommand{\lnear}{\mathcal{N}}
\newcommand{\lsurr}{\mathcal{S}}

\SetAlFnt{\small}
\SetAlCapFnt{\small}
\SetAlCapNameFnt{\small}
\SetAlCapHSkip{0pt}
\IncMargin{-\parindent}
\newcommand{\linterior}{\mathcal{I}}

\DeclareMathOperator{\leverywhere}{\mathcal{E}}
\newcommand{\lsomewhere}{\mathcal{F}}

\newcommand{\msep}{\,\mid\,}
\newcommand{\form}{{\bf F}}

\colorlet{VINC}{orange}
\colorlet{GINA}{blue}

\usepackage{subfig}
\usepackage{tikz}
\usepackage{latexsym}
\definecolor{darkgreen}{RGB}{30,120,30}

\newcommand{\RED}[1]{\textcolor{red}{#1}}
\newcommand{\BROWN}[1]{\textcolor{brown}{#1}}

\newcommand{\GREEN}[1]{\textcolor{darkgreen}{#1}}
\newcounter{dgnot}
\newenvironment{dgnot}[1][]{\refstepcounter{dgnot}\par\medskip
   \noindent \textbf{\RED{NfD~\thedgnot.}  #1} \rmfamily}{\medskip}

\newcommand{\ed}{\hfill$\bullet$}
\newcommand{\SET}[1]{\{#1\}}
\newcommand{\ZET}[2]{\SET{#1 | #2}}
\newcommand{\attrib}{A}
\newcommand{\aeval}{{\cal A}}
\newcommand{\peval}{{\cal V}}
\newcommand{\weval}{{\cal W}}
\newcommand{\lneg}{\neg}
\newcommand{\lfalse}{\bot}
\newcommand{\ltrue}{\top}
\newcommand{\lreach}{{\cal R}}
\newcommand{\lfromto}{{\cal T}}
\newcommand{\lssurr}{{\lsurr_!}}
\newcommand{\lflt}[1]{{\cal J}^{< #1}}
\newcommand{\lbsurr}[1]{\lsurr^{#1}}
\newcommand{\lNbsurr}[1]{\hat{\lsurr}^{#1}}

\newcommand{\mv}[1]{{$\tt #1$}}

\newcommand{\lssim}[7]{\triangle\!\!\!{\scriptstyle \triangle}_{#1}
{\tiny 
\left[
\begin{array}{ccc}
#2 & #3 & #4\\
#5 & #6 & #7
\end{array}
\right]
}}
\newcommand{\EDT}{{\tt EDT}}
\newcommand{\MDDT}{{\tt MDDT}}
\newcommand{\SCMP}{{\tt SCMP}}
\newcommand{\mkhis}{{\cal H}}
\newcommand{\mean}[1]{\overline{#1}}
\newcommand{\cc}{\mathbf{r}}

\renewcommand{\form}{\Phi}
\renewcommand{\arel}{R}

\usepackage{doi}
\usepackage{url}

\newcommand{\MAGENTA}[1]{\textcolor{magenta}{#1}}

\newcounter{mknot}
\newenvironment{mknot}[1][]{\refstepcounter{mknot}\par\medskip
   \noindent \textbf{\MAGENTA{NfM~\themknot.}  #1} \rmfamily}{\medskip}

\newcounter{vcnot}
\newenvironment{vcnot}[1][]{\refstepcounter{vcnot}\par\medskip
   \noindent \textbf{\BROWN{NfV~\thevcnot.}  #1} \rmfamily}{\medskip}

\newcounter{gbnot}
\newenvironment{gbnot}[1][]{\refstepcounter{gbnot}\par\medskip
   \noindent \textbf{\GREEN{NfG~\thegbnot.}  #1} \rmfamily}{\medskip}

\newcommand\Siena{(Case courtesy of Azienda Ospedaliera Universitaria Senese; image and data processing performed in compliance with EU GRDP 679/2016)}

\providecommand{\url}[1]{{#1}}

\newtheorem{definition}{Definition}

\begin{document}

\title{Spatial Logics and Model Checking for\\
Medical Imaging\thanks{This work is partially supported by  Azienda Ospedaliera Universitaria Senese.}\\
{ - Extended Version - }
}



\author{
Fabrizio Banci Buonamici \and Gina Belmonte\\
Azienda Ospedaliera Universitaria Senese, Siena, Italy\\
{\tt\small $\{$F.Banci,G.Belmonte$\}$@ao-siena.toscana.it} \and
Vincenzo Ciancia, Diego Latella ,Mieke Massink\\
CNR-ISTI, Pisa, Italy\\
{\tt\small $\{$Vincenzo.Ciancia, Diego.Latella, Mieke.Massink$\}$@cnr.it}	
}



\date{ }

\maketitle


\begin{abstract}
Recent research on spatial and spatio-tempo\-ral model checking  provides novel image analysis
methodologies, rooted in logical methods for topological
spaces. Medical Imaging (MI) is a field where such methods show
potential for ground-breaking innovation. 
Our starting point is \SLCS, the {\em Spatial Logic for Closure Spaces}--- 
Closure Spaces being a generalisation of topological spaces, covering also discrete space structu\-res--- and 
\topochecker, a model-checker for \SLCS (and extensions thereof).
We introduce the logical language \SLCSMI (``Image Query Language''). \SLCSMI extends \SLCS with logical operators describing \emph{distance} and \emph{region similarity}.
The spatio-temporal model checker \topochecker is correspondingly enhanced 
with state-of-the-art algorithms, borrowed from computational image processing, for efficient implementation of distance-based operators, namely
{\em distance transforms}. 
Similarity between regions is defined by means of a {\em statistical similarity} operator, based on  notions from {\em statistical texture analysis}.
We illustrate our approach by means of two examples
of analysis of Magnetic Resonance images: segmentation
of \emph{glioblastoma} and its \emph{oedema}, and segmentation of
\emph{rectal carcinoma}.
\end{abstract}

{\em keywords:} Spatial logics; Closure spaces; Model checking; Medical Imaging;  
  Segmentation; Magnetic Resonance Imaging; Distance Transform;  Statistical Texture Analysis


\section{Introduction}\label{sec:introduction}

Computer Science plays a fundamental role in 
the field of medical image
analysis.  Computational methods are currently in use for
several different purposes, such as: 
\emph{Computer-Aided Diagnosis} (CAD), aiming at the
  classification of areas in images, based on the presence of signs of
  specific diseases \cite{Doi2007};
\emph{Image Segmentation}, tailored to identify areas that
  exhibit specific features or functions (such as organs or
  sub-structures) \cite{Gordillo2013};
\emph{Automatic contouring} of Organs at Risk or target
  volumes for radiotherapy applications \cite{brock2014image};
Identification of \emph{indicators}, computed from the acquired images, 
	enabling early diagnosis, or understanding of microscopic characteristics of specific
  diseases, or help in the identification of prognostic factors to predict a treatment
  output 
  \cite{Chetelat2003} \cite{Toosy2003} 
  (examples of indicators are
  the \emph{mean diffusivity} and the \emph{fractional anisotropy} obtained from 
  Magnetic Resonance (MR) Diffusion-Weighted Images, or the 
  \emph{magnetisation transfer ratio} maps obtained from a Magnetisation
  Transfer acquisition 
  \cite{DeSantis2014} \cite{Li2015}).

Such kinds of analyses are strictly tied to the spatial features of images.

In this paper we focus on image segmentation, in particular to identify glioblastomas and rectal carcinomas. Glioblastomas are the most common malignant intracranial tumours whereas rectal carcinomas manifest themselves as particular tumours situated at the end of the large intestine. 
For the treatment of glioblastomas neuroimaging protocols are used before and after treatment to evaluate the effect of treatment strategies and to monitor the evolution of the disease. In clinical studies and routine treatment magnetic resonance images (MRI) are evaluated based mostly on qualitative criteria such as the presence of hyper-intense tissue appearing in the images~\cite{Menze2015}. The study and development of automatic and semi-automatic segmentation algorithms is aiming at overcoming the current time consuming practise of manual delineation of such tumours and at providing an accurate, reliable and reproducible method of segmentation of the tumour area and related tissues~\cite{Dupont2016}. 

Segmentation of medical images, and brain segmentation in particular, is nowadays an important topic on its own in many applications in neuroimaging; several automatic and semi-automatic methods have been proposed~\cite{lemieux1999fast,Despotovi2015} constituting a very active research area (see for example~\cite{Dupont2016,Fyllingen2016,Simi2015,Zhu2012}). One of the technical challenges of the development of automated (brain) tumour segmentation is that lesion areas are only defined through changes in the intensity (luminosity) in the (black \& white) images that are {\em relative} to surrounding normal tissue. Even manual segmentation by experts shows significant variations when intensity gradients between adjacent tissue structures are smooth or partially obscured~\cite{Menze2015}. Furthermore, there is a considerable variation across images from different patients and images obtained with different MRI scanners.

In this paper we propose a novel approach to image segmentation, namely an interactive, logic based method, supported by {\em spatial model checking}, tailored to loosely identify a region of interest in MRI on which to focus the analysis of glioblastoma or other types of tumours. This approach is particularly suitable to exploit the {\em relative} spatial relations between tissues of interest mentioned earlier.
 
\emph{Spatial  and Spatio-temporal model checking} are the subject of a  recent trend in Computer Science (see for instance  \cite{De+07,Gr+09,CLLM14,CLLM16,CLLM16bertinoro,NBCLM15,SPATEL}) that uses specifications written in logical languages describing \emph{space} -- called \emph{spatial logics} -- to automatically identify patterns and structures of interest in a variety of domains, ranging from signals~\cite{NBCLM15} and images~\cite{CLLM16,SPATEL} to Collective Adaptive Systems~\cite{CLMP15,CGLLM14,CLMPV16}. 


The research presented in the present paper stems from the \emph{topological} approach to spatial logics, dating back to the work of Alfred Tarski, who first recognised the possibility of reasoning on space using topology as a mathematical framework for the interpretation of modal logic (see
\cite{Ch5HBSL} for a thorough introduction). In this context, formulas
are interpreted as sets of points of a topological space, and in
particular the modal operator $\diamond$ is usually interpreted as the (logical representation of the) topological \emph{closure} operator.  A standard reference is the \emph{Handbook of Spatial Logics} 
\cite{HBSL}. Therein, several spatial logics are described, with applications far
beyond topological spaces; such logics treat not only aspects of
morphology, geometry and distance, but also advanced topics such as
dynamic systems, and discrete structures, that are particularly
difficult to deal with from a topological perspective (see, for example~\cite{Gal99}).
In recent work \cite{CLLM14,CLLM16}, Ciancia et al. pushed such theoretical development further to encompass directed graphs, resulting in the definition of the \emph{Spatial Logic for Closure Spaces} (\SLCS), and a related model checking algorithm.
Subsequently, in \cite{CGLLM15}, a spatio-temporal logic, combining \emph{Computation Tree Logic} and the newly defined spatial operators, was introduced; the (extended) model checking algorithm has been implemented in the prototype \emph{spatio-temporal model checker} \topochecker\footnote{Topochecker: \emph{a topological model checker}, see \url{http://topochecker.isti.cnr.it}, \url{https://github.com/vincenzoml/topochecker}}.


The broader scope of our research interest in the context of medical imaging is to 
enable the \emph{declarative description} and
\emph{automatic} or \emph{semi-automatic}, efficient identification of
regions in images (such as tumours, infiltrations, organs at risk,
lesions, etc.)  using {\em spatial logic} formulas specifying relevant features,
such as texture or similarity, bound together by spatial constraints,
for example, proximity, boundary properties, distance, and so on, that increase
the significance and signal-to-noise ratio of the obtained
results. This is possible by considering such images as instances of (quasi discrete) closure spaces. The tools and methods we introduce can be used both for two-dimensional (2D) and three-dimensional (3D) MI; we remark that modern MRI machines can usually provide 3D data for analysis; however, in standard practice, 3D information is often discarded in favour of 2D (slice by slice) analysis, due to the lack of well-established methods for 3D analysis. Using 3D information may lead to improved accuracy and it is therefore of high interest, in current research, to identify techniques for this purpose.

\subsubsection*{Original contributions:} This paper details and extends the ideas outlined in~\cite{BCLM16}, providing several further original contributions:
\begin{itemize}

\item extension of the spatial logic \SLCS to \SLCSMI, introducing di\-stance-\-based operators and showing their formal relation to the other spatial logic operators of \SLCS. A novel approach to model checking of distance-based operators is provided based on so-called \emph{distance transforms}, that forms the basis for the definition an efficient algorithm to solve the model checking problem. Asymptotic time complexity of the procedure we propose is linear or quasi-linear, depending on the kind of distance used. 
This result makes such procedure suitable for the analysis of higher resolution or 3D images;
\item introduction of a novel logical connective aimed at  estimating similarity between regions.
This operator is based on \emph{statistical texture analysis} and is able to classify points of the space based on the similarity between the area where they are localised, and a target region, expressed in logical terms. The connective is specific for medical image analysis. Its embedding shows how such connectives can be integrated into the spatial logic. This provides an example of how other specialised existing algorithms could be introduced and exploited within the spatial logic model checking framework; 
\item enhancement of the results in the glioblastoma case study first introduced in~\cite{BCLM16}, providing the relevant technical details on the logical specification;
\item presentation of a further case study --- namely, segmentation of \emph{rectum carcinoma} --- showing that the method can also be applied to the segmentation of other types of tumours that are situated in other parts of the body;
\item development of efficient model checking algorithms, that are competitive in computational efficiency w.r.t. state-of-the-art (semi-)automatic segmentation approaches. As an additional benefit, logical specifications are transparent, reproducible, accurate, human-readable,  and applicable  to both 2D and 3D images. 
\end{itemize}

Texture analysis, distance, and reachability in space can be freely combined as high-level logical operators with a clear and well-defined topological semantics. The interplay of these aspects is the key to obtain our experimental results. The work in~\cite{BCLM16} constituted a first \emph{proof-of-concept} study. In that study \topochecker\ was used for the declarative specification of regions in medical images. The model checker was used to automatically and efficiently identify and colour \emph{glioblastoma} and the surrounding \emph{oedema} in MRI scans, 
on the basis of a declarative definition of the two \emph{regions of interest}, given 
in terms of their visual appearance. The latter is defined by image features such as proximity, interconnection, and texture similarity. 
The input to the model checker consists of a precise, declarative, unambiguous logical specification, that besides being fairly close to the level of abstraction of an expert description of the process, is also remarkably concise, human readable, robust and reproducible.


 \subsubsection*{Related work:}
 The idea of using model checking, and in particular spatial or spatio-temporal model checking, for the analysis of medical images is relatively recent and there are only a few articles exploring this field so far.  In particular, \cite{SGBM16} uses spatio-temporal model checking techniques inspired by \cite{Gr+09} -- pursuing machine learning of the logical structure of image features -- for the detection of tumours. In contrast, our approach is more focused on human-intelligible logical descriptions that provide reproducible results. Other interesting work is that in~\cite{PG16} where spatio-temporal meta model checking is used for the analysis of biological processes, with an interesting focus on \emph{multi-scale} aspects.

Among the fully automated approaches that recently are gaining interest are those based on machine learning and deep learning (see for example~\cite{Akkus2017} for a recent review). Although manual segmentation is still the standard for {\em in vivo} images, this method is expensive and time-consuming, difficult to reproduce and possibly inaccurate due to human error. Machine learning and deep learning approaches have shown promising results in pattern recognition in areas where large, reliable datasets are available and are currently being developed for application in MRI based brain segmentation with the aim to obtain reliable automatic segmentation methods. Deep learning is based on the use of artificial neural networks, consisting of several layers, that can extract a hierarchy of features from raw input data. These methods depend heavily on the availability of large training datasets and the generation of manual ground truth labels, i.e. data sets in which segments of interest are indicated by experts manually in a standard way. This is a complicated task not only because it is very laborious, but also because of the relatively high intra-expert and inter-expert variability of 20$\pm$15\% and 28$\pm$12\%, respectively, for manual segmentations of brain tumour images~\cite{Mazzara2004}. Interactive approaches based on spatial model checking may therefore also be of help to improve the generation of manual ground truth labels in a more efficient, transparent and reproducible way.
 

\subsubsection*{Outline:} A technical introduction to spatial logics and distance-based operators
is provided in Section~\ref{sec:SLDCS}. Syntax and semantics of the fragment of \SLCS 
we will use in this paper are recalled, as well as the main notions of spatial model checking
for the fragment. The definition of a distance operator for \SLCSMI is presented as well.
In Section~\ref{sec:LogicTA} the logic framework we propose for
statistical texture analysis is presented. In Section~\ref{sec:examples}, the two case studies are presented in detail, including, where available, a first assessment of validation. Some concluding remarks are given in Section~\ref{sec:conclusions}.

\section{Logics for Closure Spaces with Distance}
\label{sec:SLDCS}

In this section, we discuss the background knowledge that we use in the technical developments of the paper and we extend it with notions of {\em distance}. 
In particular, we  briefly introduce the notion of closure spaces,  the fragment
of \SLCS~\cite{CLLM14,CLLM16} we use in this paper, the related model checking algorithm, and \topochecker. 
We detail the use of so-called \emph{distance operators} in this research line and we extend the logic fragment with a distance operator parametric on the specific notion of distance; we also give an account of the extension of the model checking algorithm necessary for dealing with the distance operator,
based on the notion of {\em distance transform} and its implementation in \topochecker for two specific notions of distance, namely the {\em Euclidean} and the {\em shortest path} distances.

In the sequel we will often make explicit reference to 2D images and their pixels;
here we point out that this is done only for the sake of simplicity and that all notions,
notations, definitions and results equally apply to 3D images and their voxels (i.e.
{\em volumetric picture elements}, the 3D counterpart of pixels).

\subsection{Closure Spaces, Spatial logics and Model Checking}
In 
\emph{spatial logics}, modal
operators are interpreted using the concept of \emph{neighbourhood} in
a topological space, enabling one to reason about \emph{points} of the
space using familiar concepts such as proximity, distance, or
reachability. A comprehensive reference for these
theoretical developments is~\cite{HBSL}. Transferring the results in the field to applications,
and in particular to model checking, requires one to use \emph{finite}
models. However, finite topological spaces are not satisfactory in
this respect; for instance, they cannot encode {\em arbitrary} graphs, including e.g. 
those with a non-transitive / non-symmetric edge relation, that may be the object of spatial reasoning in several applications (for instance, consider the graph of roads in a town, including the one-way streets). 
Extending topological spaces to \emph{closure spaces}
(see~\cite{Gal99}) is the key to generalise these results.
In this paper we use a fragment of \SLCS comprising an operator, 
called \emph{near}, interpreted as proximity, and the \emph{surrounded} connective, 
which is a spatial variant of the classical temporal \emph{weak until} operator, able to
characterise unbounded areas of space, based on their boundary properties. The \emph{surrounded} connective is similar in spirit to the spatial \emph{until}
operator for topological spaces discussed by Aiello and van Benthem
in~\cite{A02,vBB07}, although it is interpreted in closure spaces. Several derived operators may be defined, among which, notably, variants of the notion of \emph{reachability} in space. 
%
The combination of \SLCS with temporal operators from
the well-known branching time logic CTL (Computation Tree
Logic)~\cite{ClE82}, has been explored in~\cite{CGLLM15}. Some related case 
studies have been analysed in \cite{CLMPV16,Ciancia2018} where 
the logic  caters for spatio-temporal reasoning and model checking. 
In the present paper, we focus on spatial properties; therefore we restrict our attention to spatial aspects
of our framework.

\subsubsection{A fragment of \SLCS}\label{sec:FSLCS}
\SLCS is a logic for {\em space}, where the latter is modelled by means of {\em closure spaces}.
Before introducing the fragment of \SLCS we use in the present paper, we recall some basic notions of closure spaces~\cite{Gal99,Gal03}.

\begin{definition}\label{def:ClosureSpaces}
A {\em closure space} is a pair $(X,\closure)$ where $X$ is a non-empty set (of {\em points})
and $\closure: 2^X \to 2^X$ is a function satisfying the following three axioms:
\begin{enumerate}
\item $\closure(\emptyset)=\emptyset$;
\item $Y \subseteq \closure(Y)$ for all  $Y \subseteq X$;
\item $\closure(Y_1 \cup Y_2) = \closure(Y_1) \cup \closure(Y_2)$ for all $Y_1,Y_2\subseteq X$. \ed
\end{enumerate}
\end{definition}

According to the well known Kuratowski definition, adding the \emph{idempotence axiom} 
$\closure(\closure(Y))=\closure(Y)$ for all $Y \subseteq X$ in Definition~\ref{def:ClosureSpaces} makes it a definition of topological spaces~\cite{Gal03}.
Consequently, the latter are a subclass of closure spaces. 

Given any relation $\arel \subseteq X \times X$, function $\closure_{\arel}:2^X \to 2^X$ 
with $\closure_{\arel}(Y) = Y\cup \ZET{x}{\exists y \in Y. y \, \arel \,x}$ satisfies the axioms of Definition~\ref{def:ClosureSpaces} thus making 
$(X,\closure_{\arel})$ a closure space. The class
of closure spaces generated by binary relations on the set of points represent a very interesting 
subclass of closure spaces, known as {\em quasi-discrete} closure spaces. Quasi-discrete 
closure spaces include 
discrete structures like graphs, where each graph $(X,\arel)$ with set of nodes $X$ 
and set of the edges $\arel$ is in one to one correspondence with closure space $(X,\closure_{\arel})$.  Clearly, {\em finite} closure spaces are quasi-discrete closure spaces.

The following definition is instrumental for the definition of {\em paths} over quasi-discrete closure spaces.

\begin{definition}\label{def:ContFunc}
A {\em continuous function} $f : (X_1, \closure_1) \to (X_2, \closure_2)$ is a function $f : X_1\to  X_2$
such that, for all $Y \subseteq X_1$, we have $f(\closure_1(Y)) \subseteq \closure_2(f(Y))$.
\ed
\end{definition}

In the definition below $(\nats, \closure_{Succ})$ is the closure space of natural numbers
with the {\em successor} relation: $(n,m) \in Succ \Leftrightarrow m=n+1$.

\begin{definition}\label{def:path}
A {\em path} $\pi$ in  $(X,\closure_{\arel})$ is a continuous function
$\pi: (\nats, \closure_{Succ}) \to (X,\closure_{\arel})$.
\ed
\end{definition}
In the sequel we will let $\pi, \pi', \pi_1, \pi_2$ denote paths; the elements of $\nats$
will be called {\em indexes} in the context of paths.

A quasi-discrete closure space $(X,\closure_{\arel})$, can be used as the
basis for a mathematical model of a 2D digital image; $X$ represents the finite set
of {\em pixels} and $\arel$ is the reflexive and symmetric  {\em adjacency} relation between pixels~\cite{Gal14}. We note in passing that several different adjacency relations can be used.
For instance in the {\em orthogonal}  adjacency relation (sometimes called von Neumann adjacency)  only pixels which share an edge count as adjacent, so that each pixel is adjacent 
to (itself and) four other pixels; on the other hand, in the {\em orthodiagonal}  adjacency relation
pixels are adjacent as long as they share at least either an edge or a corner,
so that each pixel is adjacent  to (itself and) eight other pixels.

Pixels are usually associated
with specific attributes, such as colours and/or colour-intensity. We model this by assuming that
a set $\attrib$ of point {\em attribute names} is given and by enriching $(X,\closure_{\arel})$ with an
{\em attribute evaluation} function $\aeval:X \times \attrib \to V$ from points and their attributes to some value set $V$ 
such that $\aeval(x,a) \in V$ is the {\em value} of attribute $a$ of point $x$.  
%

For given set $\props$ of {\em atomic predicates} $p$, the syntax of the fragment of \SLCS 
we use in this paper is given in Figure~\ref{fig:fragSLCSsyn}.

\begin{figure}
\newcolumntype{L}{>{$}l<{$}}
\[
\begin{array}{r c l L}
\form & ::=  & p & \hfill \textsc{[Atomic Predicate]}\\
          & \msep  & \lneg \, \form & \hfill \textsc{[Negation]}\\
          & \msep  & \form_1 \, \land \, \form_2 & \hfill \textsc{[Conjunction]}\\
          & \msep  & \lnear \, \form & \hfill \textsc{[Near]}\\
          & \msep  & \form_1 \, \lsurr \, \form_2 & \hfill \textsc{[Surrounded]}\\
\end{array}
\]
\caption{\label{fig:fragSLCSsyn} Syntax of the fragment of \SLCS.}
\end{figure}

Informally,  it is assumed that space is modelled by a set of points; each atomic predicate  $p \in \props$ models a specific {\em feature} of {\em points} and is thus associated with the set of points
which have this feature. A point $x$ satisfies  $\lnear\, \form$ if a point satisfying $\form$
can be reached from $x$ in at most one (closure) step, i.e. if $x$ is {\em near} (or {\em close}) to
a point satisfying  $\form$. A point $x$ satisfies $\form_1 \, \lsurr \, \form_2$ if 
it satisfies $\form_1$ and in any path $\pi$ rooted in $x$ (i.e. such that $\pi(0)=x$)
and passing through a point $\pi(\ell)$ {\em not} 
satisfying $\form_1$, there is a point $\pi(j)$ before or at $\ell$  (i.e. $0 < j \leq n$) 
that satisfies $\form_2$.
In other words,  $x$ belongs to an area satisfying $\form_1$ and one cannot {\em escape} from such an area 
without hitting a point $\form_2$, i.e. $x$ is {\em surrounded} by $\form_2$. Finally, the fragment includes logical negation ($\neg$) and conjunction ($\land$). 

The above description is formalised by the definition of {\em model} and {\em satisfaction relation}:

\begin{definition}\label{def:model}
A {\em closure model}  is a tuple $((X,\closure), \aeval, \peval)$ consisting of a closure space
$(X,\closure)$, a valuation $\aeval: X \times \attrib \to V$, assigning to each  point and attribute 
the value of the attribute at that point, and a valuation $\peval: \props \to 2^X$ assigning to
each atomic predicate the set of points where it holds.
\ed
\end{definition}

In the sequel, we assume that an atomic predicate $p$ can be bound to an assertion $\alpha$, 
the latter stating a property of attributes, and we use the syntax $p:=\alpha$ for atomic predicate definitions, to this purpose. Assertions are standard Boolean expressions,
e.g. comparisons of the form $a \geq c$, for $c \in V$, and compositions thereof; 
we refrain from specifying the actual syntax of assertions, and we assume valuation $\aeval$
be extended in the obvious way in order to evaluate assertions, e.g. $\aeval(x,a \geq c) = \aeval(x,a)\geq c$.

\begin{definition}\label{def:satisfaction}
{\em Satisfaction} $\model, x \models \form$ at point $x \in X$ in model
$\model = ((X,\closure), \aeval, \peval)$ is defined by induction on the structure of formulas, 
as in Figure~\ref{fig:fragSLCSsem}. \ed
\end{definition}

\begin{figure}
\newcolumntype{L}{>{$}l<{$}}
\[
\begin{array}{r c l c l c l L}
\model,x & \models  & p \in P & \Leftrightarrow & x  \in \peval(p)\\
\model,x & \models  & \lneg \,\form & \Leftrightarrow & \model,x  \models \form \mbox{ does not hold}\\
\model,x & \models  & \form_1 \, \land \, \form_2 & \Leftrightarrow & \model,x \models  \form_1 \, \mbox{ and } \, \model,x \models  \form_2\\
\model,x & \models  & \lnear \, \form & \Leftrightarrow & x \in \closure(\ZET{y}{\model,y \models \form})\\
\model,x & \models  & \form_1 \, \lsurr \, \form_2 & \Leftrightarrow & \model,x \models  \form_1 \, \mbox{ and}\\
 & &  & & \mbox{for all paths } \pi \mbox{ and indexes } \ell\!:\\
 & &  & & 
\mbox{\hspace{0.1in}}\pi(0)=x \, \mbox{ and } \,  \model, \pi(\ell) \models \lneg \form_1\\
 & &  & & 
\mbox{\hspace{0.1in}}\mbox{implies }\\
 & &  & &
\mbox{\hspace{0.1in}}\mbox{there exists index } j \mbox{ such that:}\\
& &  & &
\mbox{\hspace{0.3in}}0 < j \leq \ell \, \mbox{ and } \,  \model, \pi(j) \models \form_2
\end{array}
\]
\caption{\label{fig:fragSLCSsem} Semantics of the fragment of \SLCS; whenever $p:=\alpha$ is a definition for $p$, we assume $x\in \peval(p)$ if and only if $\aeval(x,\alpha)$ yields the truth-value $true$.}
\end{figure}

\begin{figure}
\centering
{
\includegraphics[height=3cm]{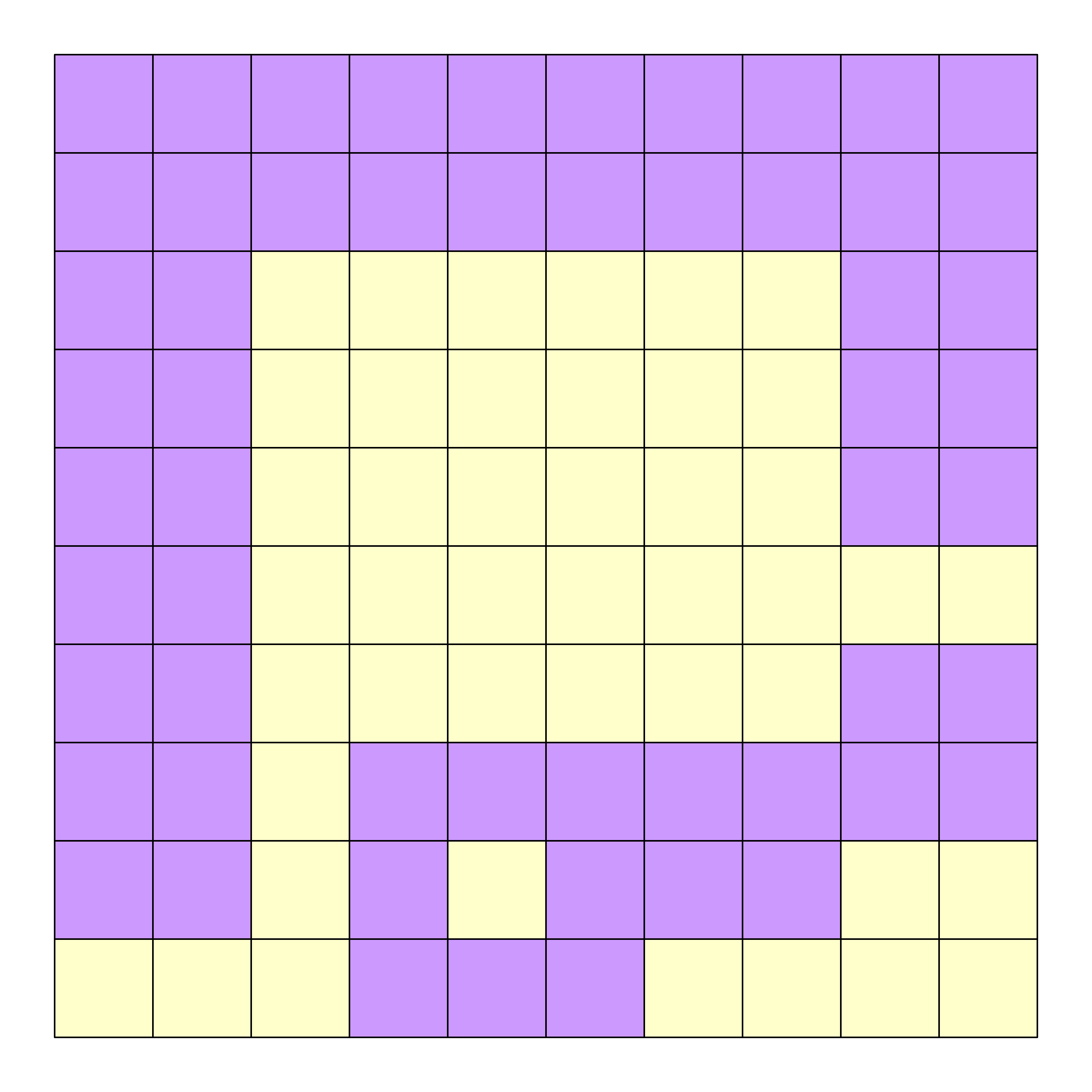}
}
\caption{An example model; the points satisfying atomic predicate $p$ are shown in violet, those satisfying $q$ are shown in yellow.}
\label{fig:exSample}
\end{figure}

\begin{figure}
\centering
\subfloat[][]
{
\includegraphics[height=3cm]{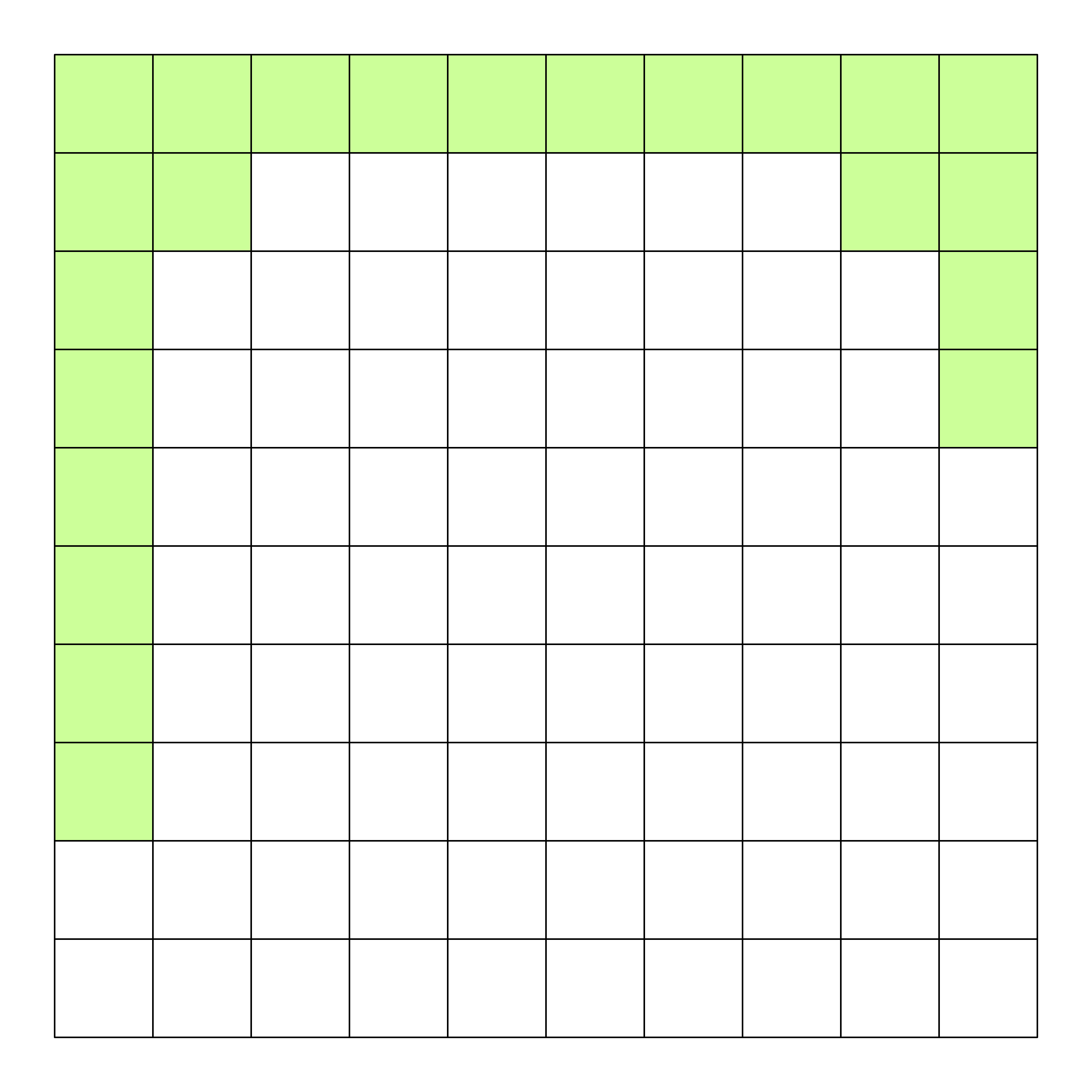}
\label{fig:exNear}
}\quad
\centering
\subfloat[][]
{
\includegraphics[height=3cm]{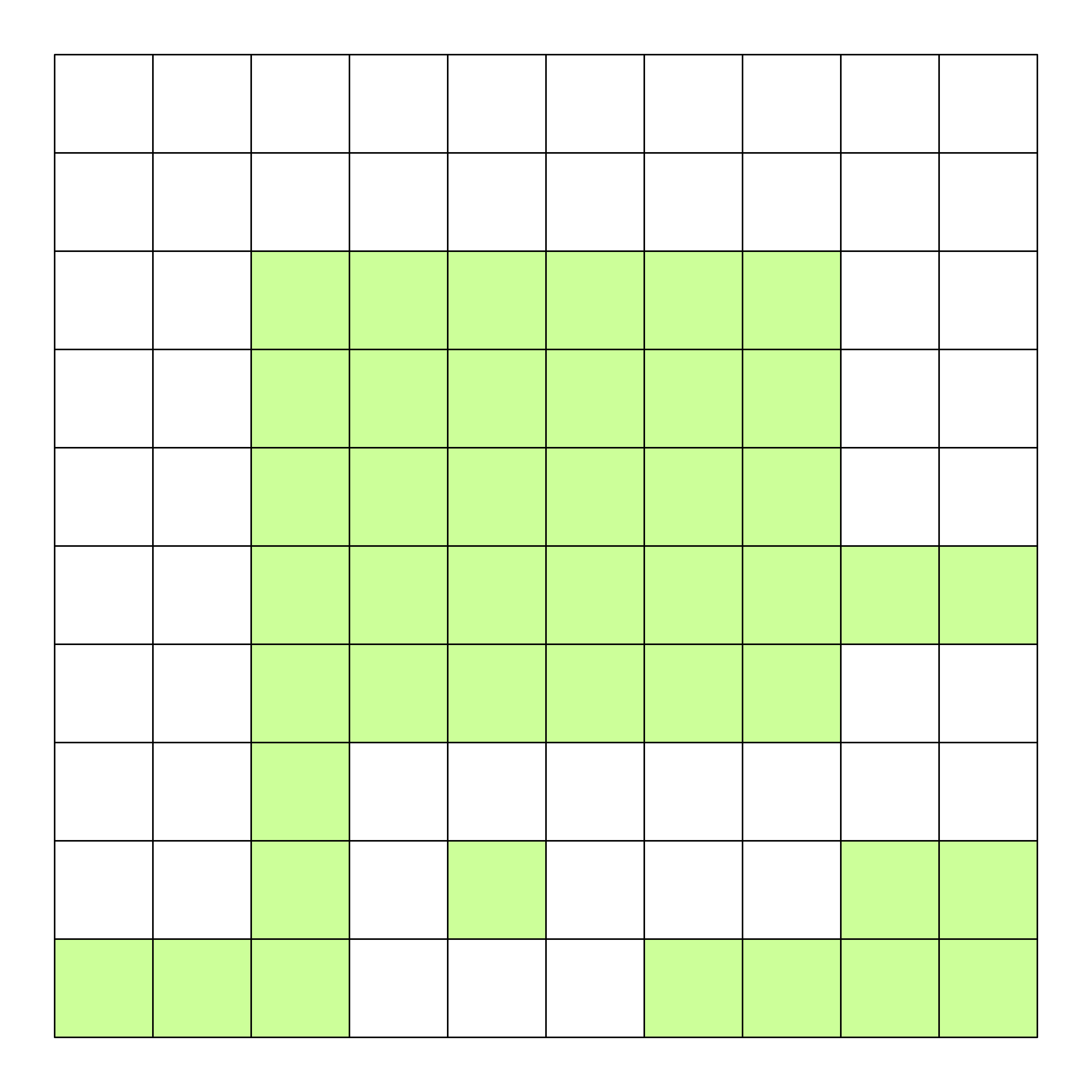}
\label{fig:exSurr}
}\\
\caption{The points in Figure~\ref{fig:exSample} satisfying $\lneg \lnear q$ (a) and those satisfying $q \, \lsurr \, p$ (b) are shown in green.}
\label{fig:exNearSurr}
\end{figure}

In Figure~\ref{fig:exSample}  a simple finite closure model is shown for which the orthogonal adjacency relation
is assumed. All the points satisfying atomic predicate $p$ are shown in violet whereas those satisfying $q$ are shown in yellow (no point satisfies $p \land q$ in this example). Figure~\ref{fig:exNear} shows in green the points that satisfy $\lneg \lnear q$, while
Figure~\ref{fig:exSurr} shows in green the points satisfying $q \, \lsurr \, p$ (i.e., all $q$-points that are
surrounded by $p$-points; note that, in the example, these are exactly {\em all} $q$-points).

\begin{figure}
\centering
{
\includegraphics[height=3cm]{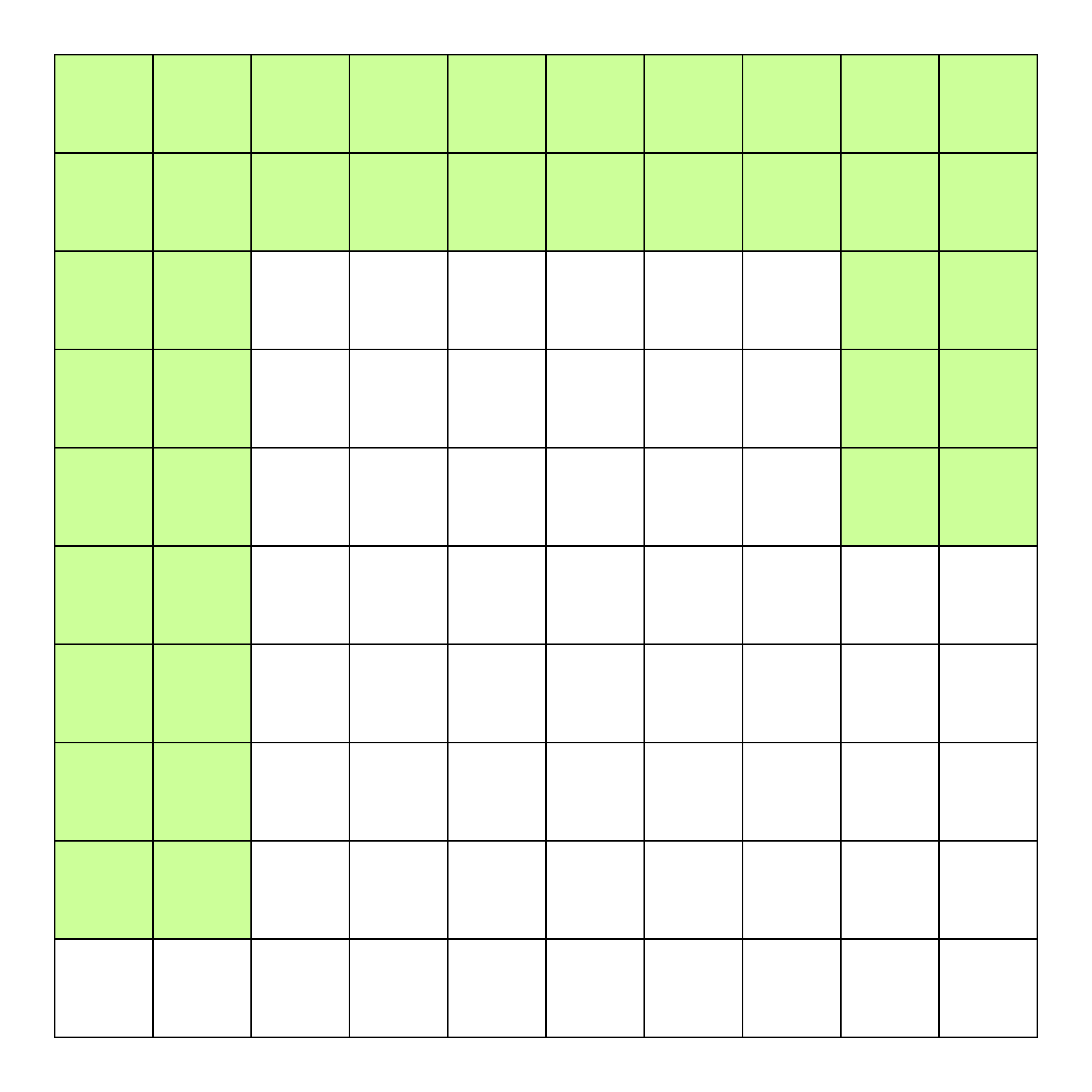}
}
\caption{The points in Figure~\ref{fig:exSample} satisfying $p \, \lfromto \lneg (\lnear \, q)$ are shown in green.}
\label{fig:exFromTo}
\end{figure}

A number of useful {\em derived operators} are defined in Figure~\ref{fig:derived}. 
A few words of explanation are worth for the $\lfromto$ operator, while we refer the reader to~\cite{CLLM16} for a general discussion on \SLCS derived operators.
A point satisfies $\form_1 \, \lfromto \, \form_2$ if and only if it lays in an area $Y_1 \subseteq X$ 
the points of which satisfy $\form_1$ and $Y_1$ ``touches'' a non-empty area $Y_2$,
the points of which satisfy $\form_2$; for this reason, sometimes we call the 
\textsc{From-To} operator ``touches''.  
With reference to Figure~\ref{fig:exSample}, Figure~\ref{fig:exFromTo} shows in green the points satisfying $p \, \lfromto \lneg (\lnear \, q)$.
Another pattern, that may be used  for filtering noise in images, is formula $\lnear \, \linterior \,\form$. The effect of such a formula is to capture the \emph{regular} region~\cite{Ch9HBSL} included in the set of points satisfying $\form$; point $x$ satisfies $\lnear \, \linterior \,\form$ if and only if it is adjacent to at least one point $y$ satisfying $\form$ which, in turn, is not adjacent to points satisfying $\lnot \,\form$. The effect of such a filter is to eliminate small regions, e.g. those consisting of a single point, when these are considered noise or artefacts. 

\begin{figure}
\newcolumntype{L}{>{$}l<{$}}
\[
\begin{array}{l c l L}
\lfalse & \triangleq  & p \, \land \, \lneg p  & \hfill \textsc{[False]}\\
\ltrue & \triangleq  & \lneg  \, \lfalse & \hfill \textsc{[True]}\\
\form_1 \, \lor \, \form_2 & \triangleq  & \lneg \,(\lneg \form_1 \,\land  \,\lneg \,\form_2) &\hfill \textsc{[Disjunction]}\\
\linterior  \,\form & \triangleq  & \lneg  \,(\lnear  \,\lneg  \,\form)&\hfill \textsc{[Interior]}\\
\form_1 \, \lreach \, \form_2 & \triangleq  & \lneg \, (\lneg \, \form_2 \, \lsurr \, \lneg \, \form_1)&\hfill \textsc{[Reachability]}\\
\form_1 \, \lfromto \, \form_2 & \triangleq  &  \form_1 \, \land \, ((\form_1 \, \lor \, \form_2) \, \lreach\, \form_2)&\hfill \textsc{[From-To]}\\
\leverywhere \form & \triangleq  & \form \, \lsurr \, \lfalse & \hfill \textsc{[Everywhere]}\\
\lsomewhere \; \form & \triangleq  & \lneg \, (\leverywhere \, \lneg \, \form) & \hfill \textsc{[Somewhere]}\\
\form_1 \, \lssurr \, \form_2 & \triangleq  & (\form_1\, \lsurr \, \form_2) \land \lneg \leverywhere \form_1 & \hfill \textsc{[Strong Surrounded]}
\end{array}
\]
\caption{\label{fig:derived} Some derived operators.}
\end{figure}

\subsubsection{Model checking \SLCS}

In this section we will briefly recall  model checking of \SLCS~\cite{CLLM14,CLLM16} over finite models.
Note that, in the context of the present paper, we are concerned with so-called {\em global} model checking,
i.e. a procedure that, given a finite model and a logic formula, returns the set of {\em all}
points in the model satisfying the formula \cite{CGP00}.
We will focus on the surrounded operator only and we will describe the related section of the model checking algorithm 
by means of an example. Model checking algorithms for the other operators of the fragment is a matter of standard routine. We will also provide a brief description of \topochecker.

\begin{figure}
{\tt\bf  Input:}\\
{\tt A model }$\model = ((X,\closure_{\arel}), \aeval, \peval)$\\ 
{\tt and a formula } $\form_1 \, \lsurr \, \form_2${\tt ;}\\\\
{\tt\bf Output:}\\
{\tt The set of points in } $X$ {\tt satisfying } $\form_1 \, \lsurr \, \form_2$;\\\\
{\tt\bf Step 1:}\\
{\tt TempBad := }$\ZET{x \in X}{\model,x \models \lneg(\form_1 \lor\form_2)}${\tt ;}\\\\
{\tt\bf Step 2:}\\
{\tt repeat}\\
{\tt Bad := TempBad;}\\
{\tt TempBad := Bad} $\cup \, ${\tt (}$\ZET{x  \in X}{\model,x \models \form_1} \cap \closure_{\arel}${\tt (Bad));}\\
{\tt until TempBad = Bad;}\\\\
{\tt\bf Step 3:}\\
{\tt return} $\ZET{x  \in X}{\model,x \models \form_1} \setminus$ {\tt Bad.}
\caption{\label{fig:MCalg} Sketch of the model checking algorithm for $\form_1 \, \lsurr \, \form_2$.}
\end{figure}

Given a finite closure model $\model = ((X,\closure_{\arel}), \aeval, \peval)$ and a formula $\form$,
the model checking algorithm returns all those points $x \in X$ such that
$\model, x \models \form$. For a formula $\form_1 \, \lsurr \, \form_2$ the algorithm, roughly speaking, first identifies  areas of {\em bad} points, that is points that can reach a 
point satisfying $\lneg \form_1$ without passing by a point satisfying $\form_2$; then
returns the points that satisfy $\form_1$  and that are not bad. A sketch of the fragment of
the model checking algorithm related to $\form_1 \, \lsurr \, \form_2$
is given in Figure~\ref{fig:MCalg}.

\begin{figure}
\centering
\subfloat[][Input model]
{
\resizebox{1.4in}{!}
{
\includegraphics{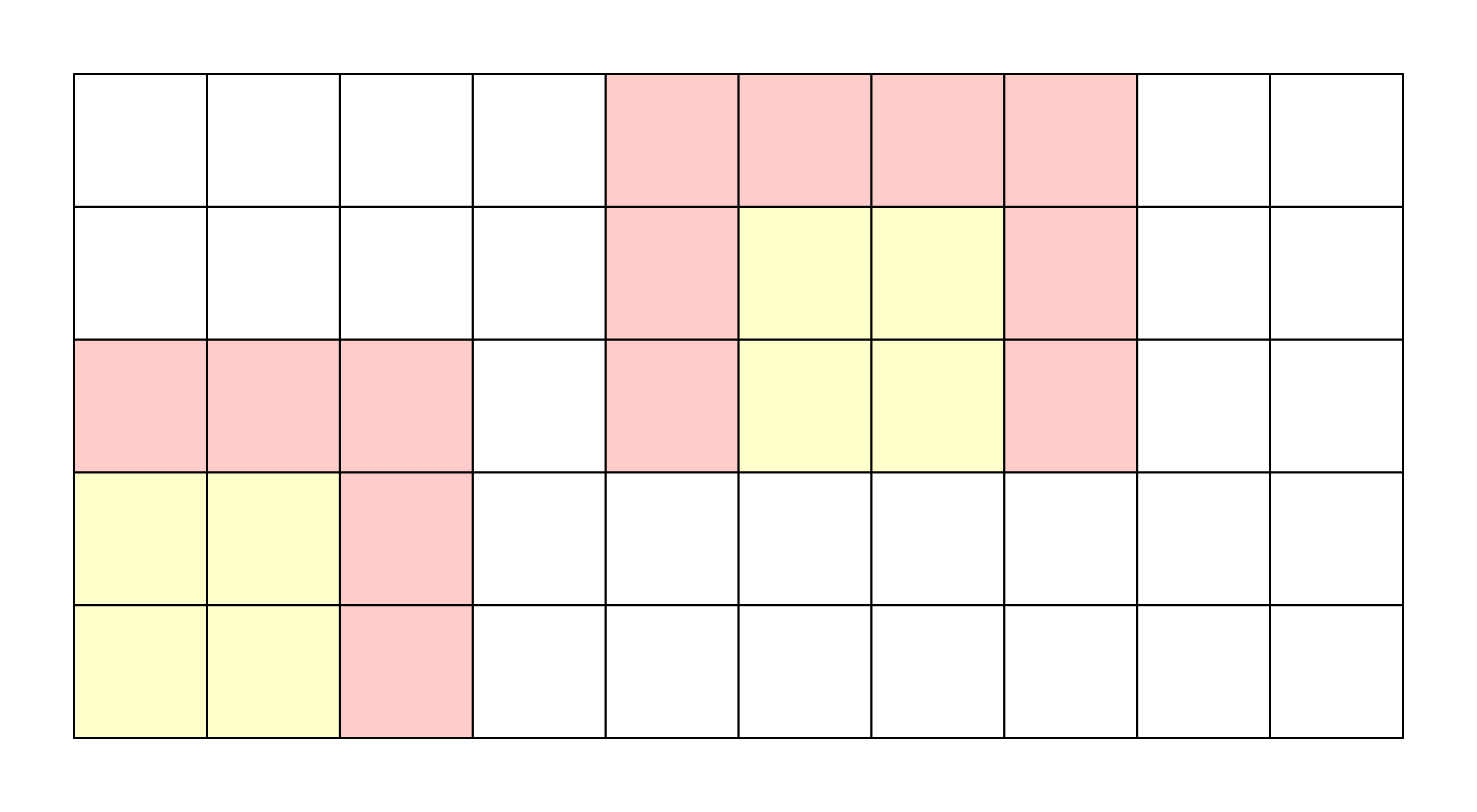}
}
\label{fig:Initialstate}
}\quad
\centering
\subfloat[][$\ZET{x}{\model,x\models \lnot(yellow \lor pink)}$]
{
\resizebox{1.4in}{!}
{
\includegraphics{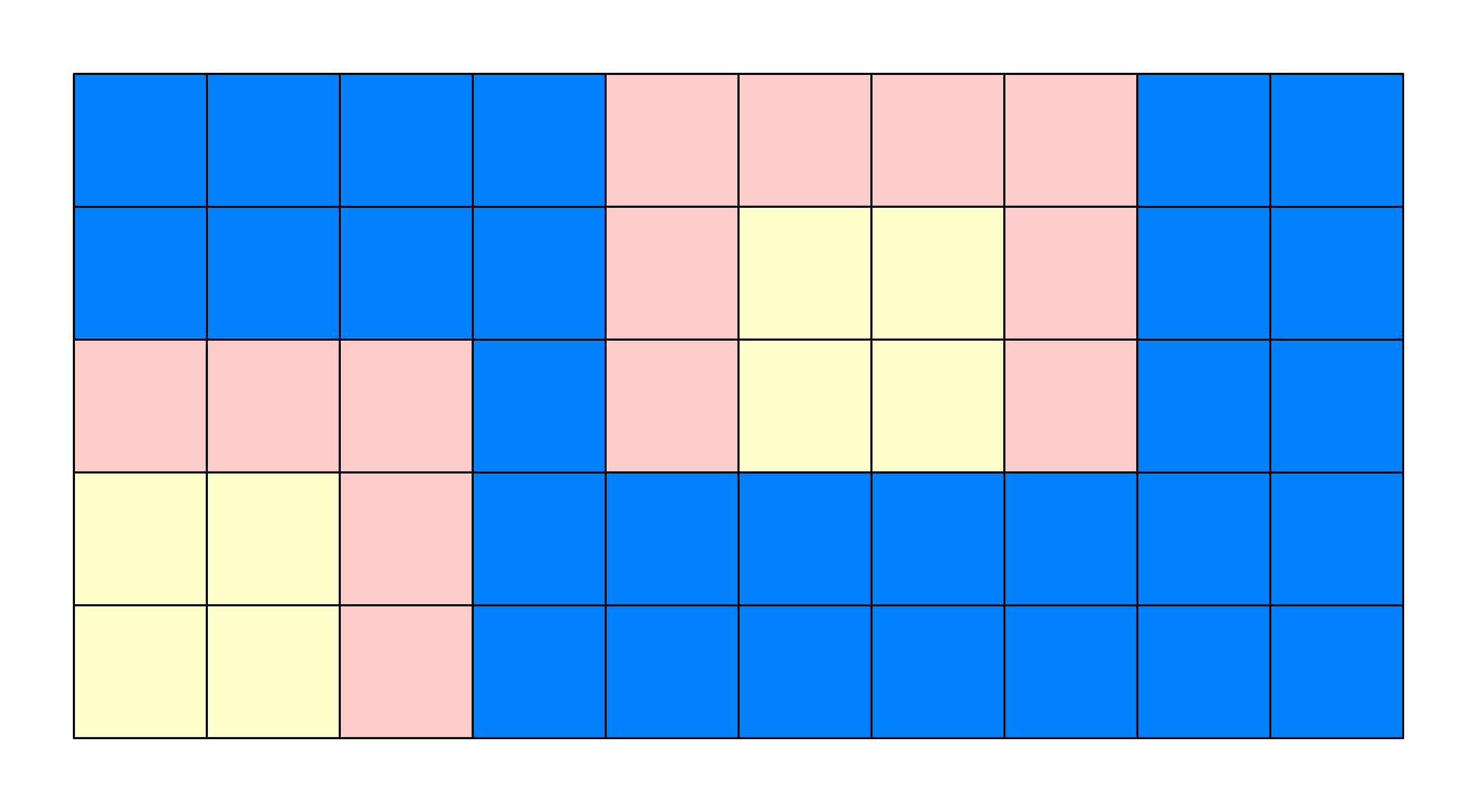}
}
\label{fig:Init}
}\\
\centering
\subfloat[][$\ZET{x}{\model,x\models yellow} \cap \closure_{\arel}(blue)$]
{
\resizebox{1.4in}{!}
{
\includegraphics{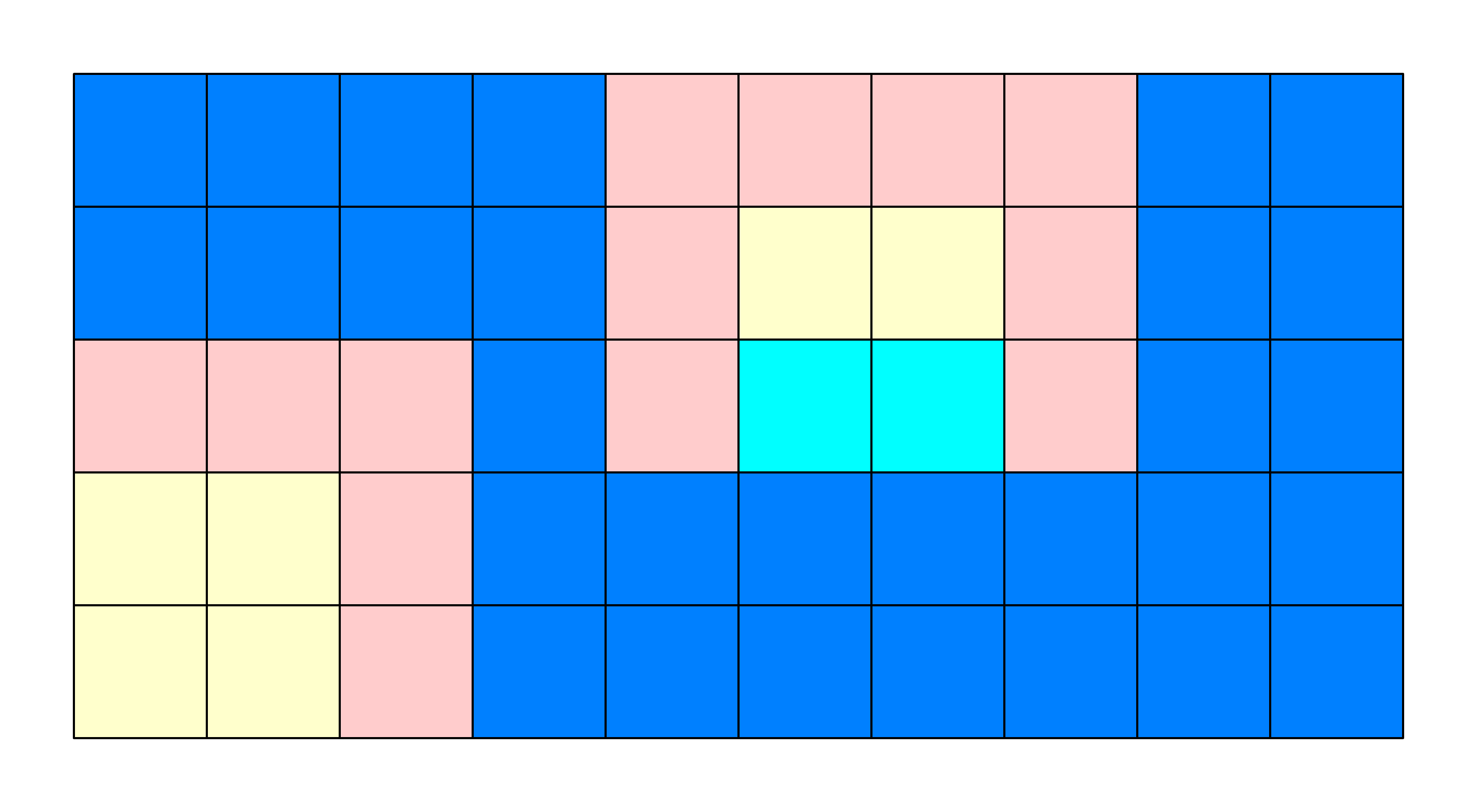}
}
\label{fig:StoneSel}
}\quad
\centering
\subfloat[][]
{
\resizebox{1.4in}{!}
{
\includegraphics{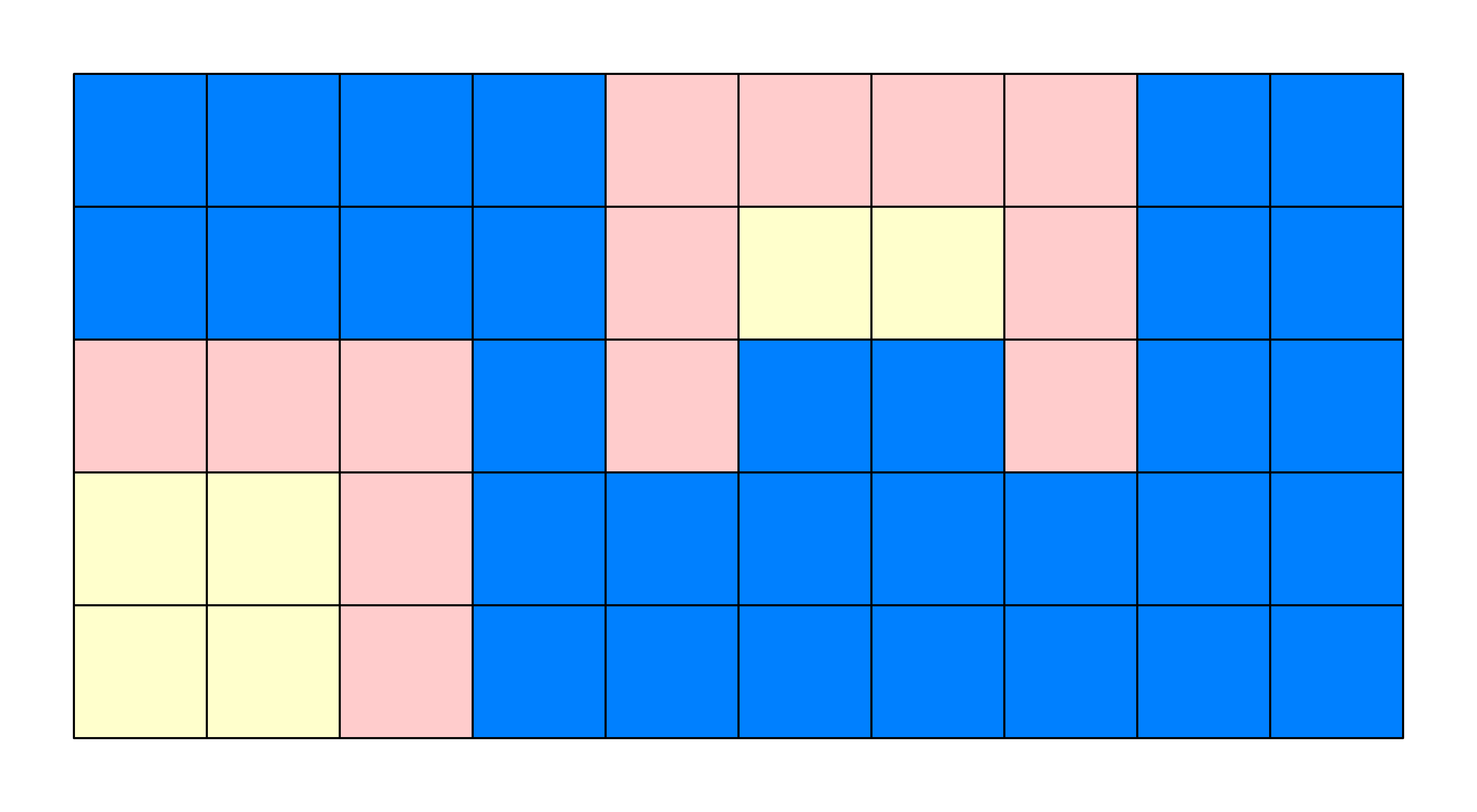}
}
\label{fig:StoneAss}
}\\
\centering
\subfloat[][$\ZET{x}{\model,x\models yellow} \cap \closure_{\arel}(blue)_{[\ref{fig:StoneAss}]}$]
{
\resizebox{1.4in}{!}
{
\includegraphics{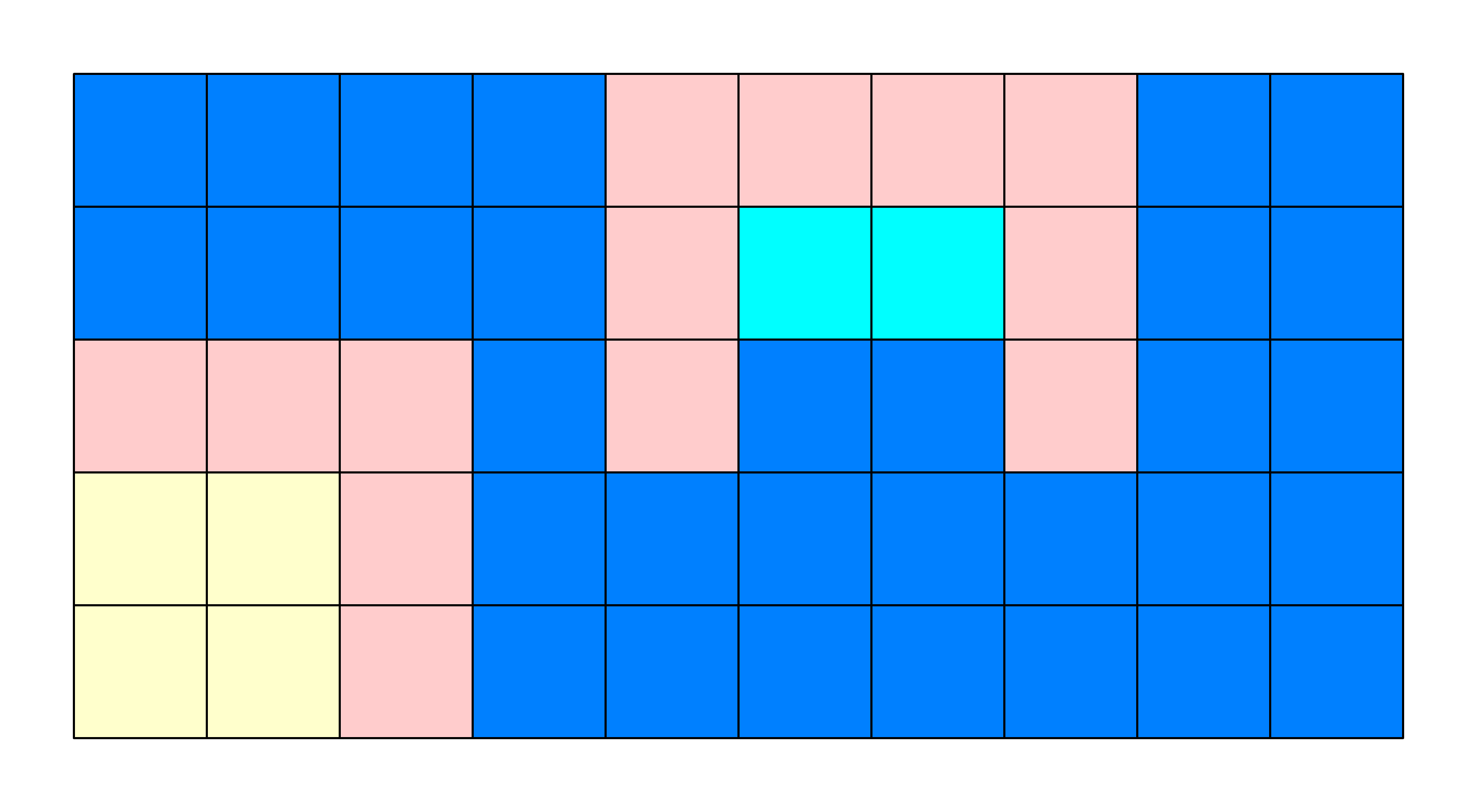}
}
\label{fig:SttwoeSel}
}\quad
\centering
\subfloat[][]
{
\resizebox{1.4in}{!}
{
\includegraphics{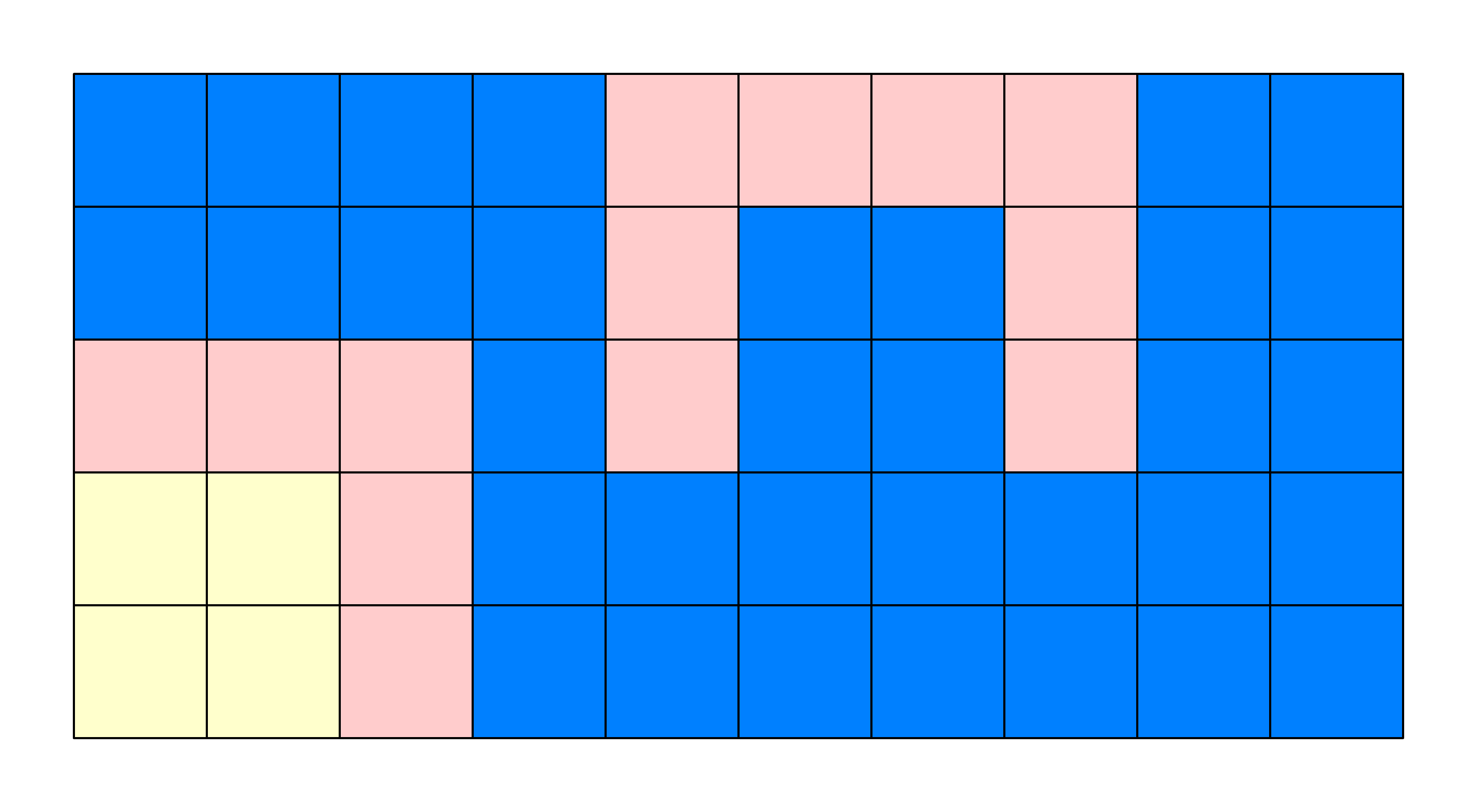}
}
\label{fig:SttwoAss}
}
\caption{Model checking $yellow \,\lsurr\, pink$.}
\label{fig:MCexample}
\end{figure}

Below, we give a brief description of how the algorithm works, using the graphs shown in
Figure~\ref{fig:MCexample}. 
Let us consider the model of Figure~\ref{fig:Initialstate} as input model, where points
are represented by coloured squares and the adjacency relation is the orthogonal one.
In this example we assume that the set of atomic predicates is the 
set $\SET{pink, yellow, white}$ --- represented
in the figure in the obvious way---and that $\peval(p) \cap \peval(p') = \emptyset$ whenever $p\not= p'$. 
Suppose the input formula is $yellow \,\lsurr\, pink$. 
The result of the assignment {\tt Bad := TempBad} of the first iteration of the {\tt repeat}
is shown in Figure~\ref{fig:Init}, where all nodes that satisfy $\lneg (yellow \lor  pink)$
are marked blue. Note that this blue-colouring is {\em not} part of the model;
we use it at a ``meta-level'' and only for describing the behaviour of the algorithm;
the same will apply to points marked in cyan in the sequel. 
In Figure~\ref{fig:StoneSel} the (only two)
$yellow$ points in the closure of the points indicated in blue,
are shown in cyan\footnote{In the caption, such a closure is abbreviated by $\closure_{\arel}(blue)$ for space reasons.}; these are the points to be selected for being added
to {\tt TempBad} in the first iteration of the {\tt repeat}. The new value of {\tt TempBad}, resulting from
the assignment, consists of all blue points of Figure~\ref{fig:StoneAss}.
The body of the {\tt repeat} is executed now with the new value of {\tt TempBad}.
In Figure~\ref{fig:SttwoeSel} the (again only two)
$yellow$ points in the closure of the set of points in blue are shown 
in cyan. Note that such a closure refers to the 
model of Figure~\ref{fig:StoneAss}; this is abbreviated in Figure~\ref{fig:SttwoeSel} as 
$\closure_{\arel}(blue)_{[\ref{fig:StoneAss}]}$. The new value of {\tt TempBad}, resulting from
the assignment, consists of all points indicated in blue in Figure~\ref{fig:SttwoAss}.
The body of the {\tt repeat} is executed now for the third time and this results in
no change in the value of {\tt TempBad}: the fixed point is reached, the 
{\tt repeat} block is exited and the points satisfying $yellow \,\lsurr\, pink$ are 
the four $yellow$ points
in the bottom-left corner of Figure~\ref{fig:SttwoAss}.
In~\cite{CLLM16} it has been shown that, for any finite closure 
model $\model = ((X,\closure_{\arel}), \aeval, \peval)$
and \SLCS formula $\form$ of size $k$, the model checking procedure terminates in 
${\cal{O}}(k\cdot(|X| + |R|))$ steps\footnote{The size of a formula is given by the number of operators in the formula: $size(p)=1; size(\lnot \form)=size(\lnear \form)=1+size(\form);
size(\form_1 \land \form_2)=size(\form_1 \lsurr \,\form_2)=1+size(\form_1)+ size(\form_2)$.}. We refer the reader to the above mentioned paper for  
further details on  \SLCS and its model checking algorithm.

The tool \topochecker is a global model checker,
capable of analysing models specified as weighted graphs, RGB images, or grayscale medical images. In the case of medical images, which is of interest in this work, the tool takes as input a file describing the spatial model to be analysed, the formulas to be checked, and a colour associated with each formula. The spatial model is described in the form of a set of images, whose intensity values are associated by the user to differently named attributes for subsequent usage in formulas. The output of the tool consists of a region of interest (ROI) for each formula to be checked, that is, an image where the specific region where such formula holds is coloured according to the user-specified colour.

Table~\ref{table:syncorr} shows the correspondence between \SLCS operators (top) 
and their syntax in \topochecker (bottom).
\begin{table}
\begin{center}
\begin{tabular}{| c | c | c | c | c | c | c | c | c | c | c |}\hline
$p$ & $\lneg$ & $\land$ & $\lnear$ & $\lsurr$ & $\lfalse$ & $\ltrue$ & $\lor$ & $\linterior$ & $\lreach$ & $\lfromto$\\\hline
\tt{[p]}& \tt{!} & \tt{\&} & \tt{N} & \tt{S} & \tt{FF} & \tt{TT} & \tt{|} & \tt{I} & \tt{R}  & \tt{T}\\\hline
\end{tabular}
\end{center}
\caption{\label{table:syncorr} \topochecker syntax.}
\end{table}
The syntax for assertions (extending the syntax for atomic predicates) is {\tt [a} $\bowtie$ {\tt c]} where  $\tt{a}$ is an attribute name, $\bowtie$ is a comparison operator  (one of \verb!=!, \verb!<!, \verb!>!, \verb!<=!, \verb!>=!) and \verb!c! is a (floating point) constant.
In \topochecker, (unnamed) assertions can be used in place of 
atomic predicates; moreover names can be given to complex formulas, by means of {\em formula definitions} as in the example shown below:
\begin{verbatim}
Let adipose = N (N [FLAIR>1.7]);
\end{verbatim}
formula  {\tt N (N [FLAIR>1.7])} is given  the name
{\tt adipose} that can be used in other formulas; the formula
exemplifies using assertion {\tt FLAIR>1.7} in place of an atomic predicate.

In general, names introduced by formula definitions may also have parameters such as
\begin{verbatim}
Let f(f1, f2,..., fn) = F
\end{verbatim}
where {\tt F}  is a formula that can use names {\tt f1,f2, \ldots, fn}, 
that are instantiated to formulas when {\tt f} is invoked. 

The spatial model checking algorithm is entirely run in central memory, aiming at fast \emph{interactive} usage. The algorithm proceeds by induction on the structure of formulas, and uses memoization to cache the intermediate result on each sub-formula, so that when the same sub-formula is used more than once, results are reused. The cache is also stored on-disk, leveraging incremental design of complex formulas across different model checking sessions. The tool is implemented in the functional programming language OCaml\footnote{See \url{http://www.ocaml.org}.}, which provides a good balance between declarative features and computational efficiency. The main loop of the algorithm has been carefully written to avoid memory allocation and garbage collection in most cases, via the use of the ``bigarray'' data type, providing direct access to memory locations and memory-mapping of large files stored on the hard drive (such as medical images). All the arrays needed for the computation are statically allocated prior to model checking execution. Such optimisations result in  competitive execution times, such as those described in Section \ref{subsec:validation}. Indeed, since no memory swap takes place, one should take into account the memory requirements of each analysis, that could render in-memory execution unfeasible; however, the size of a typical medical image is in the order of some megabytes, which is orders of magnitude smaller than the available memory on modern computers (some gigabytes); in our experiments, the algorithm never ran out of memory.

\subsection{Incorporating Distance}

Models of space as well as spatial logics can be  extended with notions of {\em distance}
(see e.g. \cite{Ch9HBSL,KWSSZ03,NB14,NBCLM15}).  Distances are very often
expressed using the non-negative real numbers $\reals_{\geq 0}$, like the Euclidean distance on continuous space. 

For quasi-discrete closure spaces, especially when used as a representation of finite graphs, it is 
natural to consider \emph{shortest path distance}, where a path between two nodes 
is a sequence of consecutive edges connecting the first node to the second, and its length
is given by the sum of the lengths of such edges. The length of an edge is often taken to be $1$; however, other notions of distance can be more
appropriate. For example, \emph{sampling} a multi-dimensional Euclidean space is often
done using a \emph{regular grid}, that is, a graph in which the nodes are arranged
on multiples of a chosen \emph{unit interval} that may vary along each dimension of the space. 
Nodes are connected by edges using a chosen notion of adjacency (e.g. the orthogonal or orthodiagonal adjacency relations discussed before, but any choice may be reasonable, depending on the application context). Such graphs can then be \emph{weighted} by associating to each edge the Euclidean distance between the nodes it connects. Graphs with nodes in an Euclidean space and weighted by Euclidean distance are known as \emph{Euclidean graphs} and are naturally equipped with both Euclidean distance between nodes and (weighted) shortest-path distance---which is also called \emph{Chamfer distance} in the particular case of Euclidean graphs with nodes arranged on a regular grid, which is the case of interest for MI.
In two-dimensional imaging, pixels---with an application-dependent choice of adjacency---form 
an Euclidean graph,  and Euclidean distance is the reference distance between (the centres of) two pixels.

\begin{figure}
\centering
\includegraphics[height=2.8cm]{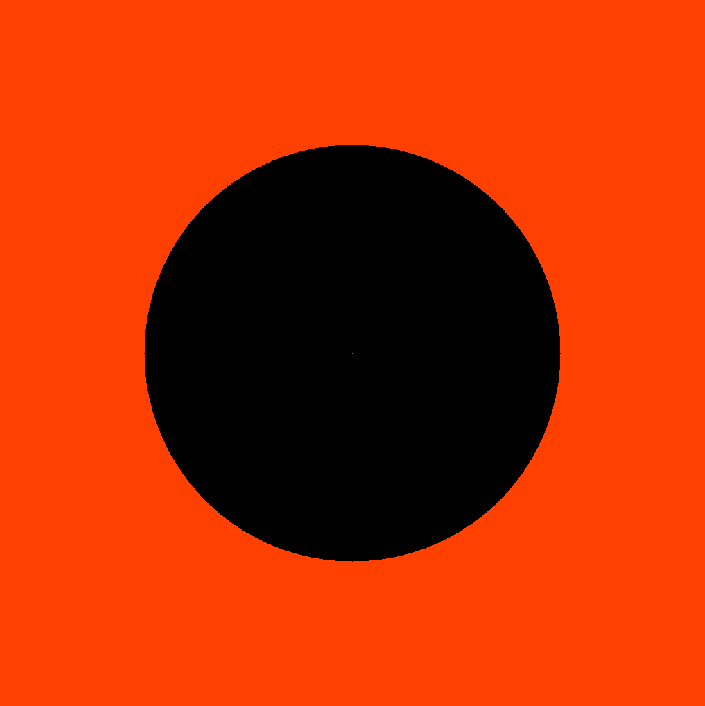}
\caption{Threshold imposed on Euclidean distance from a  point in the centre of image.}
\label{fig:EDST}
\end{figure}

\begin{figure}
\centering
\subfloat[][Chamfer distance with 8 adjacent pixels per node ($3\times 3$ square centred on node).]
{
\includegraphics[height=2.8cm]{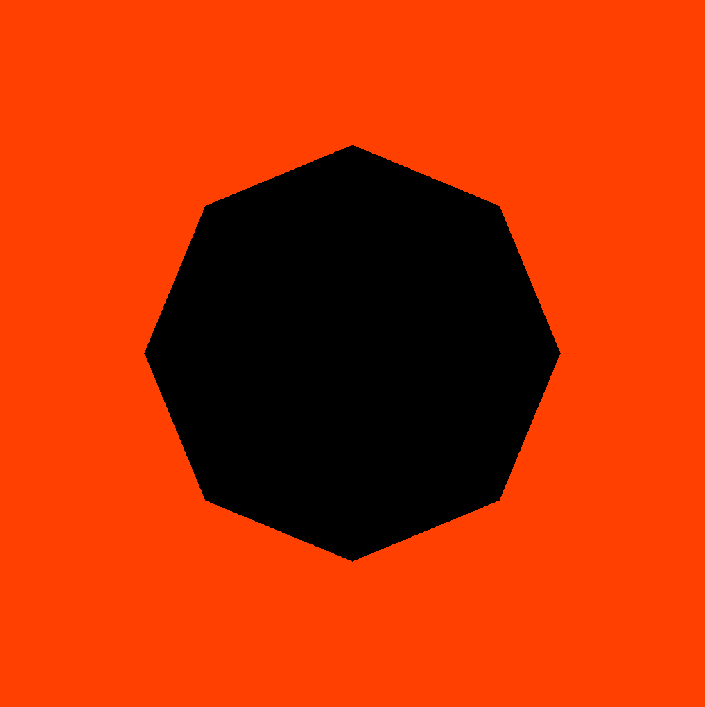}
\label{fig:MDDTw1}
}\quad
\centering
\subfloat[][Percentage error. Scale: 0-10.]
{
\includegraphics[height=2.8cm]{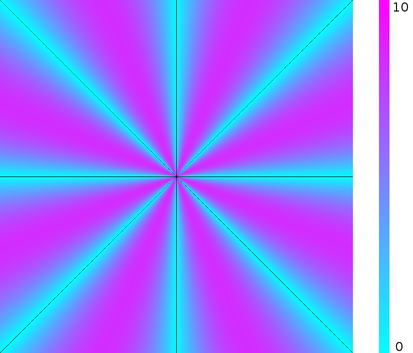}
\label{fig:MDDTw1-err}
}\\
\subfloat[][Chamfer distance with 24 adjacent pixels per node ($5\times 5$ square centred on node).]
{
\includegraphics[height=2.8cm]{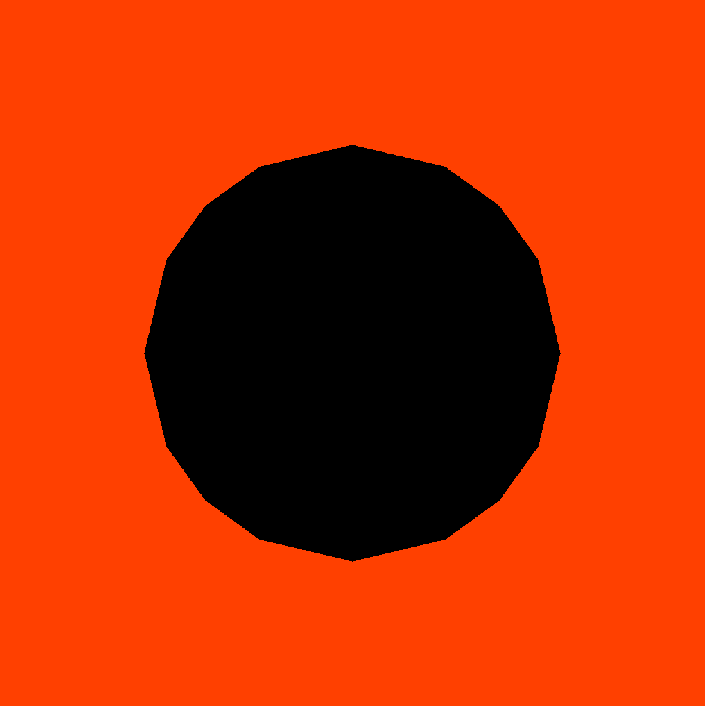}
\label{fig:MDDTw2}
}\quad
\centering
\subfloat[][Percentage error. Scale: 0-2.]
{
\includegraphics[height=2.8cm]{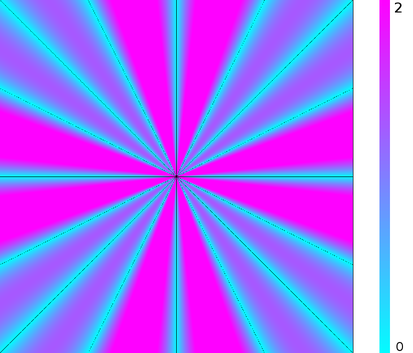}
\label{fig:MDDTw2-err}
}
\caption{Percentage error of Chamfer distance.}
\label{fig:CDST}
\end{figure}

Euclidean and Chamfer distances obviously divert, no matter how fine is the grid or how many neighbours are chosen in the adjacency relation, unless \emph{all} pairs of nodes are linked by an edge (labelled with the Euclidean distance between the
end-points of the edge).
Therefore, in this context, Euclidean distance is considered \emph{error-free}, and Chamfer 
distance is considered an approximation of the former. 
The chosen  adjacency relation determines the precision-efficiency trade-off of the computed distance:
the more pixels are considered adjacent, the more precise is the approximation, at the expenses of
generating graphs with larger out-degrees. 
This is illustrated in Figures~\ref{fig:EDST}  and~\ref{fig:CDST}. In the first figure we show a two-dimensional, rectangular image where   all and only points  at a  Euclidean distance larger than a given threshold $k$ from the centre of the figure are coloured in red. In Figure~\ref{fig:MDDTw1} the points in red are those at a Chamfer distance larger than $k$ from the centre; in particular, orthodiagonal adjacency has been used (each pixel has 8 other adjacent pixels). Figure~\ref{fig:MDDTw1-err} shows the percentage of error for each pixel with respect to the Euclidean distance, in a scale 
from $0$ to $10\%$. 
Finally, in Figure~\ref{fig:MDDTw2} we use Chamfer distance, the same threshold $k$ and an adjacency relation where each pixel has 24 other adjacent pixels (i.e. the pixels that are adjacent to
any pixel form a $5 \times 5$ square centred in the pixel). Figure~\ref{fig:MDDTw2-err} shows the percentage of error w.r.t. the Euclidean distance, in a scale from $0$ to $2\%$.
The percentage error $\delta(x)$ for Chamfer distance $d_C$ is
defined for each pixel $x$ as
$\delta(x)=\frac{\left|d_E(y,x)-d_C(y,x)\right|}{d_E(y,x)}$, where $y$ is the central point of the image
and $d_E$ denotes the Euclidean distance.

\subsubsection{\SLCS with distance operators}

In this section we extend the fragment of \SLCS presented in Section~\ref{sec:FSLCS}
with logic {\em distance} operators. We first introduce the notion of {\em distance
closure spaces} and, to that purpose we recall the well-known notion of metric space:

\begin{definition}\label{def:MetricSpace}
A {\em metric space}  is a pair $(X,d)$ where $X$ is a non-empty set (of {\em points}) and 
$d: X \times X \to \reals_{\geq 0}$ is function that satisfies the following axioms, for all $x,y,z \in X$:
\begin{enumerate}
\item $d(x,y)=0$ iff $x=y$ \textsc{[identity of indiscernible]};
\item $d(x,y)=d(x,y)$ \hfill \textsc{[symmetry]};
\item $d(x,z) \leq d(x,y) + d(y,z)$ \hfill\textsc{[triangle inequality]}.
\end{enumerate}

Whenever $X$ is a closure space $(X,\closure)$,  $((X,\closure),d)$ is called a {\em metric closure space}\ed
\end{definition}

Metric functions are easily extended to sets as follows:

\begin{definition}
Given metric space $(X,d)$, $x \in X$ and $Y, Z \subseteq X$ we let
\begin{enumerate}
\item $d(x,Y) = \inf \ZET{d(x,y)}{y \in Y}$
\item $d(Y,Z) = \inf \ZET{d(y,z)}{y \in Y \mbox{ and } z \in Z}$
\end{enumerate}
Note that if $Y\not=\emptyset$ is  finite, then $\inf \ZET{d(x,y)}{y \in Y} = \min\ZET{d(x,y)}{y \in Y}$; we let $d(x,\emptyset) =\infty$ by definition.
\ed
\end{definition}


In the case of quasi-discrete closure spaces, symmetry 
may turn out to be too much restrictive. This is for instance the case when the relation $\arel$ underlying
the closure operator $\closure_{\arel}$ is not symmetric. Similarly,  the triangle inequality 
is not fitting well when more qualitative distance ``measures'' are used, for instance  when the 
codomain of $d$ is composed of only three values, representing {\em short}, {\em medium}, and {\em large} distance
respectively. For all these reasons, for quasi-discrete closure spaces we often use the less restrictive 
notion of {\em distance space}, where only Axiom 1 of Definition~\ref{def:MetricSpace} above is required. 

\begin{definition}
\hspace{0.15in} A {\em distance closure space}  is a tuple $((X,\closure),d)$ where 
$(X,\closure)$ is a  closure space and $d: X \times X \to \reals_{\geq 0} \cup \SET{\infty}$ is function satisfying $d(x,y)=0$ iff $x=y$.\\
A {\em  quasi-discrete distance closure space} is a  distance closure space
$((X,\closure_{\arel}),d)$ where $(X,\closure_{\arel})$ is a quasi-discrete closure space.
\ed
\end{definition}

Distance operators can be added to spatial logics in various ways (see
\cite{Ch9HBSL} for an introduction). For the purposes of the present paper, we extend \SLCS with 
the operator $\dist{I}$, where $I$ is an interval of $\reals_{\geq 0}$.
The intended meaning is that a point $x$ of a distance  closure model
satisfies $\dist{I} \, \form$ if its distance from the set of points satisfying $\form$
falls in interval $I$. Below we provide the necessary formal definitions.

\begin{definition}\label{def:dmodel}
\hspace{0.15in} A {\em   distance closure model}  is a tuple 
$(((X,\closure),d), \aeval, \peval)$ consisting of a  distance closure space
$((X,\closure),d)$, a valuation $\aeval: X \times \attrib \to V$, assigning to each  point and attribute  the value of the attribute of the point and a valuation $\peval: \props \to 2^X$ assigning to
each atomic predicate the set of points where it holds.\\
A {\em  quasi-discrete distance closure model} $(((X,\closure_{\arel}),d), \aeval, \peval)$ is a distance closure model where $((X,\closure_{\arel}),d)$ is a 
quasi-discrete distance closure space.
\ed
\end{definition}

As the definition of $d$ might require the elements of $\arel$ to
be weighted---as in the case of Euclidean graphs---quasi-discrete distance closure models
are often enriched with a $\arel$-weighting function $\weval:\arel \to \reals$
assigning the weight $\weval(x,y)$ to each $(x,y)\in \arel$.  

The satisfaction relation of our fragment of \SLCS is extended as expected:

\begin{definition}\label{def:dsatisfaction}
{\em Satisfaction} $\model, x \models \form$ at point $x \in X$ in distance closure model
$\model = (((X,\closure),d), \aeval, \peval)$ is defined by induction on the structure of formulas, by adding the equation below to those in Figure~\ref{fig:fragSLCSsem}:
$$
\model, x \models \dist{I} \, \form \, \Leftrightarrow\,
d(x, \ZET{y}{\model, y \models \form}) \in I
$$
\end{definition}

Note that the definition of the \SLCS distance operator is {\em parametric} on 
the specific distance used. The particular meaning of the distance operator
is fully characterised by the specific distance $d$ of the underlying distance closure model.
In this paper, we will use the Euclidean distance $d_E$ and the Chamfer distance $d_C$.

We close this section with the definition of an additional set of derived operators  shown in Figure~\ref{fig:morederived}. 

\begin{figure}[h!]
\newcolumntype{L}{>{$}l<{$}}
\[
\begin{array}{l c l L}
\dist{\, < c} \, \form & \triangleq  &   \dist{[0,c)} \, \form \\
\dist{\, \leq c} \, \form & \triangleq  &   \dist{[0,c]} \, \form \\
\dist{\, = c} \, \form & \triangleq  &   \dist{[c,c]} \, \form \\
\dist{\, \geq c} \, \form & \triangleq  &   \dist{[c,\infty)} \, \form \\
\dist{\, > c} \, \form & \triangleq  &   \dist{(c,\infty)} \, \form \\
\lflt{c} \form & \triangleq & \dist{< c}(\lnot \dist{<c} \lnot \form)\\
\form_1 \lbsurr{I} \form_2 & \triangleq & ((\form_1 \, \land \, \lneg\form_2) \, \lssurr \, \form_2) \land \dist{I}\, \form_2 
\end{array}
\]
\caption{\label{fig:morederived} Additional derived operators.}
\end{figure}

\begin{figure}
\centering
{
\includegraphics[height=3cm]{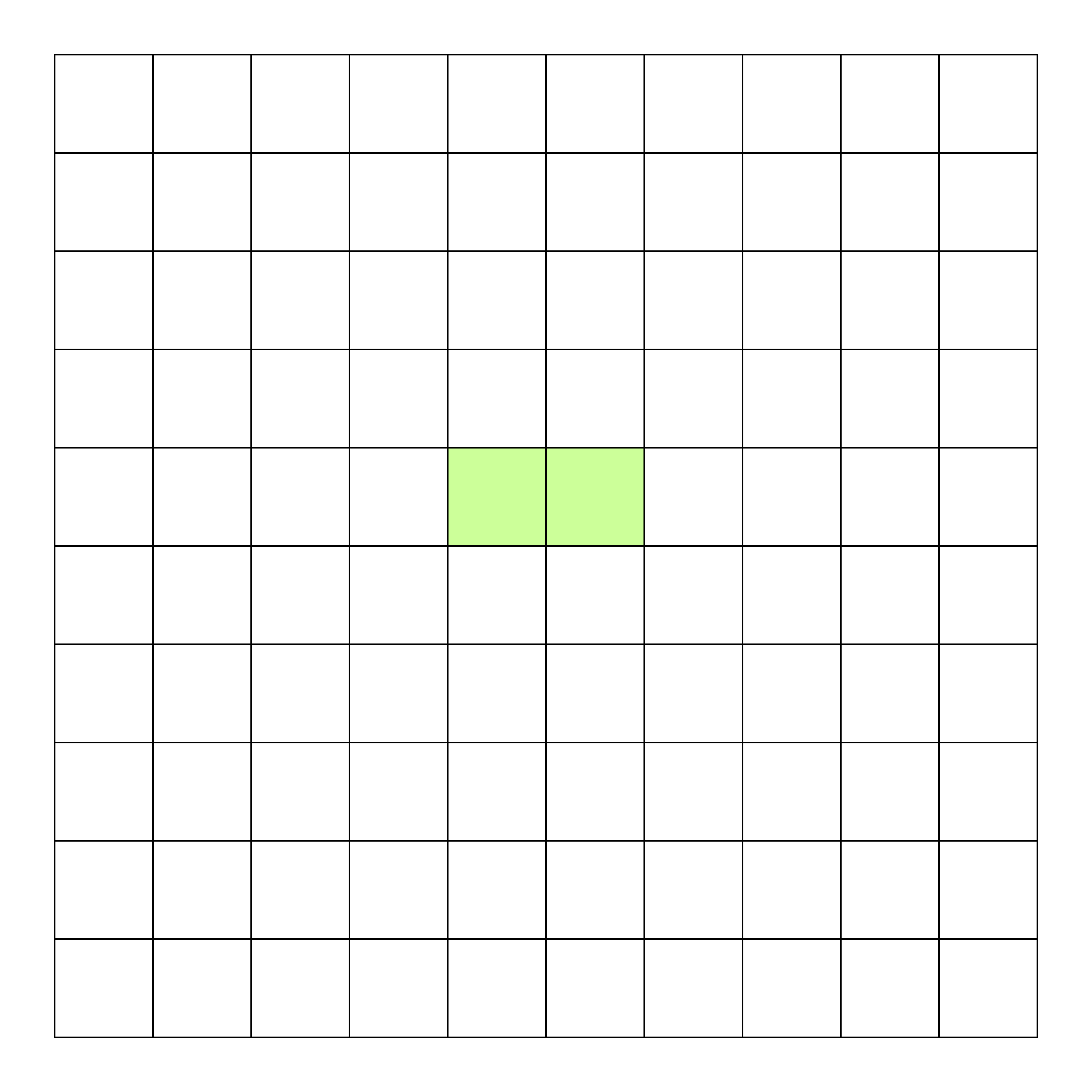}
}
\caption{The  points in Figure~\ref{fig:exSample} satisfying $\dist{\, > 2} \, p$ are shown in green.}
\label{MDDT_g_2}
\end{figure}

Again, with reference to Figure~\ref{fig:exSample}, Figure~\ref{MDDT_g_2} shows in green all the points satisfying $\dist{\, > 2} \, p$, according
to the Chamfer distance. 

Intuitively, the $\lflt{c}$ operator can be used as a form of filtering, eliminating small details caused by noise in the fine-scale structure of an image; this method is akin to 
the nested application of $\lnear$ and $\linterior$ described in Section \ref{sec:FSLCS}, parameterised with respect to a 
chosen maximum size $c$ of details to be suppressed. To see this, recall that $\linterior \Phi$ is defined as $\lnot \lnear \lnot \Phi$, therefore $\lnear \linterior \Phi$ is the same as $\lnear (\lnot \lnear \lnot \Phi)$, which is very similar to the definition 
of $\lflt{c}$, with $\lnear$ replaced by $\dist{<c}$.

\begin{figure}
\centering
\subfloat[][]
{
\includegraphics[height=3cm]{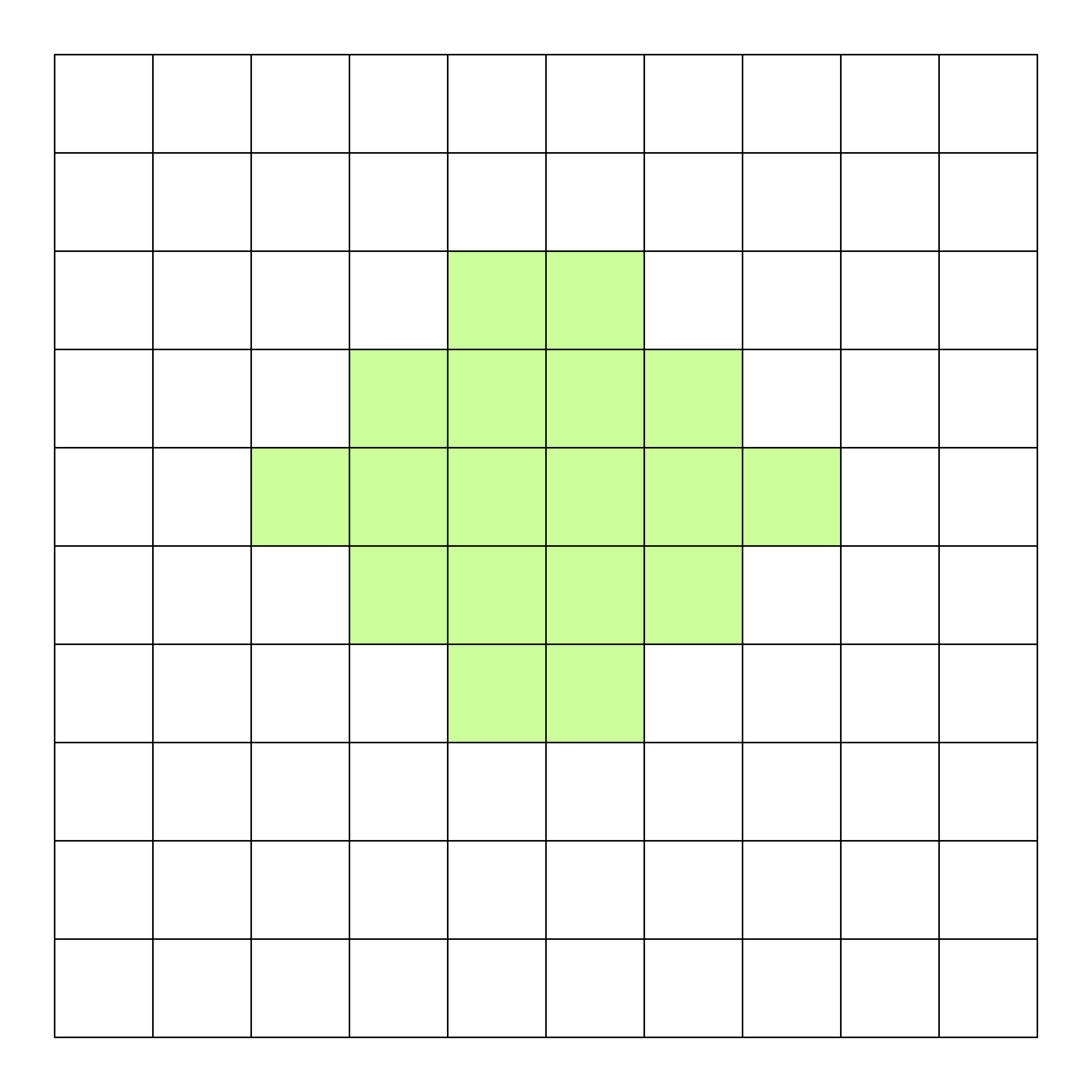}
\label{fig:exFltThree}
}\quad
\centering
\subfloat[][]
{
\includegraphics[height=3cm]{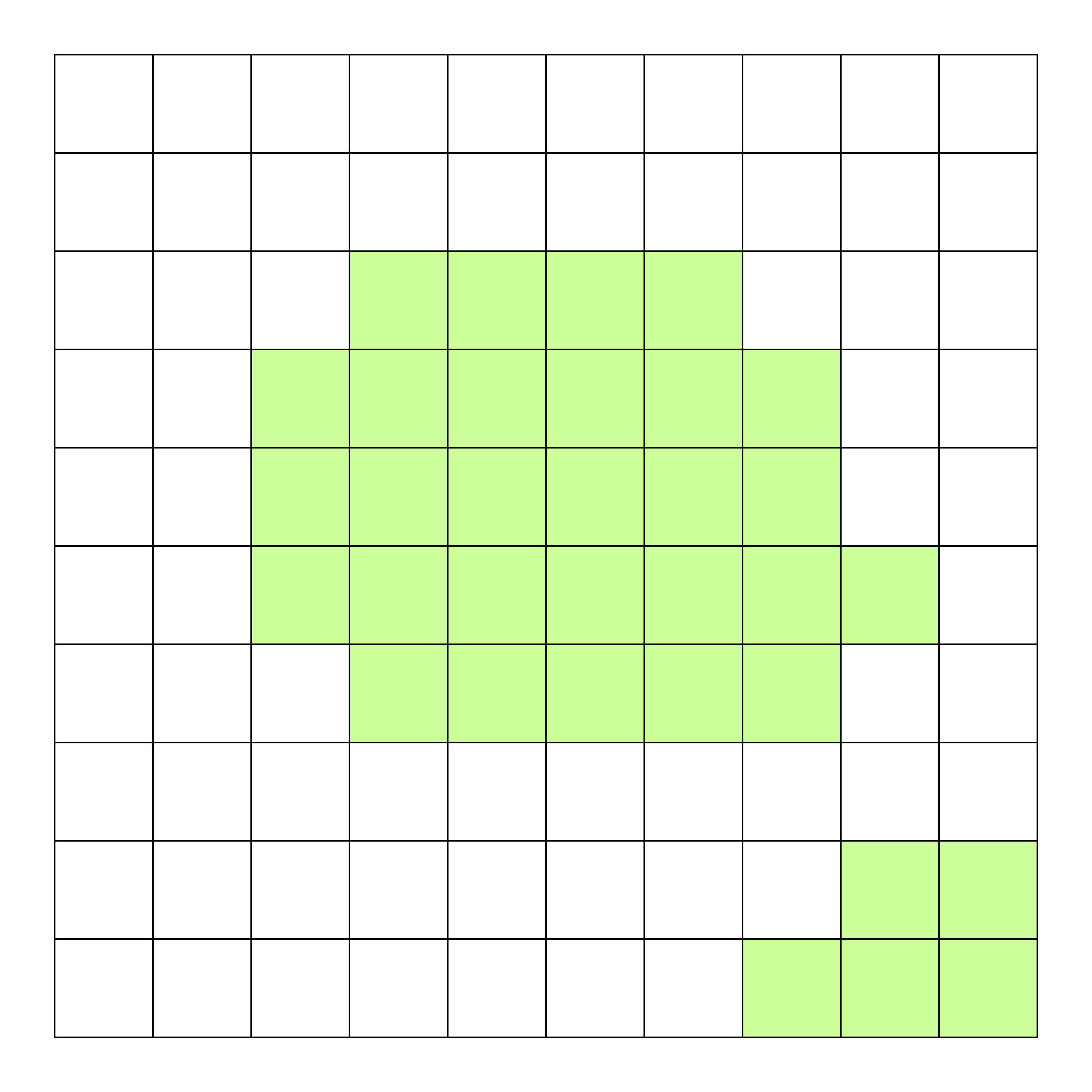}
\label{fig:exFltTwo}
}\\
\caption{The points in Figure~\ref{fig:exSample} satisfying $\lflt{3} q$ (a) and those satisfying $\lflt{2} q$ (b) are shown in green.}
\label{fig:exFlt}
\end{figure}
With reference to Figure~\ref{fig:exSample}, Figure~\ref{fig:exFltThree} shows in green the points satisfying $\lflt{3} q$, whereas those satisfying $\lflt{2} q$ are given in Figure~\ref{fig:exFltTwo}.

The {\em bounded surrounded operator} $\form_1 \lbsurr{I} \form_2 $ is satisfied by a point $x$ if and only if 
$x$ satisfies $\form_1$, is strongly surrounded by points satisfying $\form_2$ and its distance from such points 
falls in interval $I$. Note that, in the first argument of  $\lssurr$, it is required that $\lneg\form_2$ holds as well;
this ensures that all $\form_2$-points are at a distance of at least $\inf I$ from $x$.

\begin{figure}
\centering
\subfloat[][]
{
\includegraphics[height=3cm]{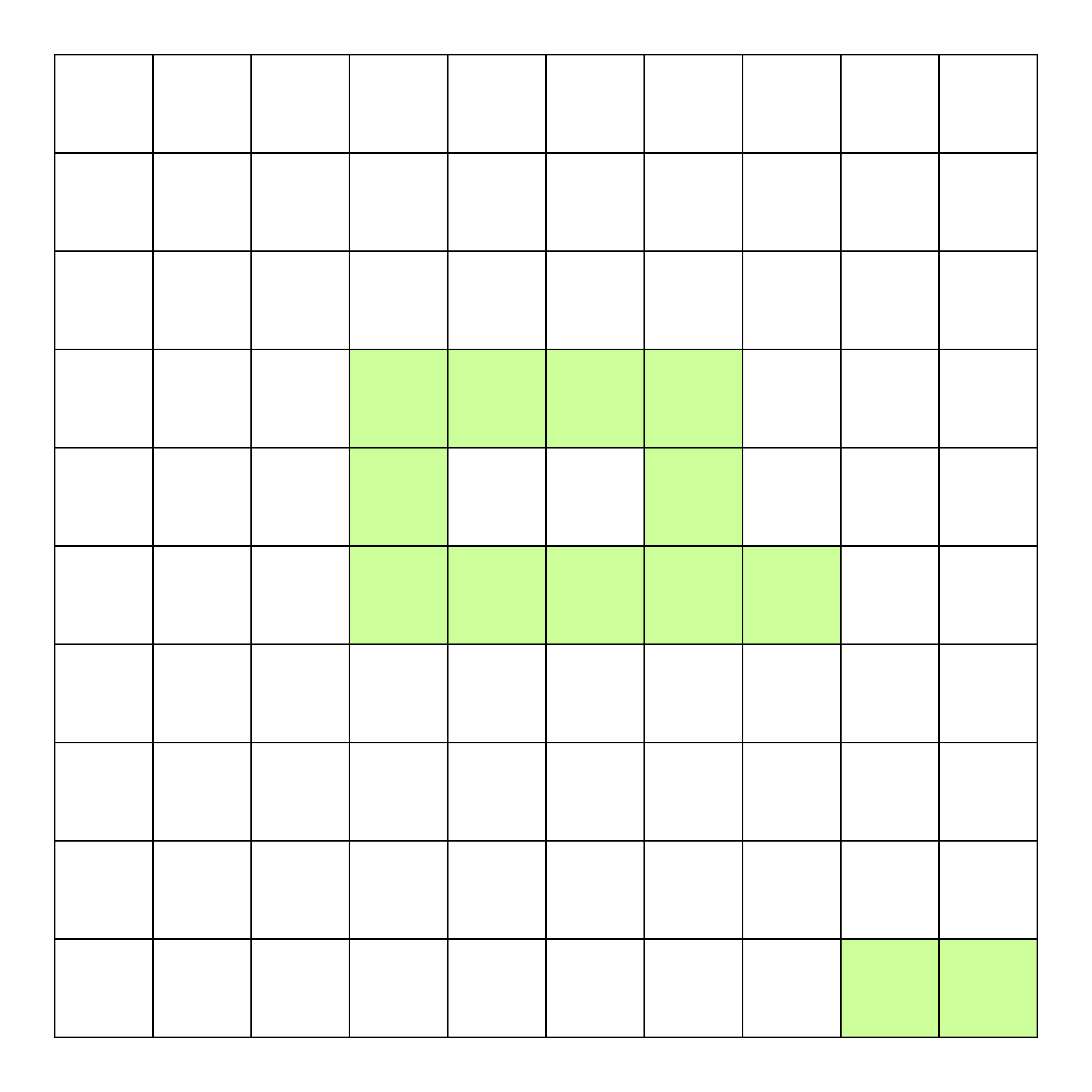}
\label{fig:exbSurTwoTwo}
}\quad
\centering
\subfloat[][]
{
\includegraphics[height=3cm]{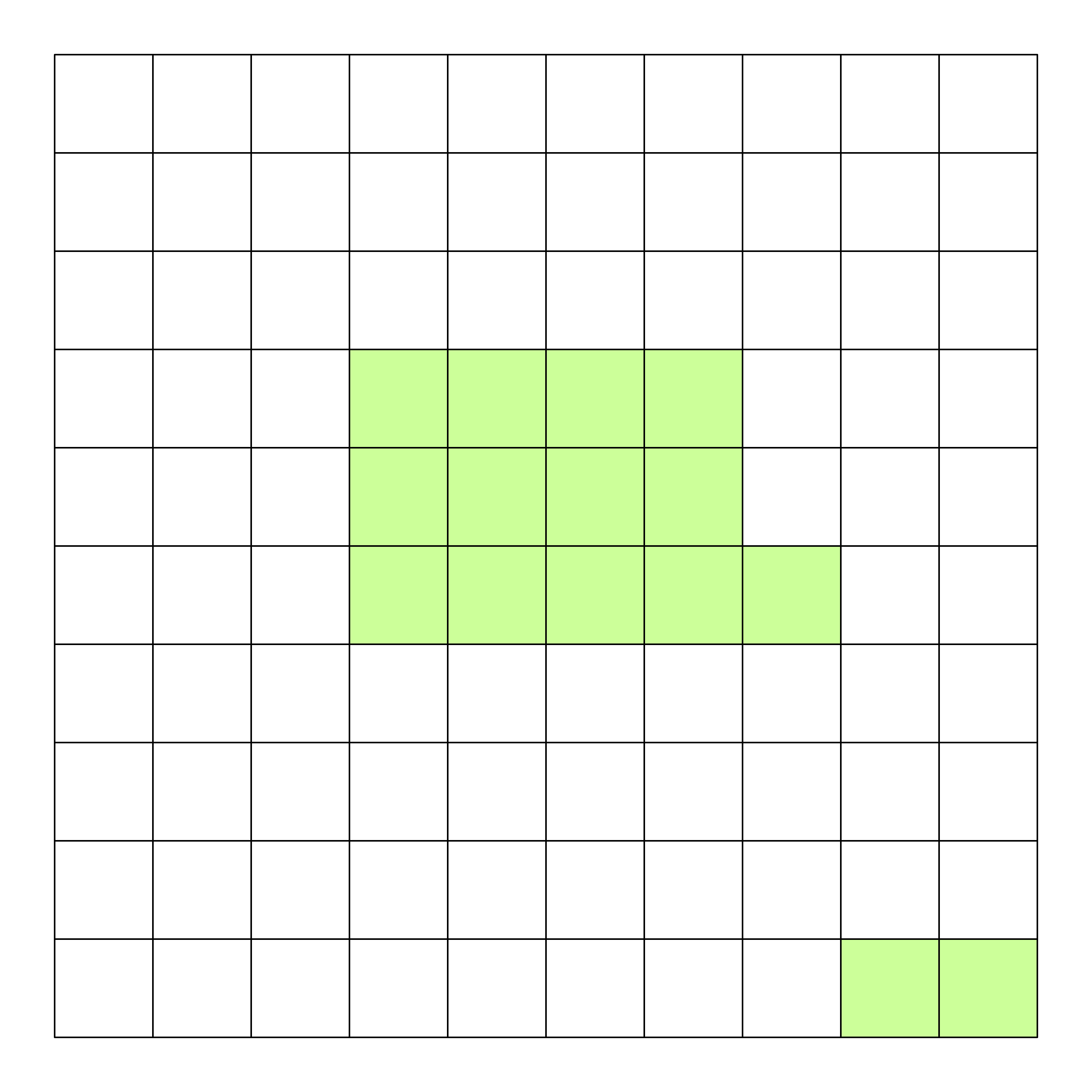}
\label{fig:exbSurTwoThree}
}\\
\caption{The points in Figure~\ref{fig:exSample} satisfying $q\, \lbsurr{[2,2]} \, p$ (a) and those satisfying $q\, \lbsurr{[2,3]} \, p$ (b) are shown in green.}
\label{fig:exbSur}
\end{figure}

In Figure~\ref{fig:exbSurTwoTwo} (Figure~\ref{fig:exbSurTwoThree}, respectively) the points satisfying $q\, \lbsurr{[2,2]} \, p$ ($q\, \lbsurr{[2,3]} \, p$, respectively) are shown in green.
%
Note that  a similar operator has been defined in~\cite{NBCLM15}, which turns out to be stronger than $\lsurr^{[a,b]}$, i.e., denoting the former operator by $\hat{\lsurr}^{[a,b]}$, we have that, for all formulas $\form_1, \form_2$, 
$
(\form_1 \land \lneg\form_2) \, \lNbsurr{[a,b]} \, \form_2
$
implies
$
\form_1 \lbsurr{[a,b]} \form_2
$.
%

We close this section pointing out that, for finite models, the operator 
$\dist{\leq c}$ coincides with the operator $\exists^{\leq c}$ proposed in~\cite{SWZ10}:
$$
\model,x \models \exists^{\leq c} \form \Leftrightarrow 
\exists y. d(x,y) \leq c \mbox{ and } \model,y \models \form
$$
and similarly for $\dist{< c}$ and $\exists^{< c}$.
Note that, this coincidence does not hold in general, e.g. for Euclidean spaces.
Our choice of the specific distance operator is motivated by its natural compatibility 
with distance transforms, as we illustrate in Section~\ref{subsec:distance} below.

\subsubsection{Model checking \SLCS with distance operators}
\label{subsec:distance}

For distance-based operators, generally speaking, the time complexity of naive model checking algorithms is quadratic in the size of the space (see \cite{NBCLM15} for an example). 
However, given a Euclidean graph representing a multi-dimensional image, spatial model checking of formulas $\dist{I} \form$ for Euclidean or Chamfer distance 
can be done in linear time or quasi-linear time, respectively, with respect to the
number of points of the space. This is achieved via so-called \emph{distance transforms}, 
that are one of the subjects of topology and
geometry in computer vision \cite{KKB96}, and are extensively used
in modern image processing and computer graphics~\cite{Ciesielski2010}.  
In particular, effective linear-time
algorithms have been recently introduced in the literature~\cite{MQR03,SURVEYDT}. 
Basically, a {\em distance transform} takes
a model $\model_{in}$ as input and produces a model $\model_{out}$ as output as follows.
Let $\model_{in}$ be a model $(((X,\closure_{\arel}),d), \aeval_{in}, \peval,\weval)$ such that every point $x \in X$ has
an attribute, say $a_{in}$, defined on a binary domain, say $\SET{0,1}$---the value of such an attribute
may represent the fact that the point satisfies a given formula $\form$ or not. The 
output model will be $\model_{out} = (((X,\closure_{\arel}),d), \aeval_{out}, \peval,\weval)$
such that every point $x \in X$ has an attribute, say $a_{out}$, and 
$\aeval_{out}(x,a_{out})=d(x,\ZET{y\in X}{\aeval_{in}(y,a_{in})=1})$. 
The fragment of the model checking algorithm related to $\dist{I} \form$ 
is  sketched in Figure~\ref{fig:DMCalg}.
During the first step, the points satisfying $\form$ are marked, so that in the second step
each point is annotated with its distance from the set of marked points and, finally,
the set of the points with a distance laying in the interval $I$ is returned.

\begin{figure}
{\tt\bf  Input:}\\
{\tt A model }$\model = ((X,\closure_{\arel}), \aeval, \peval)${\tt ;}\\ 
{\tt and a formula } $\dist{I} \; \form${\tt ;}\\\\
{\tt\bf Output:}\\
{\tt The set of points in } $X$ {\tt satisfying } $\dist{I} \; \form${\tt ;}\\\\
{\tt\bf Step 1:}\\
{\tt Compute intermediate model } $\model' = ((X,\closure_{\arel}), \aeval', \peval)$ {\tt such that for all } $x \in X${\tt :}\\
$\aeval'(x,a') = 1$ {\tt if } $\model, x \models \form$ {\tt and }\\
$\aeval'(x,a') = 0$ {\tt if } $\model, x \not\models \form$\\\\
{\tt\bf Step 2:}\\
{\tt Compute intermediate model } $\model'' = ((X,\closure_{\arel}), \aeval'', \peval)$ {\tt such that for all } $x \in X${\tt :}\\
$\aeval''(x,a'') = d(x,\ZET{y\in X}{\aeval'(y,a')=1})${\tt ;}\\\\
{\tt\bf Step 3:}\\
{\tt return} $\ZET{x\in X }{\aeval''(x,a'') \in I}${\tt .}
\caption{\label{fig:DMCalg} Sketch of the model checking algorithm for $\dist{I} \; \form$.}
\end{figure}

In \topochecker, two of the standard algorithms for distance transform are currently implemented;
one for the Euclidean distance $d_E$ and the other for the Chamfer distance $d_C$.
Correspondingly, two distance operators are provided, with syntax as in Table~\ref{table:syndist}.
\begin{table}
\begin{center}
\begin{tabular}{| c | c | c | c | c | c | c | c | c | c | c |}\hline
$\dist{\bowtie c}$ & $\dist{\bowtie c}$\\
based on $d_E$ & based on $d_C$\\\hline
\EDT($\_, \bowtie$ {\tt c})& \MDDT($\_, \bowtie$ {\tt c}) \\\hline
\end{tabular}
\end{center}
\caption{\label{table:syndist} \topochecker syntax for distance operators.}
\end{table}

For Euclidean distances, \topochecker uses the linear algorithm that
was proposed by Maurer in \cite{MQR03}. Such algorithm computes
Euclidean distance transforms on anisotropic multi-dimensional grids
(such as 2D and 3D medical images); it has linear complexity, its run-time is
predictable, and it is among the most efficient algorithms for the
purpose \cite{FCDTO08}. The general idea of the algorithm is to
proceed by induction on the number of dimensions of the image. The
distance transform problem in $n+1$ dimensions is reduced to the
problem in $n$ dimensions by a technique that relies on
multi-dimensional Voronoi maps. We refer the interested reader to \cite{Ciesielski2010},
where a theoretical study of the algorithm is provided.
The specification described therein was closely followed in our
implementation.

For shortest-path distances over arbitrary
directed graphs, \topochecker employs a variant of the well-known
Dijkstra shortest-path algorithm, called ``modified Dijkstra distance
transform'' in \cite{Grevera2007}. 
The standard Dijkstra algorithm uses a
priority queue sorted by distance from a root node. The queue is
initialised to the root node of the considered graph, whose priority
is set to $0$. In the modified version, when computing the distance
transform from a set of nodes identified by formula $\form$, the
queue is initialised with all the nodes that satisfy $\form$ and
have an outgoing edge reaching a node not satisfying $\form$; all
such nodes have priority $0$. The algorithm then proceeds as the
standard algorithm. As a result, after termination, each node of the
graph is labelled with the  shortest-path distance from the the set of nodes
satisfying $\form$, as required by the specification.  The asymptotic
run-time of this procedure is not linear but
quasi-linear due to the usage of a priority queue. In this respect,
research is still active to optimise the procedure in specific cases
(see e.g. \cite{tho99}). However, the effective
run-time behaviour of the algorithm is highly dependent on the structure of the
considered graph and the chosen implementation of data structures; in
our tests on Euclidean graphs, this procedure is typically faster than
computing the Euclidean distance transform using Maurer's algorithm,
although a precise comparison of efficiency between the two algorithms is
obviously implementation dependent, and also depends on the
precision-efficiency trade-off given by the chosen adjacency relation.

\section{A Logical Framework for Texture Analysis}
\label{sec:LogicTA}

In this section we define an additional logic operator that, when incorporated in
the spatial logic presented in the previous sections, provides a logical framework for 
Texture Analysis (TA).

TA can be used for finding and analysing patterns in (medical) images, 
including some that are imperceptible to the human visual system. Patterns in images are
entities characterised by distinct combinations of features, such as brightness, colour, shape, size.  
TA includes several techniques and has proved promising in a large number
of applications in the field of medical imaging
\cite{Kassner2010,Lopes2011,Castellano2004,Davnall2012}; in particular
it has been used in \emph{Computer Aided Diagnosis}
\cite{Woods2007,Han2014,Heinonen2009} and for classification or
segmentation of tissues or organs
\cite{Chen1989,Sharma2008,RodriguezGutierrez2013}.
In TA, image textures are usually characterised by estimating some
descriptors in terms of quantitative features. Typically, such
features fall into three general categories: syntactic, statistical,
and spectral \cite{Kassner2010}. Our work is mostly focused on
\emph{statistical} approaches to texture analysis. 
For two-dimensional images, statistical methods
consist of extracting a set of statistical \emph{descriptors} from the
distributions of local features in a neighbourhood of each pixel.
 
In this paper, we explore the use of \emph{first order} statistical methods, that are
statistics based on the probability distribution function of the
intensity values of the pixels of parts, or the whole, of an image. The classical first-order statistical approach to TA makes use of 
statistical indicators of the local distribution of image intensity around each pixel, 
such as \emph{mean}, \emph{variance}, \emph{skewness}, \emph{kurtosis}, \emph{entropy}
\cite{Srinivasan2008}.  
Although such indicators ignore the relative spatial placement of adjacent pixels, 
statistical operators are useful in MI because their application is invariant under
transformations of the image, in particular \emph{affine transformations}
(rotation and scaling), which is necessary when analysing several
images acquired in different conditions. It is worth mentioning that current research also 
focuses on constructing features using first order operators, keeping some
spatial coherence, but loosing at least partially the aforementioned
invariance \cite{Tijms2011}. The method we propose is an attempt to improve over the classical setting described above, by analysing (the histograms of) statistical distributions directly.


In the following, we introduce a spatial logic operator that compares
image regions in order to classify points that belong to sub-areas in
the image where the statistical distribution of the intensity of
pixels is {\em similar} to that of a chosen reference region. 
Several similarity measures exist (see
\cite{Meshgi2015}), that can be used to compare distributions in images.
In particular, as a starting point, we use the
\emph{cross-correlation} function (also called \emph{Pearson's
  correlation coefficient}), that is often used in the context of
\emph{image retrieval}, but is also popular in other computer vision
tasks. In MI, cross-correlation is also frequently used in the case of
\emph{image co-registration} (\cite{B92})\footnote{In image
  processing, the problem of \emph{co-registration} is that of mapping
  two images coming from different sources to the same spatial domain,
  by finding transformations of the considered images that make given
  image features coincide.}.

%

\subsection{A logical operator for statistical similarity}
\label{sec:loss}

The statistical distribution of an area $Y$ of a black and white  image is
approximated by the \emph{histogram} $h$ of the grey levels of points (pixels or voxels,
for two- and three-dimensional images) belonging to $Y$, defined as follows. 
Given a minimum value $m$, a maximum value $M$, and a positive number of \emph{bins} $k$, let
$\Delta = (M - m)/k$ and define the histogram $h$ as a function associating to each
\emph{bin} $i \in [1,k]$ the number of points that have intensity in
the (half-open) interval $[(i-1) \cdot \Delta + m,i \cdot \Delta + m)$. 
The minimum value $m$ and the maximum value $M$ are aimed at improving
the resolution of histograms, by excluding rare peaks in the image,
that may be due to artefacts in acquisition and would result in a high
number of empty bins.
A formal definition  is given below:

\begin{definition}
Given  closure model $\model = ((X,\closure), \aeval, \peval)$,
we define
function $\mkhis : A \times 2^X \times \reals \times \reals \times \nats \to (\nats \to \nats)$
such that for all values $m,M \in \reals$, with $m<M$, and $k,i \in \nats$, with $k>0$ and 
$i\in [1,k]$, letting $\Delta=\frac{M-m}{k}$:\\
$
\mkhis(a,Y,m,M,k)(i) =
$\\
$
\left|\ZET{y\in Y}{(i-1) \Delta \leq \aeval(y,a) - m < i \Delta}\right|
$
\ed
\end{definition}
So $\mkhis(a,Y,m,M,k)$ is the histogram of the distribution of the values of attribute $a$ of the points in $Y$, in the interval $[m,M]$ with step $\Delta$.
The above definition applies also to quasi-discrete / distance closure models.

In the sequel, for histogram $h: [1,k] \to \nats$ we let 
$\mean{h}=\frac{1}{k}\sum_{i=1}^k h(i)$ denote the {\em mean}  of $h$.

The definition of {\em cross correlation} between two histograms follows:
\begin{definition}
Let $h_1, h_2: [1,k] \to \nats$ be two histograms. The {\em cross correlation} of
$h_1$ and $h_2$ is defined as follows:
$$
\cc(h_1,h_2) = 
\frac
{\sum_{i=1}^k\left( h_1(i) - \mean{h_1} \right) \left(h_2(i) - \mean{h_2} \right)}
{
\sqrt{ 
\sum_{i=1}^k \left(h_1(i) - \mean{h_1} \right)^2
}
\sqrt{
\sum_{i=1}^k \left(h_2(i) - \mean{h_2} \right)^2
}
}
$$\ed
\end{definition}

The value of $\cc{}{}$ is \emph{normalised} so that
$-1\le \cc{}{}\le 1$; $\cc(h_1,h_2)=1$ indicates that $h_1$ and $h_2$
are \emph{perfectly correlated} (that is, $h_1 = ah_2+b$, with $a>0$); $\cc(h_1,h_2) =-1$
indicates \emph{perfect anti-correlation} (that is, $h_1=ah_2+b$, with $a<0$). On the other
hand, $\cc(h_1,h_2) = 0$ indicates no correlation. Note that normalisation
makes the value of $\cc{}{}$ undefined for constant histograms, having
therefore standard deviation of $0$; in terms of statistics, a
variable with such standard deviation is only (perfectly)
correlated to itself. This special case is handled by letting
$\cc(h_1,h_2)=1$ when both histograms are constant, and $\cc(h_1,h_2)=0$ when only one
of the $h_1$ or $h_2$ is constant.
We are now ready for embedding the statistical similarity operator $\lssim{\bowtie c}{m}{M}{k}{\rho}{a}{b}$ in  \SLCSMI.

\begin{definition}\label{def:sdsatisfaction}
{\em Satisfaction} $\model, x \models \form$ at point $x \in X$ in distance closure model
$\model = (((X,\closure),d), \aeval, \peval)$ is defined by induction on the structure of formulas, by adding the following equation, where $m,M \in \reals$, with $m<M$, and $k\in \nats$, with $k>0$:
$$
\model, x \, \models \,
\lssim{\bowtie c}{m}{M}{k}{\rho}{a}{b} \form \Leftrightarrow
\cc(h_a,h_b) \bowtie c
$$
with:\\
$h_a(i)=\mkhis(a,S(x,\rho),m,M,k)(i)$, \\
$h_b(i)=\mkhis(b,\ZET{y}{\model,y \models \form},m,M,k)(i)$,\\
$\bowtie \, \in \, \SET{=,<,>,\leq,\geq}$, and\\
$S(x,\rho)=\ZET{y\in X}{d(x,y) \leq \rho}$ is the {\em sphere} of radius $\rho$ centred in $x$, to the equations in Figure~\ref{fig:fragSLCSsem}. \ed
\end{definition}

So $\lssim{\bowtie c}{m}{M}{k}{\rho}{a}{b} \form$ compares the region of the
image constituted by the sphere of radius $\rho$ centred in $x$ against the 
region characterised by $\form$. The comparison is  based on the cross correlation
of the histograms of the chosen attributes of (the points of) the two regions, namely $a$ and $b$
and both histograms share the same domain ($[m,M]$) and the same bins ($[1,k]$). 
In summary, the operator allows to check {\em to which extent} the {\em sphere around the point of interest} is {\em statistically similar} to a given region (specified by) $\form$.

\subsection{Model checking \SLCSMI with statistical similarity operators}

In \topochecker the statistical similarity operator \SCMP{} is provided, with the following 
syntax 
\[
\SCMP\verb!(a,fa,R,! \bowtie \verb!c, m, M, k)(b,fb)!
\]
corresponding to 
$
\form_a \land \lssim{\bowtie c}{m}{M}{k}{\rho}{a}{b} \form_b
$
where {\tt fa}, {\tt fb} and {\tt R} represent $\form_a, \form_b$ and $\rho$, respectively, in the language of the tool.
For example, formula 
\[
\SCMP\verb!(a,TT,10.0,>= 0.7, 200, 2000, 100)(b,TT)!
\]
is true at voxels centred in a region---of radius $10.0$---where the
distribution of the values of {\tt a} has cross-correlation greater
than $0.7$ with the distribution of the values of {\tt b} in the whole
image. In this case, cross-correlation is computed using $100$
bins, and taking into account only values between $200$ and $2000$.


The algorithm we use for implementing the \SCMP{} operator is straightforward. An array $v_b$, sized to the number of bins $k$, is allocated, initialised to $0$ at each index, and the histogram $h_b$ is stored in it, by iterating over all the points $y$ of satisfying $\Phi$, finding the index $i$ of the bin corresponding to the grey level of $y$, and increasing the corresponding value of $v_b[i]$. An array $v_a$, sized to the number of bins $k$, is allocated.  For each pixel $x$, $v_a$ is (re-)initialised to $0$ at each index, and all the points $y$ laying in the sphere of radius $\rho$ centred in $x$ are examined; for each $y$, the index $i$ of its bin is identified, and the value of $v_a[i]$ is increased, so that when all the $y$ have been examined, $v_a$ represents the histogram $h_a$. The cross-correlation value $\cc(h_a,h_b)$ is then computed by simple calculations that are linear in the number of bins $k$. This algorithm has time complexity proportional to $v \cdot \rho^n \cdot k$, where $v$ is the number of pixels in the image and $n$ the number of dimensions (indeed the number of pixels in an $n$-dimensional sphere is proportional to $\rho^n$). Since $\rho$ and $k$ are usually fixed along a given analysis, such algorithm can still be considered ``linear'' in the size of the image. This basic procedure is amenable to optimisation, for instance by observing that the spheres centred around two different points of the image may share some pixels, therefore the histogram of each one could be computed starting from the histogram of the other, at the expenses of more memory needed to store the histogram of different points. We leave the study of similar optimisations for future work as the algorithm we discussed has proved to be sufficiently fast for our experiments.

\section{Using \topochecker for MI analysis with \SLCSMI}
\label{sec:examples}

MR images are produced using different kinds of \emph{sequences} of magnetic field gradients and radio-waves. Images so obtained are called {\em weighted} images; these can be further post-processed in various ways.
For instance, typical weighted images are 
those produced using \emph{Fluid-attenuated inversion recovery} pulse sequence (MR-FLAIR),
\emph{T2 weighted pulse sequence} (MR-T2w), or \emph{diffusion weighted images}, whereas the 
\emph{Apparent Diffusion Coefficient} maps (ADC) are obtained via post-processing of diffusion weighted images. A standard reference for such matters is \cite{BCCHTV14}.
In this section, we  illustrate our approach on some of these kinds of medical images, using a tutorial-like, step-wise style.
We do this by means of two examples
of segmentation in MRI, introduced below:
\begin{enumerate}
\item Glioblastoma tumour and oedema segmentation in images
  obtained using  MR-FLAIR; this analysis is carried out in two dimensions.
\item Rectal cancer segmentation in images obtained using  MR-T2w and ADC.
This analysis is carried out in three dimensions.
\end{enumerate}

\noindent For glioblastoma, our procedure was successfully tested on five images from different sources, that were 
acquired in very different conditions. However, validation of the methodology for actual clinical usage requires
extensive clinical research. We refer to Section~\ref{subsec:validation} for preliminary validation results and a 
more detailed discussion.

Model definitions using medical images are introduced in \topochecker by associating an arbitrary number of attribute names to files containing medical images\footnote{The model loader of \topochecker currently supports the NIfTI (Neuroimaging Informatics Technology Initiative) file
format (\url{https://nifti.nimh.nih.gov/}, version 1 and 2). In this work, images
downloaded from \emph{Radiopaedia.org} in \emph{jpeg} format, and
\emph{dicom} images have been converted to NIfTI-1.
}, 
as follows:
\begin{verbatim}
Model "med:img1=file1.nii,img2=file2.nii,...";
\end{verbatim}
where \verb!med! is a keyword indicating the file type,
and \verb!img1!, \verb!img2! \ldots are names of pixel/voxel attributes, for which the related values are drawn from  \verb!file1.nii!, \verb!file2.nii! \ldots, respectively.

For this to work, all the loaded images must have the same voxel coordinates and orientation (e.g., coming from the same machine and type of acquisition, or after manual resampling). No resampling is currently done in \topochecker. 


\subsection{Example: segmentation of glioblastoma}
\label{subsec:GBM}

In this example we detail the specification of an analysis aimed at
the segmentation of glioblastoma (GBM) and oedema in MR-FLAIR images. 
Being able to segment tumour and oedema in medical images can be of
immediate use for \emph{automatic contouring} applications in
radiotherapy and, in perspective, it can be helpful in detecting the
invisible infiltrations in Computer-Aided Diagnosis applications.
The
procedure is non-trivial, but every step is justified by
morphological and spatial considerations on the arrangement of parts
of the head and the brain.

Normal tissues of the head can be divided into several classes. The outer layer of the head consists of adipose tissue (and skin) surrounding the skull that in turn consists of bone and bone-marrow. The skull encloses the brain tissues. The brain itself is suspended in cerebrospinal fluid (CSF) and isolated from the blood stream. 
Thresholds in the grey levels of images can be used to single out specific tissues in a medical image; however, in doing so, noise is generated in the form of (small, scattered) regions not belonging to the tissue. The relative positioning of tissues --- the so-called \emph{topological information} of the image --- plays an important role in suppressing such noise. We will see in the following how such information is encoded by logic formulas in the methodology we propose.

GBMs are intracranial tumours composed of typically poorly-marginated,
diffusely infiltrating necrotic masses. Even if the tumour is totally
resected, it usually recurs, either near the original site, or at more
distant locations within the brain. GBMs are localised to the cerebral
hemispheres and grow quickly to various sizes, from only a few
centimetres, to lesions that cover a whole hemisphere. Infiltration
beyond the visible tumour margin is always present. In MR-FLAIR 
images GBMs appear \emph{hyperintense} and surrounded by
\emph{vasogenic oedema}\footnote{Vasogenic oedema is an abnormal
  accumulation of fluid from blood vessels, which is able to disrupt
  the blood-brain barrier and invade extracellular space.}.

Segmentation of GBM according to our method is performed in three steps:

\begin{enumerate}
\item a preprocessing step (not using \topochecker), aimed at
  normalisation of images, to make the choice of thresholds in our
  experiment applicable to different images;
\item brain segmentation, to limit the area of the image where the
  tumour is searched for;
\item tumour and oedema segmentation, which is the stated goal of this
  example.
\end{enumerate}

\subsubsection{Preprocessing}
\begin{figure}
\centering
\subfloat[][Case courtesy of Dr.  Ahmed Abd Rabou, Radiopaedia.org, rID: 22779.]
{
\includegraphics[height=3.8cm]{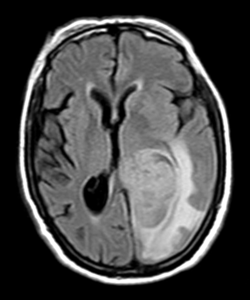}
\label{subfig:GBM2}
}\hfill
\centering
\subfloat[][A different slice of the acquisition in Figure~\ref{subfig:GBM2}.]
{
\includegraphics[height=3.8cm]{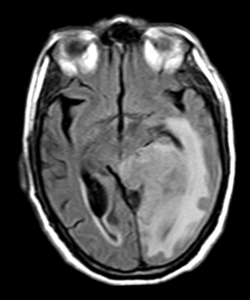}
\label{subfig:GBM2-2}
}\\
\subfloat[][Case courtesy of A.Prof Frank Gaillard, Radiopaedia.org, rID: 5292.]
{
\includegraphics[height=3.8cm]{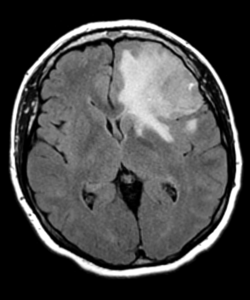}
\label{subfig:GBM49}
}\hfill
\centering
\subfloat[][Histograms of Figure \ref{subfig:GBM2} (blue), \ref{subfig:GBM2-2} (green), \ref{subfig:GBM49} (red).]
{
\includegraphics[height=2.8cm]{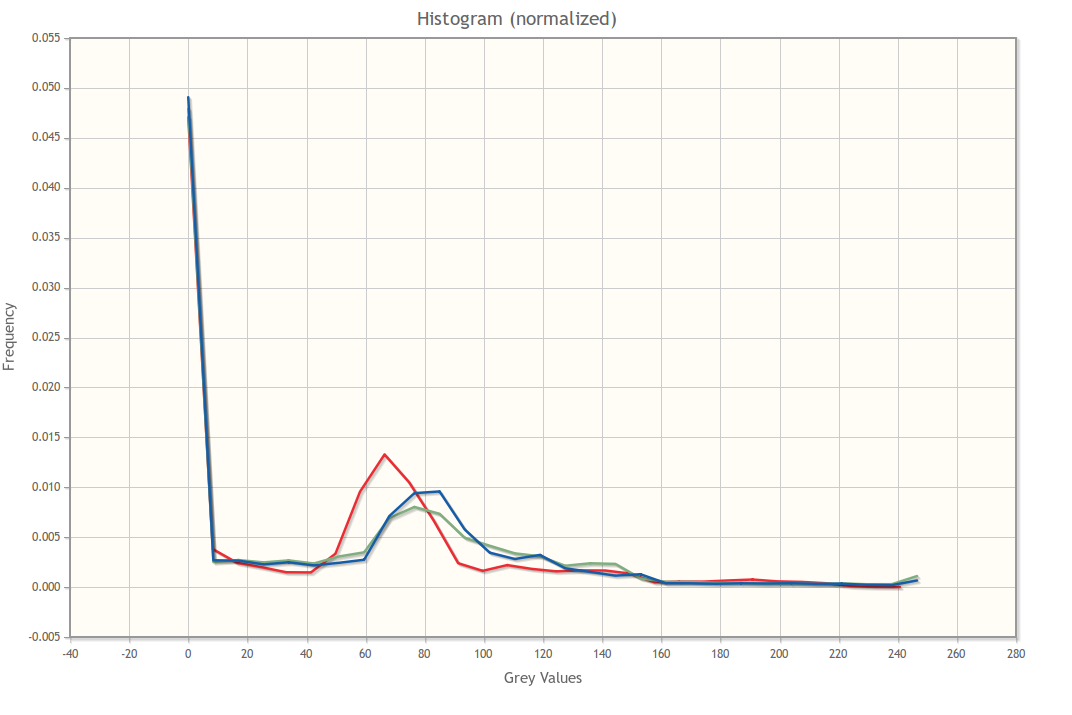}
\label{histGBM1}
}
\caption{Slice of MR-FLAIR brain acquisition of different patients and corresponding histogram.}
\label{fig:histOrig}
\end{figure}

\begin{figure}
\centering
\subfloat[][The pixels in Figure~\ref{subfig:GBM49} with grey levels below a given threshold are shown in green.]
{
\includegraphics[height=3.5cm]{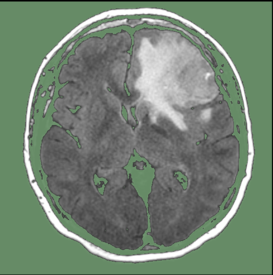}
\label{subfig:mask1}
}\hfill
\centering
\subfloat[][The sub-mask that touches the border of the image is shown in orange.]
{
\includegraphics[height=3.5cm]{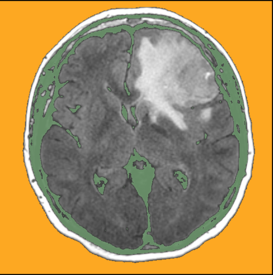}
\label{subfig:mask2}
}\\
\centering
\subfloat[][The  mask of the image excluding the background is shown in green.]
{
\includegraphics[height=3.5cm]{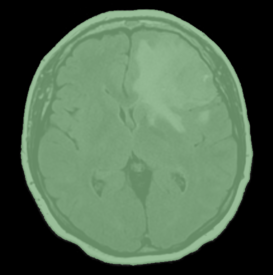}
\label{fig:GBM49mask}
}\hfill
\centering
\subfloat[][Histograms of normalised version of images in
    Figure~\ref{subfig:GBM2} (blue), Figure~\ref{subfig:GBM2-2}
    (green), Figure~\ref{subfig:GBM49} (red) and
    Figure~\ref{subfig:eqIm} (orange).]
{
\includegraphics[height=2.8cm]{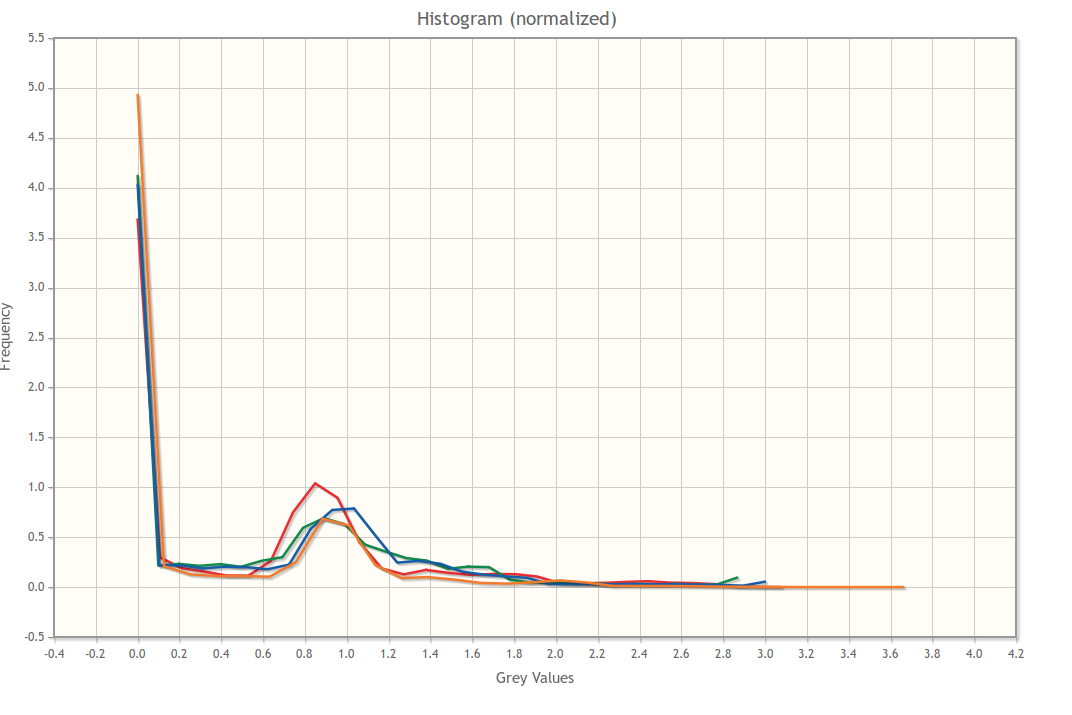}
\label{fig:histNorm}
}
\caption{Finding the mask for normalisation.}
\label{fig:normMask}
\end{figure}

\begin{figure}
\centering
\subfloat[][A slice of MR acquisition of brain (on the left) and its equalised version (on the right) \Siena]
{
\includegraphics[height=3.8cm]{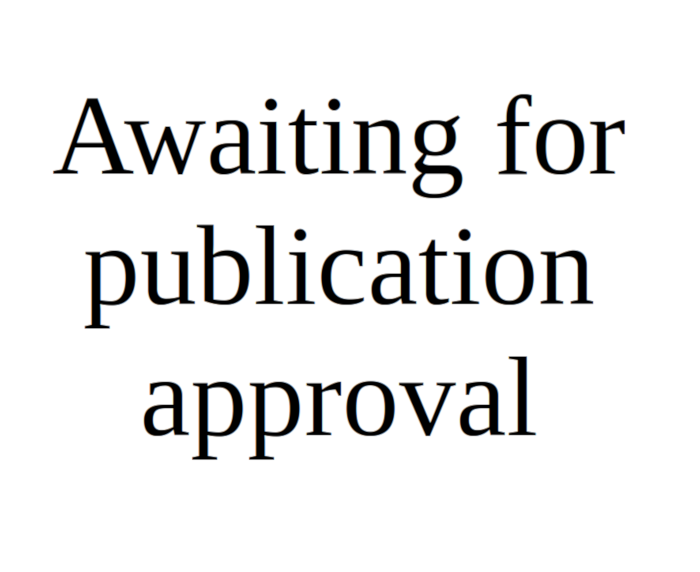}
\label{subfig:eqIm}
}\hfill
\centering
\subfloat[][Histograms of grey levels of the original (red) and equalised (green) version of image in \ref{subfig:eqIm}.]
{
\includegraphics[height=2.8cm]{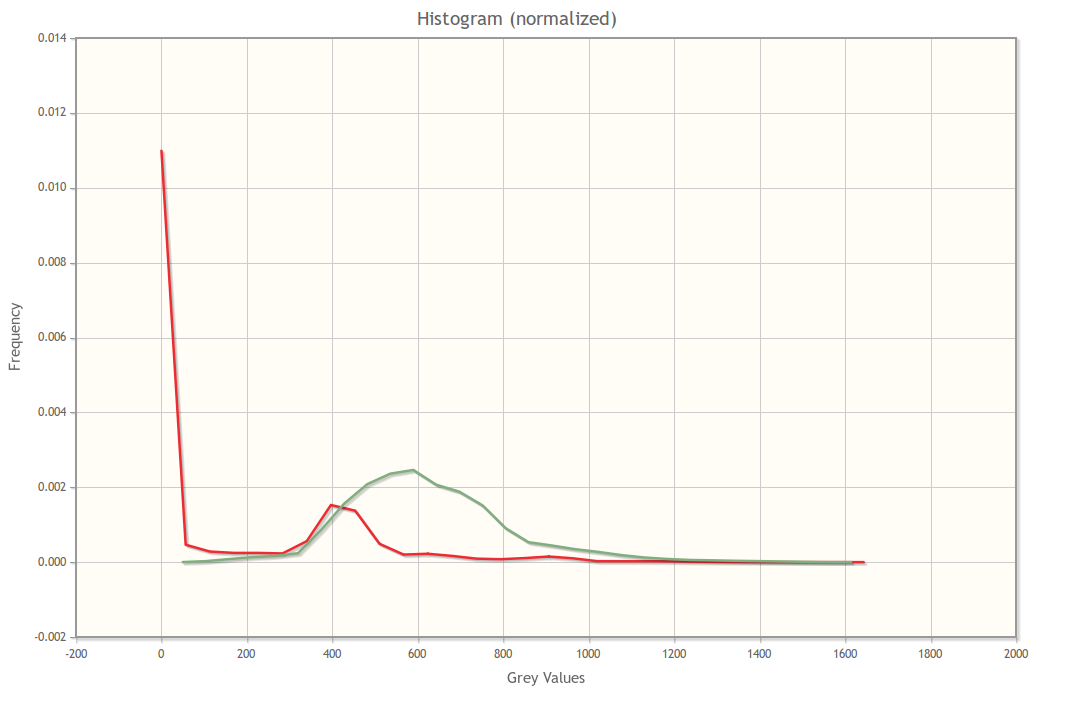}
\label{subfig:eqHist}
}
\caption{Effect of histogram equalisation.}
\label{fig:equalH}
\end{figure}

Histograms of grey levels of images\footnote{To ease visual comparison, in Figg.\ref{histGBM1}, \ref{fig:histNorm}, \ref{subfig:eqHist}, the histograms that we show 
  are normalised so that the measure of  the area below the curve is $1$.} of the same
body part may differ from each other due to inter-patient or
inter-scanner differences or depending on the actual acquisition
volume (Figure~\ref{fig:histOrig}) or the file format used to store
the image\footnote{For instance, \emph{jpeg} images, as downloaded
  from \emph{Radiopaedia.org}, typically use 8-bit precision (typical
  range 0-255) (see Figure~\ref{fig:histOrig}) whereas \emph{dicom}
  images saved by scanners typically use 12 or 16-bit (for MR images, the
  typical range is 0-4096 or 0-65536, respectively) (see
  Figure~\ref{fig:equalH}).}.

More uniform results, on different images, can be obtained by  
dividing the
intensity of each pixel by the average of the intensity levels of all
the \emph{significant} pixels in the image. A pixel is considered
significant when it does not belong to the background. 
Significant pixels are selected using a Boolean mask (indicated by the green area in 
Figure~\ref{fig:GBM49mask}).  In order to compute such a mask,
we start from the observation that the background (corresponding to the air surrounding
the head of a patient) is darker than the rest of the image, so it
mostly contributes to the initial part of its histogram. This
situation is witnessed in the histogram by a peak close to 0. A threshold 
is thus selected for each image as the value immediately
following such peak. Using this threshold, it is possible
to isolate the background, by separating it from the head
(Figure~\ref{subfig:mask1}). Note that the obtained mask also includes
cerebrospinal fluid (CSF) and bone. The part of the mask that touches
the boundary of the whole image is then selected
(Figure~\ref{subfig:mask2}) and its complement, that is, the green area 
in Figure~\ref{fig:GBM49mask}, is finally used to select the significant pixels to compute the mean value for normalisation. Figure~\ref{fig:histNorm} shows the histograms of images after normalisation.

We remark that \emph{equalisation of histograms} is
another form of normalisation, frequently used for texture analysis
(\cite{Har73}). We do not use this method as it changes the
relationship between grey levels of different structures in the image
(as shown in Figure~\ref{fig:equalH}), that we use rather prominently
for differentiating different tissues; normalisation of image
intensity is sufficient for our purposes.

\subsubsection{Brain segmentation}
In this second phase of our method, \topochecker is used
to perform a segmentation of the brain in order to limit
the search area of the tumour. This improves the accuracy of the
output (e.g., avoiding areas in bone marrow or skull) and
reduces computing time.
%

In the process below, we  fix some thresholds for identifying different tissues in the brain; 
note that, thanks to the preprocessing step described above, these 
can be kept uniform across different images.

Intuitively, the general model of a patient head that we use to segment the
brain in MR-FLAIR images is defined as follows:

\begin{itemize}
\item Darker pixels in the head belong to CSF and bones;
\item Brighter pixels belong both to adipose tissue surrounding the head, and to
  bone marrow; 
\item Also pixels belonging to the tumour and oedema are brighter than the
  surrounding tissues;
\item The brain region is composed of white matter, grey matter,
  tumour and oedema;
\item The brain (excluding the tumour) has intermediate intensities and is
  mainly surrounded by CSF.
\end{itemize}

\noindent
The model definition in \topochecker is as follows. 

\begin{verbatim}
Model "med:FLAIR=GBM-NORM.nii";
\end{verbatim}

\noindent
\texttt{GBM-NORM.nii} is the normalised NIfTI image of the MR-FLAIR
acquisition shown in Figure~\ref{subfig:GBM49}. 
By the above declaration, in the rest of the analysis, the relevant attribute, 
i.e. the normalised grey level, of each pixel in 
this image is referred to as \mv{FLAIR}. 
Formula definitions for general derived operators {\tt reach} and {\tt touch} are given according to 
Figure~\ref{fig:derived} ($\lreach$, $\lfromto$);
similarly for operator $\lflt{c}$, with reference to Figure~\ref{fig:morederived}:

{\small
\begin{verbatim}
Let reach(a,b) = !(!b S !a);
Let touch(a,b) = a & reach(a|b,b);
Let flt(a) = MDDT(!(MDDT(!a,<1)),<1);
\end{verbatim}
}

Furthermore, we define also a few operators that serve as macros and that are specifically useful in the segmentation procedure that follows:
{\small
\begin{verbatim}
Let grow(a,b) = (a|b) S (!b);
Let denoise(a) = touch(a,MDDT(!a,>=2));
Let closeTo(a) = MDDT(a,<3.0);
\end{verbatim}
}


Formula \verb!grow(a,b)! is inspired by the image segmentation method of \emph{seeded region growing}~\cite{Adams1994}. This method starts from a number of \emph{seed points} in the region of interest and examines neighbouring points to decide whether they should be added to the region. We start from points that satisfy property {\tt a} and to which all points satisfying property {\tt b} are added that, together with those satisfying {\tt a}, form a common region that is surrounded by points that do not satisfy {\tt b}. Formula {\tt grow(a,b)} can be used  only when it is guaranteed that all points satisfying {\tt a} are also satisfying {\tt b} (but not the other way around).

Let $A$ be the set of points  satisfying formula {\tt a}.
The formula {\tt denoise(a)} is used to remove small areas from $A$, as follows:
first $A$ is shrunk by $2.0$ units; in doing so, some subareas of $A$ may disappear;
the areas that do not disappear are restored to their original shape by means of the {\tt touch} operator. 
This operation is  similar to {\tt flt}, but it preserves the contours of the original area $A$.
The formula {\tt closeTo(a)} denotes the points that lay at a distance less than 3.0 units from the set of points satisfying {\tt a}. For this analysis the {\tt MDDT} Chamfer distance operator uses 4 adjacent pixels per node and the distance units are in millimetres with respect to the actual dimension of the head, i.e. the {\em real-world} coordinates. 



Next we define a number of useful thresholds for the grey levels of the image that are used to obtain a first approximation of different kinds of tissue of interest:

{\small
\begin{verbatim}
Let lowIntsty = [FLAIR < 0.5];
Let medIntsty = [FLAIR > 0.5] & [FLAIR < 1.3];
Let highIntsty = [FLAIR > 1.7];
Let tumIntsty = [FLAIR > 1.17] & [FLAIR < 1.53];
Let oedIntsty = [FLAIR >= 1.47] & [FLAIR < 2.4];
\end{verbatim}
}

We distinguish three general levels of intensity (low, medium, and high), and two specific intensities that are typical for tumour and oedema, respectively.

We are now ready to start the segmentation procedure. First we identify the points that are part of the background of the image. These all have a very low intensity, but there are other points in the image that have low intensity as well. What distinguishes the points of the background from the other low intensity points is that the background area touches the border of the image. In \topochecker{,} when loading images, a special atomic proposition named  {\tt border} is defined, that is satisfied by points that form the border of an image.
This way points of the background are exactly those that satisfy the property {\tt background}:

{\small
\begin{verbatim}
Let background = touch(lowIntsty,[border]);
\end{verbatim}
}

The points that satisfy {\tt background} are shown in red in Figure~\ref{subfig:bkg_adi}. The original image is shown in Figure~\ref{subfig:GBM49-1}.

The next step is to look for the external border of the head, consisting of skin and adipose tissue.
For our purposes, it is sufficient to identify the  adipose tissue, since the brain is surrounded by the
adipose tissue, which separates it from the skin.
Adipose tissue in the normalised MR-FLAIR images has intensity above $1.7$, so of high intensity. As before, there may be other points with high intensity in the image, but we exploit the knowledge that adipose is at the external border of the head, and thus close to the background. These points can be found with the following formula:

{\small
\begin{verbatim}
Let adipose = touch(highIntsty,closeTo(background));
\end{verbatim}
}

The points that satisfy {\tt adipose} are shown in green in Figure~\ref{subfig:bkg_adi}.

Using the properties {\tt background} and {\tt adipose} it is not difficult to specify the points that are part of the head. These are all those points that are not part of the background or close to adipose tissue.

{\small
\begin{verbatim}
Let head = !(closeTo(adipose) | background);
\end{verbatim}
}


The points that satisfy {\tt head} are the union of the green and red points in Figure~\ref{subfig:head_csf} (see below).

In the next steps we show how we can distinguish the various tissues within the area of the head, namely the brain and the cerebrospinal fluid (CSF) that contains it. We start from the identification of points that are part of CSF. These are points that are within the head and that have low intensity:

{\small
\begin{verbatim}
Let CSF = lowIntsty & head;
\end{verbatim}
}

The points that satisfy {\tt CSF} are shown in red in Figure~\ref{subfig:head_csf}.

We proceed with segmentation of the brain in four subsequent steps. As a first approximation we look for the points of the brain with medium intensity within the head (and that are not belonging to CSF). Within this approximation we select some inner areas that are most certainly part of brain tissue and that can serve as a seed from which to `grow' in a more precise way points belonging to the brain. Finally, we remove pieces that have been erroneously identified as part of the brain, but that are actually relatively small areas that are part of the skull or bone, having a similar intensity as that of the brain. This way we obtain all pixels that are actually part of the brain. The four steps of the specification are given below.

{\small
\begin{verbatim}
Let brainApprox = head & (!CSF) & medIntsty;
Let brainSeed = MDDT(!brainApprox,>10);
Let noisyBrain = grow(brainSeed,head & (!CSF));
Let brain = touch(noisyBrain,brainSeed);
\end{verbatim}
}

The points that satisfy {\tt brainApprox} ({\tt brainSeed}, {\tt noisyBrain}, respectively) are shown in green in Figure~\ref{subfig:brainApprox} (Figure~\ref{subfig:brainSeed}, Figure~\ref{subfig:noisyBrain}, respectively). The final result of the brain is shown in Figure~\ref{subfig:brain}.

\subsubsection{GBM segmentation}
\begin{figure*}
\centering
\subfloat[][Original image.]
{
\includegraphics[height=2.65cm]{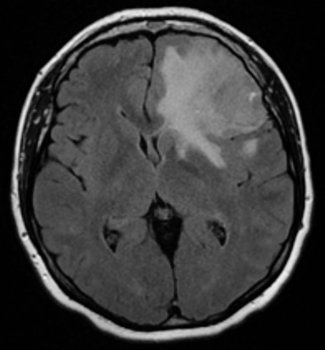}
\label{subfig:GBM49-1}
}~~~
\centering
\subfloat[][Background (red) and adipose (green).]
{
\includegraphics[height=2.65cm]{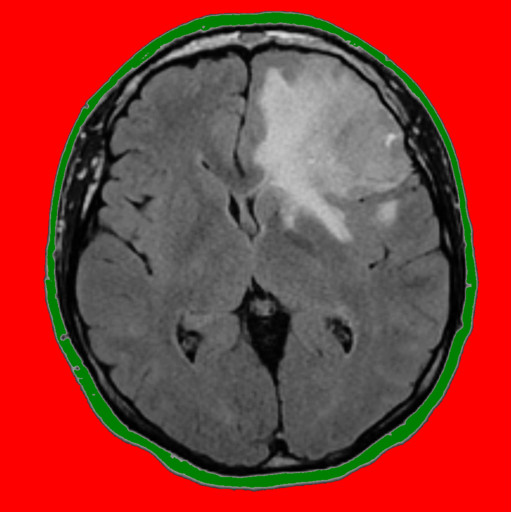}
\label{subfig:bkg_adi}
}~~~
\centering
\subfloat[][Output showing {\tt head} (red+green) and {\tt CSF} (red).]
{
\includegraphics[height=2.65cm]{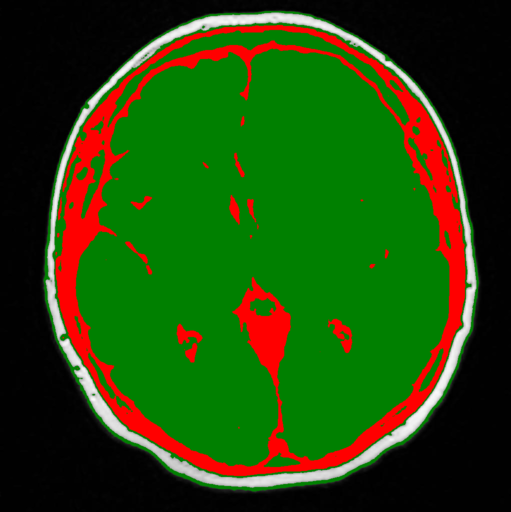}
\label{subfig:head_csf}
}~~~
\centering
\subfloat[][Output of {\tt brainApprox}.]
{
\includegraphics[height=2.65cm]{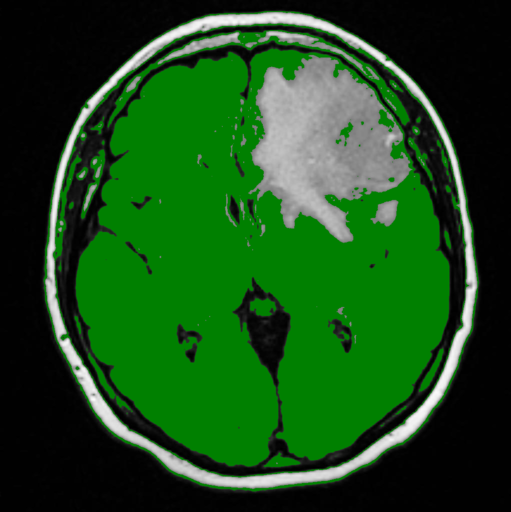}
\label{subfig:brainApprox}
}\\
\centering
\subfloat[][Output of {\tt brainSeed}.]
{
\includegraphics[height=2.65cm]{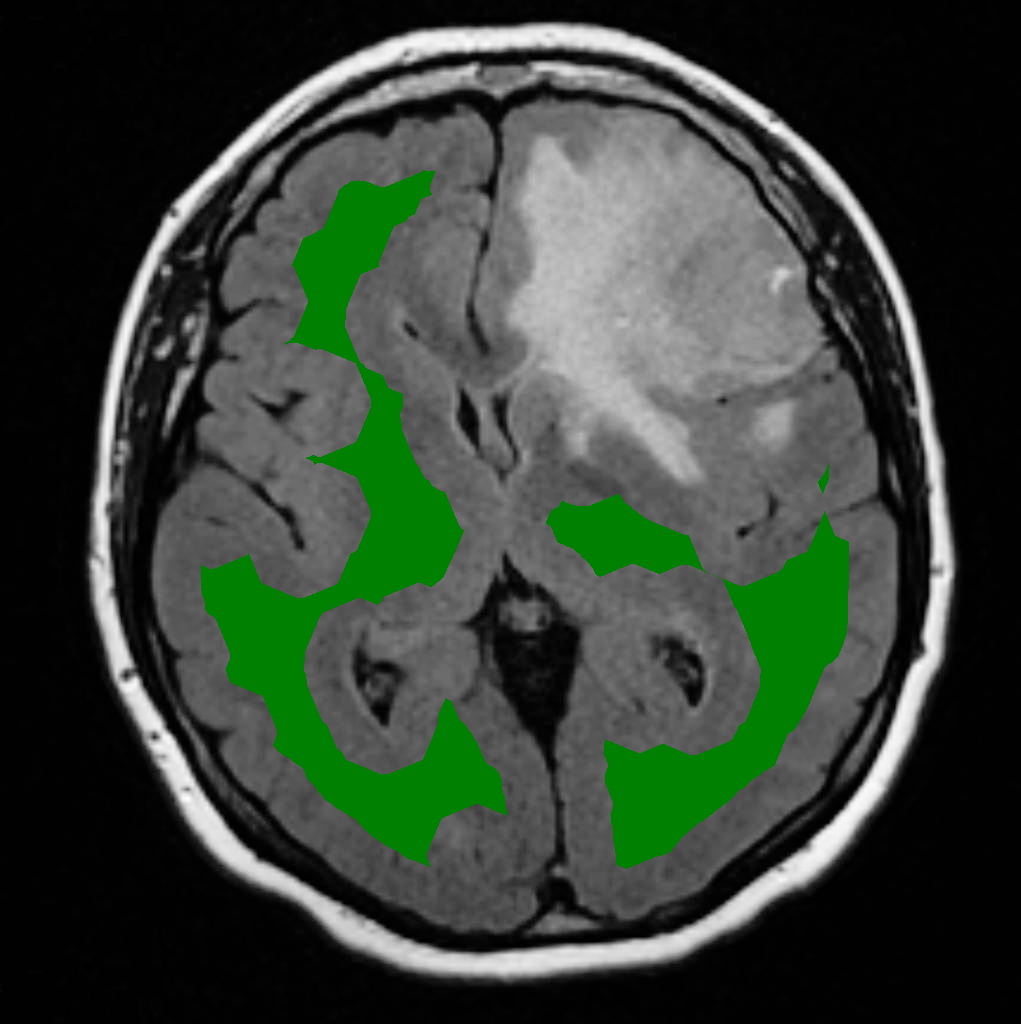}
\label{subfig:brainSeed}
}~~~
\centering
\subfloat[][Output of {\tt noisyBrain}.]
{
\includegraphics[height=2.65cm]{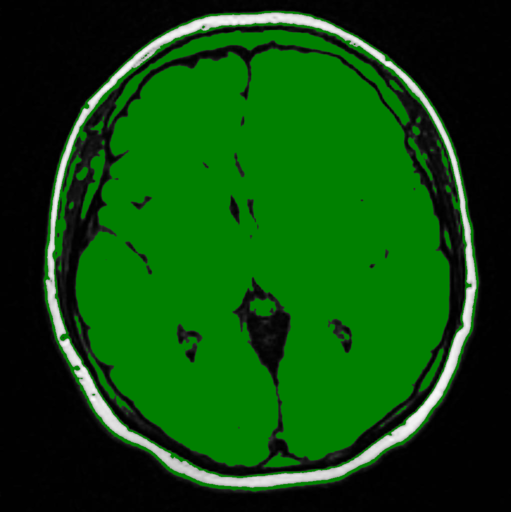}
\label{subfig:noisyBrain}
}~~~
\centering
\subfloat[][Output of {\tt brain}.]
{
\includegraphics[height=2.65cm]{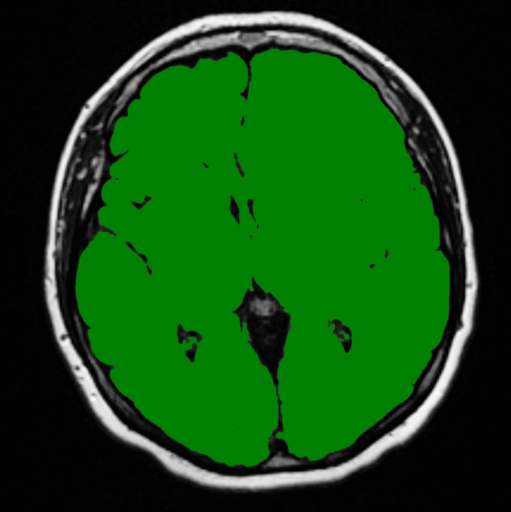}
\label{subfig:brain}
}~~~
\centering
\subfloat[][Output of {\tt tum0} (green) and {\tt oed0} (red).]
{
\includegraphics[height=2.65cm]{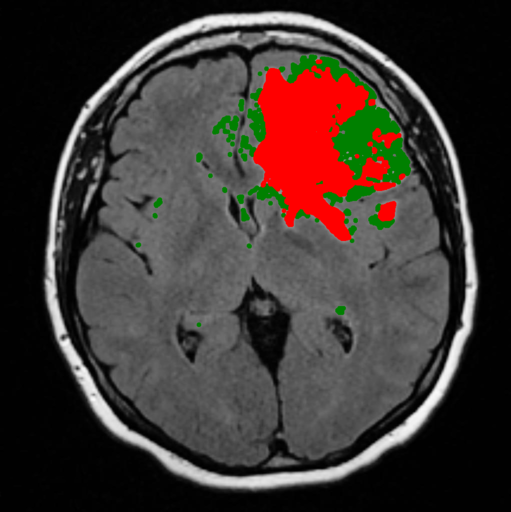}
\label{subfig:tum0_oed0}
}\\
\centering
\subfloat[][Output of {\tt tum1} (green) and {\tt oed1} (red) .]
{
\includegraphics[height=2.65cm]{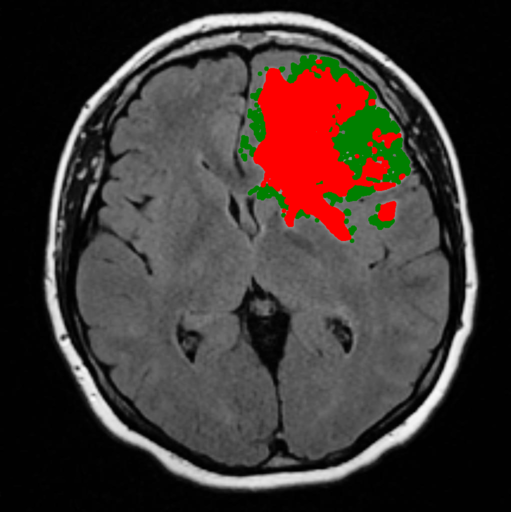}
\label{subfig:tum1_oed1}
}~~~
\centering
\subfloat[][Output of {\tt tum2} (green) and {\tt oed2} (red).]
{
\includegraphics[height=2.65cm]{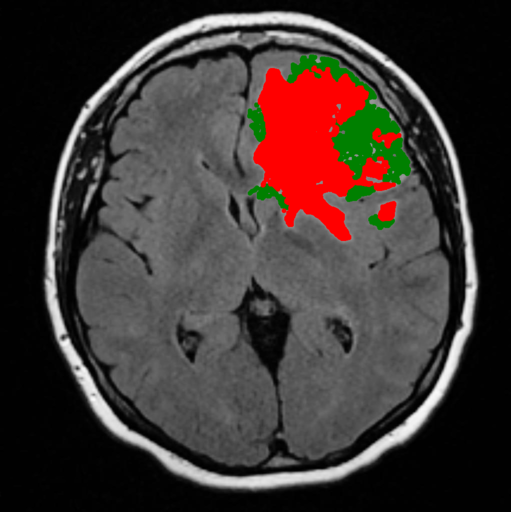}
\label{subfig:tum2_oed2}
}~~~
\centering
\subfloat[][Output of {\tt tumour} (green) and {\tt oedema} (red).]
{
\includegraphics[height=2.65cm]{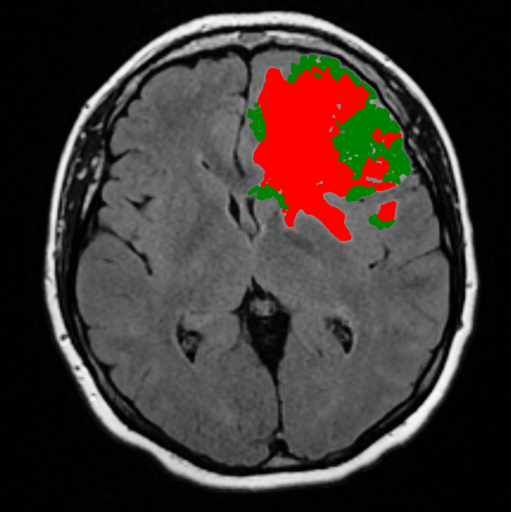}
\label{subfig:tumour_oedema}
}
\caption{Experimental results of using \topochecker for segmentation of glioblastoma (green) and oedema (red) (case courtesy of A.Prof Frank Gaillard, Radiopaedia.org, rID: 5292).}
\label{fig:GBM}
\end{figure*}

In the final part of our analysis, we identify tumour and oedema
regions. Since in MR-FLAIR, GBM and oedema are hyperintense areas, and the oedema is
brighter than the tumour, we start by using the thresholds we introduced before that provide a
rough segmentation of the image shown in Figure~\ref{subfig:GBM49-1}:

{\small
\begin{verbatim}
Let tum0 = flt(tumIntsty S (brain|CSF));
Let oed0 = flt(oedIntsty S (brain|CSF));
\end{verbatim}
}

In Figure~\ref{subfig:tum0_oed0} we show in red the points that satisfy formula {\tt oed0}
referring to the oedema, and in green those satisfying formula
{\tt tum0}  referring to the tumour. These are points that have the selected intensity ({\tt oedIntsty} and {\tt tumIntsty}, respectively) and are part of the brain tissue, i.e. they are surrounded by {\tt brain} or {\tt CSF}. Note that the regions {\tt oed0} and  {\tt tum0} are partially overlapping.
Moreover, we remove from these identified regions areas  whose
radius is smaller than 1mm using the {\tt flt} operator defined above.

An important constraint, that drastically reduces noise in the output
of our analysis, is the \emph{a priori} knowledge that oedema and tumour are very close to each
other. We exploit this knowledge using the distance operator {\tt MDDT} as follows:

{\small
\begin{verbatim}
Let oeddst = MDDT (oed0,<=2.0);
Let tum1 = touch(tum0,oeddst);
Let oed1 = oed0 & reach(oeddst,tum1);
\end{verbatim}
}



We first define the region {\tt oeddst} at distance less than 2mm from {\tt oed0}; then select sub-regions of
{\tt tum0} that \emph{touch} {\tt oeddst} (formula {\tt tum1}) and sub-regions of {\tt  oed0} that can reach points satisfying {\tt tum1} by passing only through points satisfying {\tt oeddst} (formula {\tt oed1}). 
The result is shown in Figure~\ref{subfig:tum1_oed1}. Comparing the latter with Figure~\ref{subfig:tum0_oed0} we can observe that some green areas, located in the left half of the brain, have disappeared. These were points with a similar intensity as that of tumour tissue, but not actually part of it since they were not connected to the tumour.
In this example, we used shortest-path distance as an approximation of
Euclidean distance, for the sake of execution speed, as high accuracy
for the distance is less important in this particular case.


%
%


{\small
\begin{verbatim}
Let tum2 = denoise(tum1);
Let oed2 = denoise(oed1);
\end{verbatim}
}

Figure~\ref{subfig:tum2_oed2} illustrates the areas
defined by {\tt tum2} (green) and
{\tt oed2} (red). Compared to Figure~\ref{subfig:tum1_oed1} this removes a number of small detached areas of oedema that were located in the tumour area and should be considered as noise.  

Finally, {\tt tumor} and {\tt oedema} are defined as being inter-reachable. This part could remove some separate areas that have tumour or oedema intensity but should not be considered as such since they are too far apart. In this particular case no such areas were present apparently as can be observed comparing Figure~\ref{subfig:tum2_oed2}  with the final output of the segmentation in Figure~\ref{subfig:tumour_oedema}.

{\small
\begin{verbatim}
Let tumor = touch(tum2,oed2);
Let oedema = touch(oed2,tum2);
\end{verbatim}
}

\begin{figure*}
\centering
\subfloat[][Slice of rID: 22779.]
{
\includegraphics[height=2.8cm]{img/GBM-2.png}
\label{subfig:input1}
}~~~
\centering
\subfloat[][Output of {\tt tumour} (green) and {\tt oedema} (red).]
{
\includegraphics[height=2.8cm]{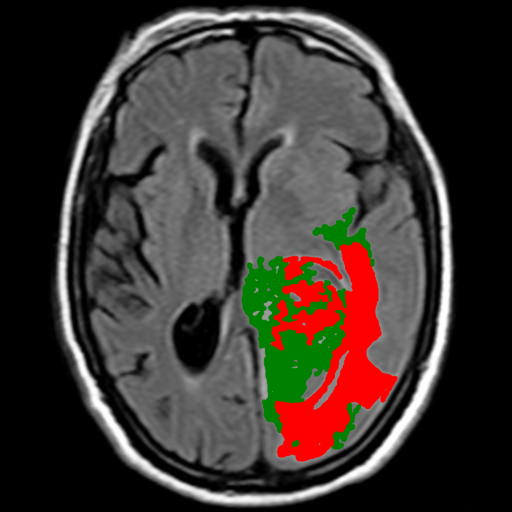}
\label{subfig:tumour_oedema_2}
}~~~
\centering
\subfloat[][Another slice of rID: 22779.]
{
\includegraphics[height=2.8cm]{img/GBM-2-2.png}
\label{subfig:input2}
}~~~
\centering
\subfloat[][Output of {\tt tumour} (green) and {\tt oedema} (red).]
{
\includegraphics[height=2.8cm]{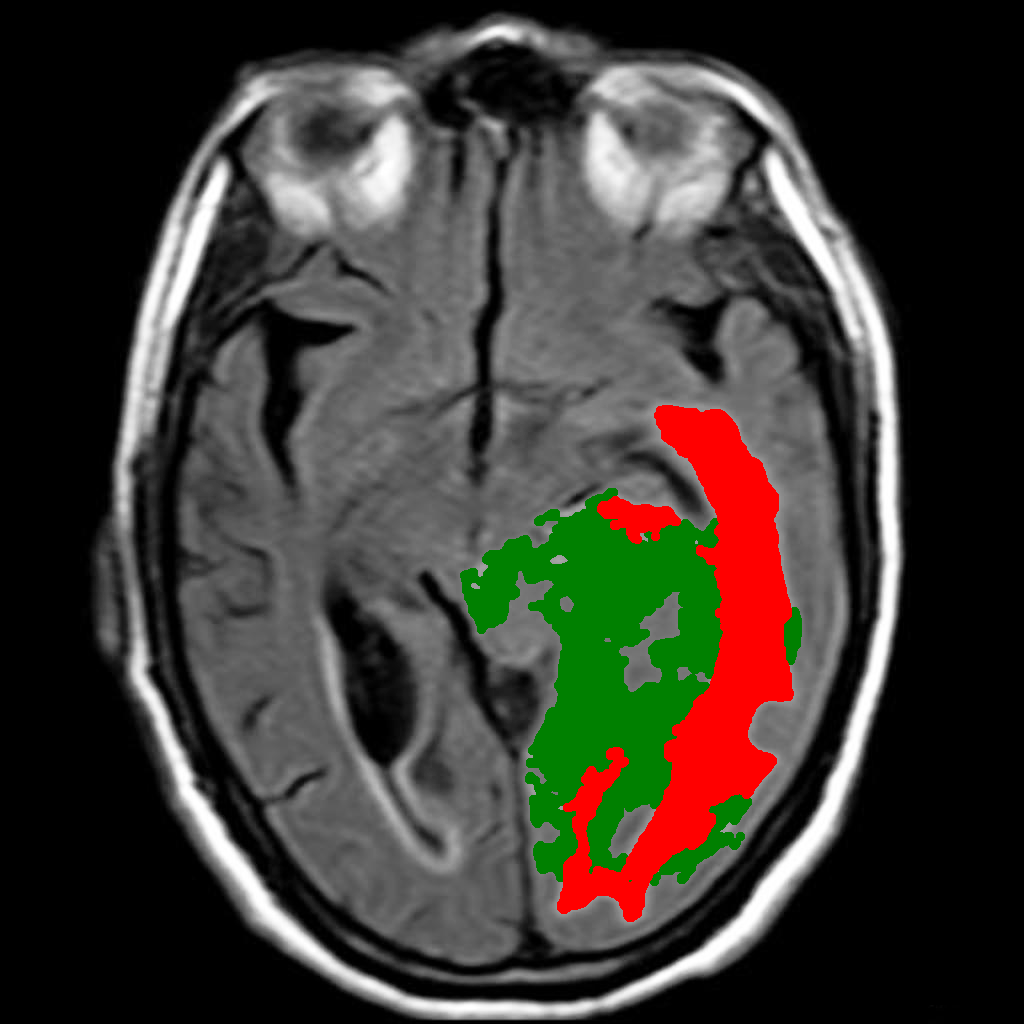}
\label{subfig:tumour_oedema_3}
}
\caption{Additional results of using \topochecker for segmentation of glioblastoma (green) and oedema (red) (Case courtesy of Dr.  Ahmed Abd Rabou, Radiopaedia.org, rID: 22779).}
\label{fig:GBMmore}
\end{figure*}

As a further result, in Figure~\ref{fig:GBMmore} we show the final segmentation of tumour and oedema on two other images from a different patient applying the same specification. Fig~\ref{subfig:tumour_oedema_2} shows the segmentation of the image in Figure~\ref{subfig:GBM2}, and Figure~\ref{subfig:tumour_oedema_3} shows the one of the image in Figure~\ref{subfig:GBM2-2}. The original images are also shown aside of the result in Figure~\ref{fig:GBMmore} for more convenient comparison.


For completeness, we show the code outputting the resulting images in
\topochecker. Output is saved in image {\tt GBM-seg.nii}, colours for
formulas are as specified in the first parameter of the {\tt Check}
instructions; a \emph{colour palette} (mapping $8$ to red and $7$ to green) has been applied to display the images.

{\small
\begin{verbatim}
Output GBM-seg.nii
Check "8" oedema;
Check "7" tumor;
\end{verbatim}
}

\begin{figure*}
\centering
\subfloat[][]
{
\includegraphics[height=3.8cm]{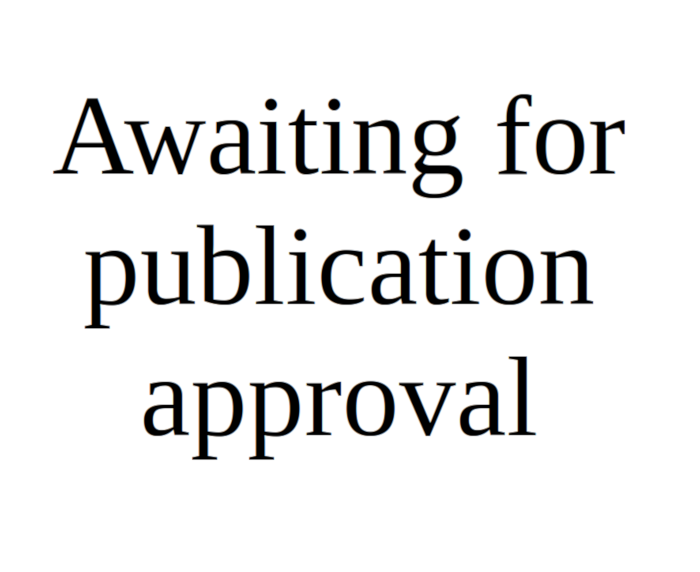}
\label{subfig:GBMpat2}
}~~~
\centering
\subfloat[][]
{
\includegraphics[height=3.8cm]{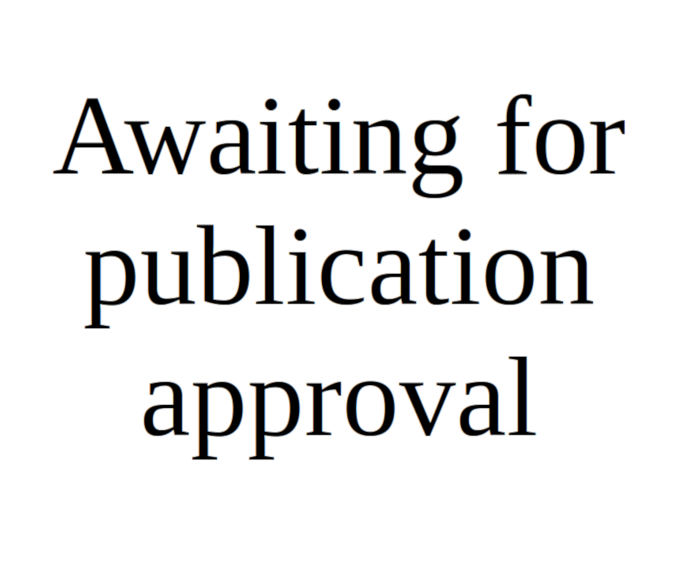}
\label{subfig:GBMpat3}
}~~~
\centering
\subfloat[][]
{
\includegraphics[height=3.8cm]{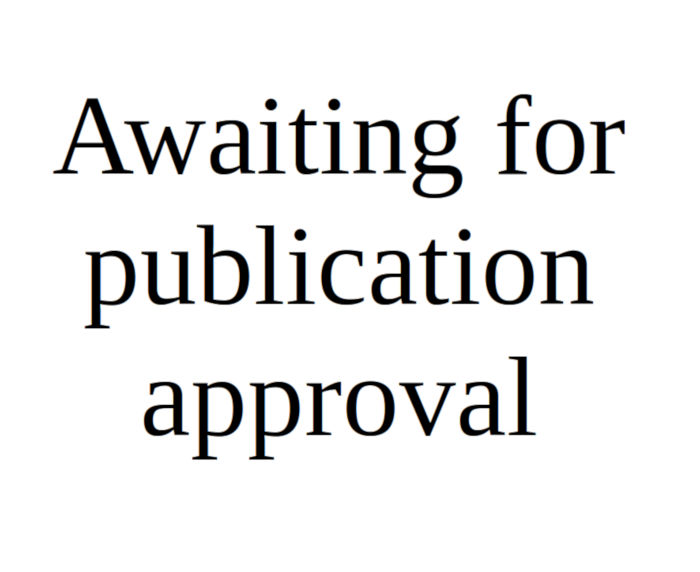}
\label{subfig:GBMpat4}
}\\
\centering
\subfloat[][]
{
\includegraphics[height=3.8cm]{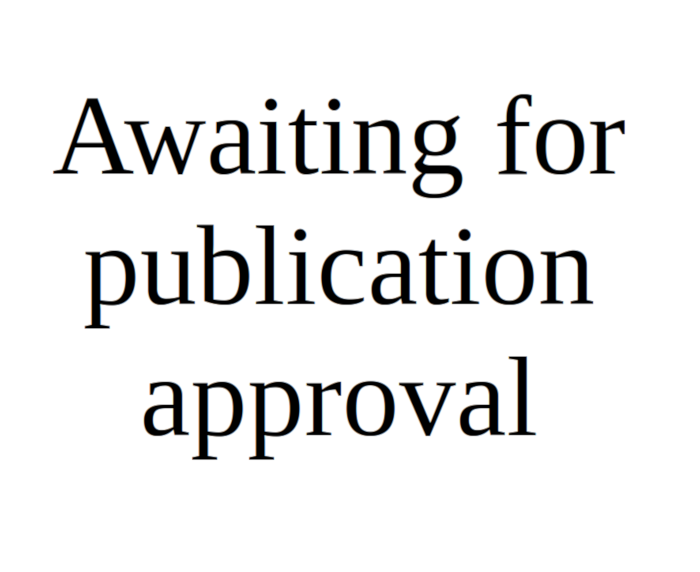}
\label{subfig:GBMpat2-seg}
}~~~
\centering
\subfloat[][]
{
\includegraphics[height=3.8cm]{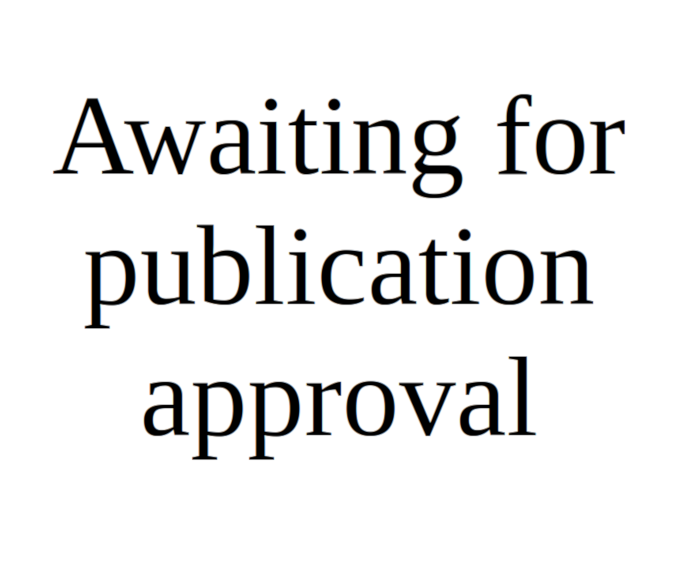}
\label{subfig:GBMpat3-seg}
}~~~
\centering
\subfloat[][]
{
\includegraphics[height=3.8cm]{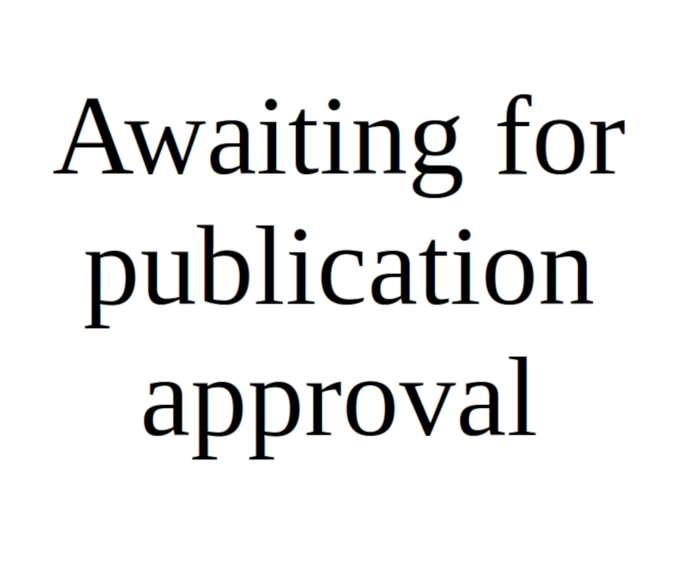}
\label{subfig:GBMpat4-seg}
}
\caption{Slices of an image obtained using the method we presented for segmentation of glioblastoma (green) and oedema (brown) on a 3D volume. The top row shows the original slices. The bottom row is the output of segmentation \Siena.}
\label{fig:GBMpatSeg}
\end{figure*}

The whole analysis presented in this section has been carried out in 2D. The same approach also works in 3D, with
minor modifications.  Figure~\ref{fig:GBMpatSeg} shows some slices of
the segmentation of MR-FLAIR acquisition of the patient in
Figure~\ref{subfig:eqIm}, using \topochecker on the whole 3D volume
image. Minor modifications to the model checking session presented in
this section were required. For space reasons, we omit the
details. However, in Section~\ref{subsec:rectum} we detail rectum
carcinoma segmentation, that has been carried out in 3D for
accuracy reasons.

In the GBM case we have not used the statistical comparison operator, which will instead be used in the example in Section~\ref{subsec:rectum}. In general there may be many different ways to obtain an accurate segmentation. Ideally, these should be robustly working for many different images, both in 2D and in 3D. In future work we plan to investigate this in more detail and compare various variants from the point of view of robustness, accuracy and computational efficiency.  Regarding the latter, the 2D analysis of GBM was performed on a standard laptop (equipped with an Intel CORE i7 CPU, and 8 gigabytes of RAM) and performed in a little less than 1 minute, which as a first indication is in line with the current state-of-the-art.
 
%

\subsection{Example: segmentation of rectal carcinoma}
  \label{subsec:rectum}

  As a further example we detail an analysis aimed at segmentation of
  rectal cancer. Rectal carcinoma is a frequent pathology
  \cite{Siegel2016} and the survival rate after radical surgery is
  much improved in case of early diagnosis. Therefore, identifying the
  tumour by diagnostic imaging has a key role in the output of the
  treatment. Furthermore, segmenting the tumour in images is
  an important step in preparation for radiotherapy.
  Rectal cancer MR imaging protocols usually include T2w images of the
  pelvic district, which is considered the key sequence for the diagnosis of
  rectal cancer. However, several studies have underlined the importance of
  using DWI (\emph{Diffusion-weighted imaging}) sequences for a more
  detailed study of the disease \cite{Sun2014}.
  Briefly, MR-DWI images measure the degree of diffusion of water molecules
  through imaged tissues. Changes in tissues caused by the growth of a
  tumour (apoptosis, necrosis, increased vascularity) modify the
  effective diffusive capacity of water molecules in that area, and
  DWI is useful to capture this phenomenon. The properties of
  diffusion are quantified out of DWI images building so-called ADC
  (\emph{apparent diffusion coefficient)} maps. ADC maps are
  \emph{hyperintense} in areas where water diffusion is \emph{free} and
  \emph{hypointense} in areas where water diffusion is \emph{restricted} due to
  the presence of obstacles.
  Rectal carcinomas have intermediate grey levels in T2w and are
  \emph{hypointense} in ADC maps.

  Since positioning of \emph{regions of interest} (ROIs) has a
  considerable influence on tumour ADC values \cite{Lambregts2011},
  instead of using the T2w images and then co-registering segmentation
  output to ADC maps, we perform the segmentation of rectal cancer
  directly on ADC maps for more accurate results.

  Differently from Section~\ref{subsec:GBM}, segmentation of rectal
  cancer is performed using the 3D volume of the image as a whole,
  rather than considering separate slices. In our experimentation, 3D
  analysis has yielded better results, as the considered regions are
  rather small and reasoning simultaneously on different slices
  maximises the information which is available to each analysis
  pass. The segmentation process is composed of four steps:

  \begin{enumerate}
  \item preprocessing (performed without use of \topochecker), aimed at
    normalisation of images;
  \item rectum segmentation in T2w images, to limit the area of the
    image where the tumour is searched for; rectum segmentation is done in
    T2w as the contrast of ADC is not sufficient to properly
    distinguish organs;
  \item co-registration of rectum segmented in T2w to ADC, using patient positioning information that is stored in images by the scanner;   
  \item tumour segmentation in the ADC map, which is the stated goal of this example.
  \end{enumerate}

\subsubsection{Preprocessing}
\begin{figure*}
\centering
\subfloat[][T2w axial slice.]
{
\includegraphics[height=3.8cm]{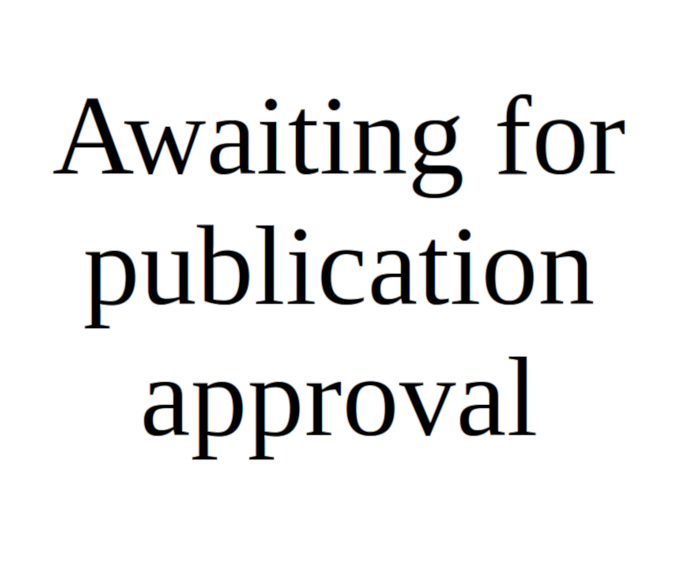}
\label{subfig:rectumT2ax}
}~~~
\centering
\subfloat[][T2w sagittal slice.]
{
\includegraphics[height=3.8cm]{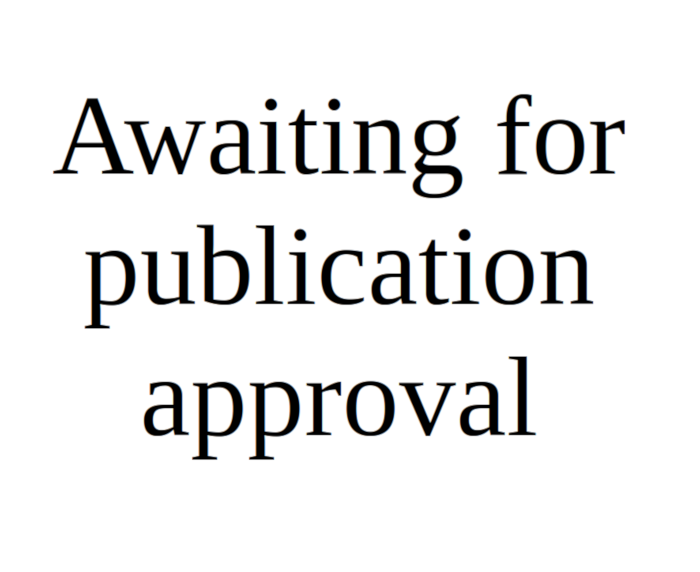}
\label{subfig:rectumT2sag}
}\\
\subfloat[][ADC map axial slice.]
{
\includegraphics[height=3.8cm]{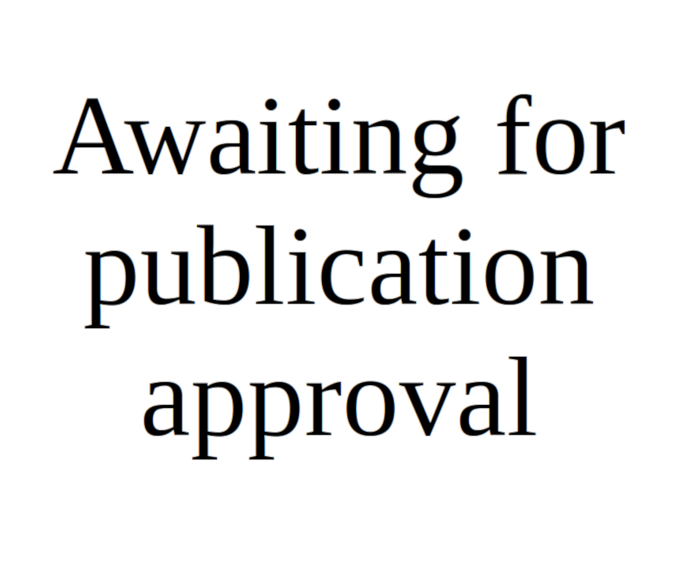}
\label{subfig:rectumADCax}
}~~~
\centering
\subfloat[][ADC map sagittal slice.]
{
\includegraphics[height=3.8cm]{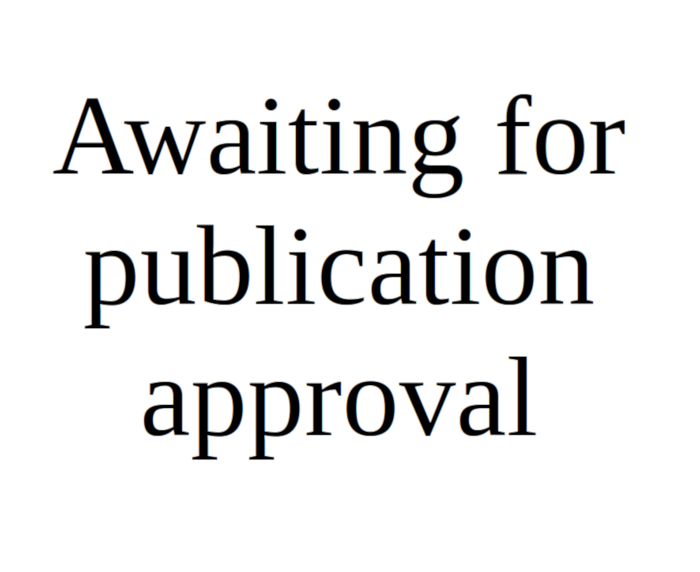}
\label{subfig:rectumADCsag}
}
\caption{Rectum acquisition \Siena.}
\label{fig:rectumAcq}
\end{figure*}

Figure~\ref{fig:rectumAcq} shows one axial and one sagittal view of
T2w and ADC acquisitions.

\begin{figure*}
\centering
\subfloat[][DWI slice.]
{
\includegraphics[height=3.5cm]{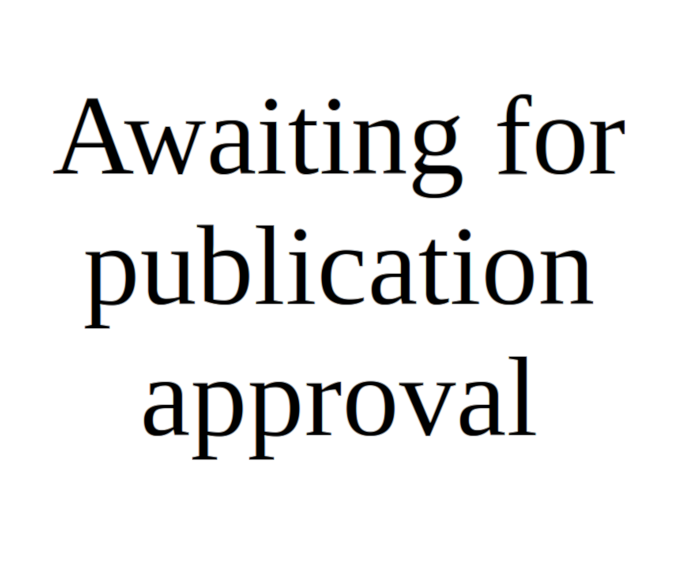}
\label{subfig:DWI}
}~~~
\centering
\subfloat[][ADC slice.]
{
\includegraphics[height=3.5cm]{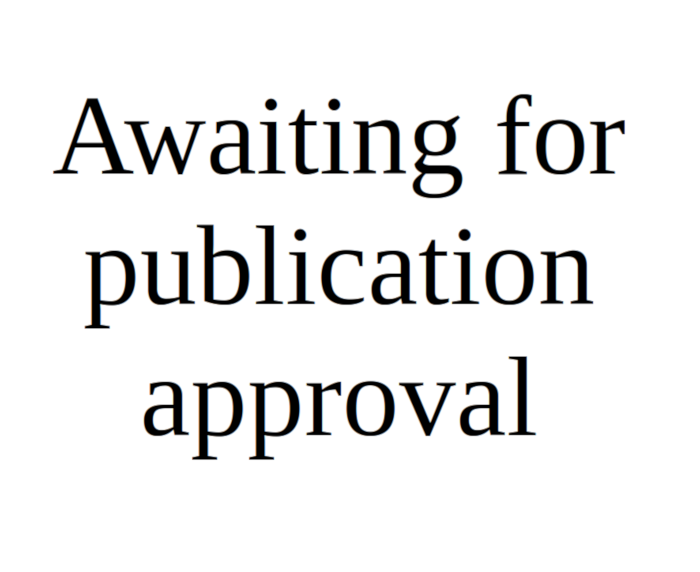}
\label{subfig:ADC}
}~~~
\subfloat[][Mask on ADC.]
{
\includegraphics[height=3.5cm]{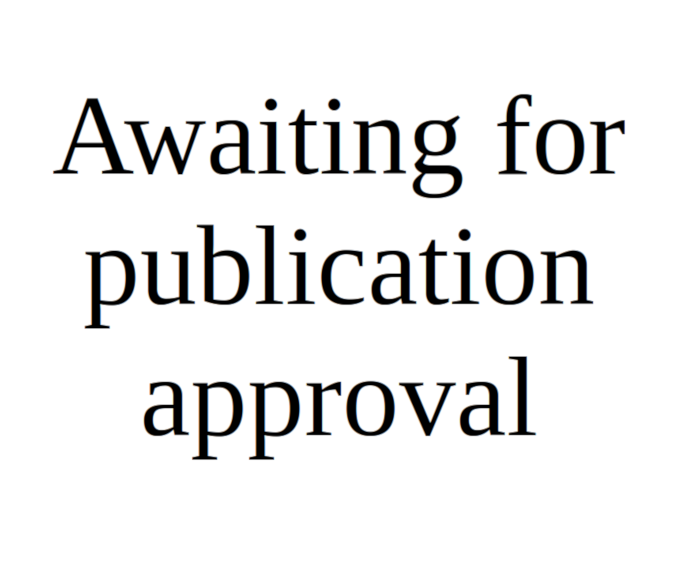}
\label{subfig:ADCmask}
}
\caption{ADC mask \Siena}
\label{fig:ADCmask}
\end{figure*}

Since the FOV (\emph{Field of View}) of the T2w acquisition lies entirely
within the patient body
(Figure~\ref{subfig:rectumT2ax} and Figure~\ref{subfig:rectumT2sag}),
normalisation of the T2w volume is obtained dividing the grey level of
each voxel by the average of voxel intensities. For ADC maps instead,
a mask is created using a procedure similar to that described in
Section~\ref{subsec:GBM}. However,
  we used the DWI images to obtain the mask, because in the ADC maps the
  background is very noisy (see Figure~\ref{fig:ADCmask} -- note that DWI and ADC masks have the same coordinate system).

\subsubsection{Rectum segmentation}
The model definition for rectum segmentation defines the relevant attribute, i.e. the normalised grey level of each pixel in this image, as \mv{T2}, which is used throughout the analysis.

{\small
\begin{verbatim}
Model "med:T2=T2-NORM.nii";
\end{verbatim}
}

We slightly change the definition of \verb!flt!
(defined in Figure~\ref{fig:morederived}   ($\mathcal{J}$),
and used in Section~\ref{subsec:GBM} 
to remove small regions attributed to noise). We consider regions that only appear on one slice as noise, even when these are not filtered out by the previous definition. In order to 
remove such regions, we employ nested application of \verb!I! and \verb!N! on top of the definition
of \verb!flt!.

{\small
\begin{verbatim}
Let flt3D(a) = N(I(flt(a)));
\end{verbatim}
}

The area corresponding to the rectum in T2w images is identified as the union of a
\emph{hyperintense} region, called \verb!hyperT2r!, a \emph{hypointense}
region, called \verb!hypoT2r!, and a region having an intermediate intensity, called
\verb!intermT2r!, that are close to each other. This is detected using the \verb!touch! operator.


The aforementioned hyperintense region is defined as \verb!hyperT2r! below.

{\small
\begin{verbatim}
Let hyperT2=flt3D([T2>1.6]);
Let hyperT2Super = flt3D([T2>2.5]);
Let hyperT2r = touch(hyperT2,hyperT2Super);
\end{verbatim}
}


The hypointense region \verb!hypoT2r! is defined below as being within 5mm from \verb!hyperT2r!.

{\small
\begin{verbatim}
Let hypoT2 = flt3D([T2>0.17] & [T2<0.5]);

Let hyperT2rS = MDDT(hyperT2r,<5);
Let hypoT2r = touch(hypoT2,hyperT2rS);
\end{verbatim}
}

Finally, the region of intermediate intensity \verb!intermT2r! is defined as follows.

{\small
\begin{verbatim}
Let rectum1S = MDDT(hyperT2r |  hypoT2r,<5);

Let intermT2 = flt3D([T2>0.9] & [T2<1.4]);
Let intermT2r = touch(intermT2,rectum1S);
\end{verbatim}
}

\begin{figure*}
\centering
\subfloat[][T2w axial slice.]
{
\includegraphics[height=4cm]{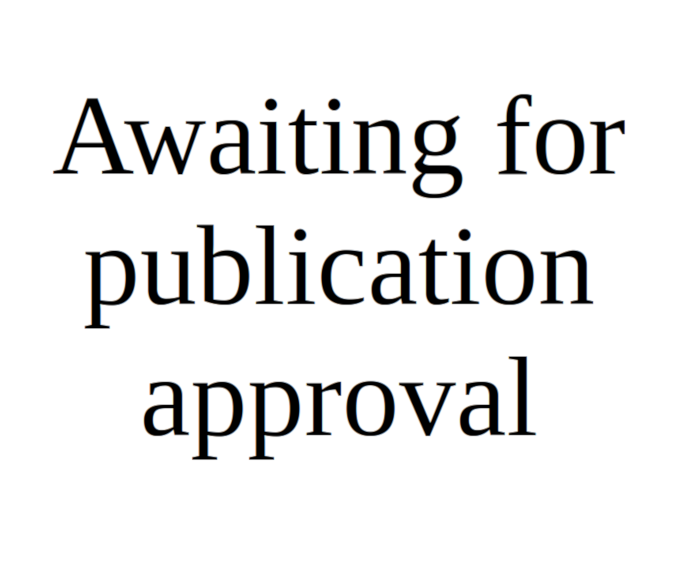}
\label{subfig:rectumSeg1ax1}
}~~~
\centering
\subfloat[][T2w axial slice.]
{
\includegraphics[height=4cm]{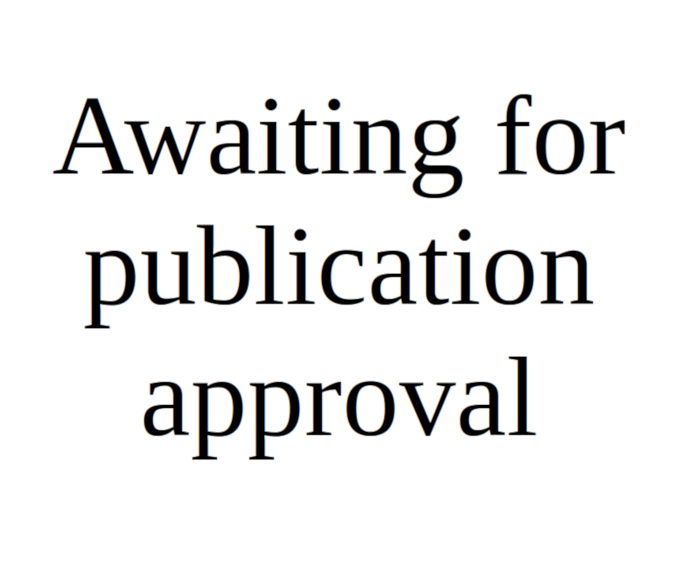}
\label{subfig:rectumSeg1ax2}
}\\
\subfloat[][T2w sagittal slice.]
{
\includegraphics[height=3.55cm]{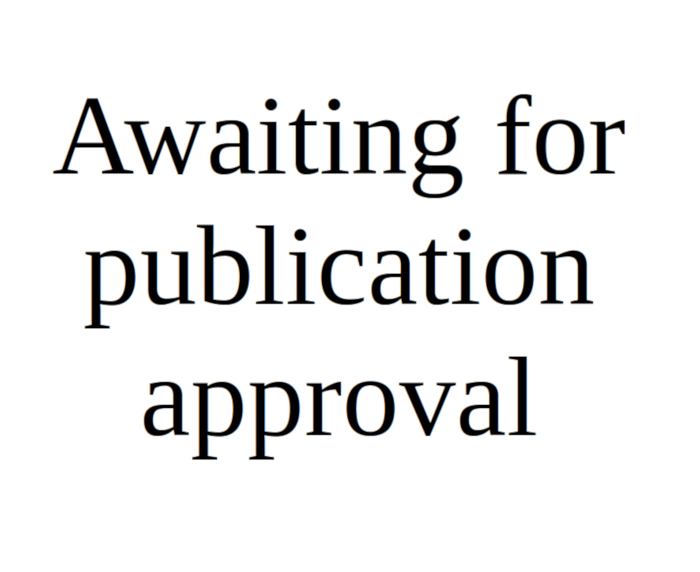}
\label{subfig:rectumSeg1sag}
}~~~
\centering
\subfloat[][T2w coronal slice.]
{
\includegraphics[height=3.55cm]{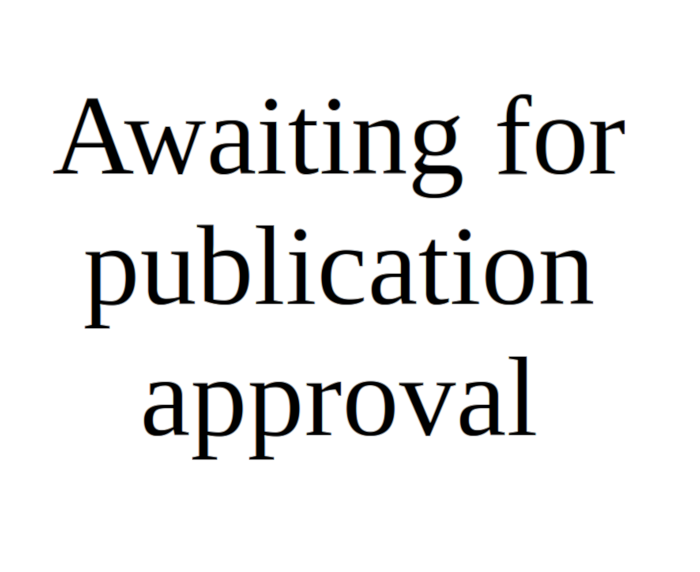}
\label{subfig:rectumSeg1cor}
}
\caption{Hyperintense (green), hypointense (brown) and intermediate
    intensity (red) regions used to segment rectum in T2w \Siena.}
\label{fig:RectumSeg1}
\end{figure*}

\begin{figure*}
\centering
\subfloat[][T2w axial slice.]
{
\includegraphics[height=4cm]{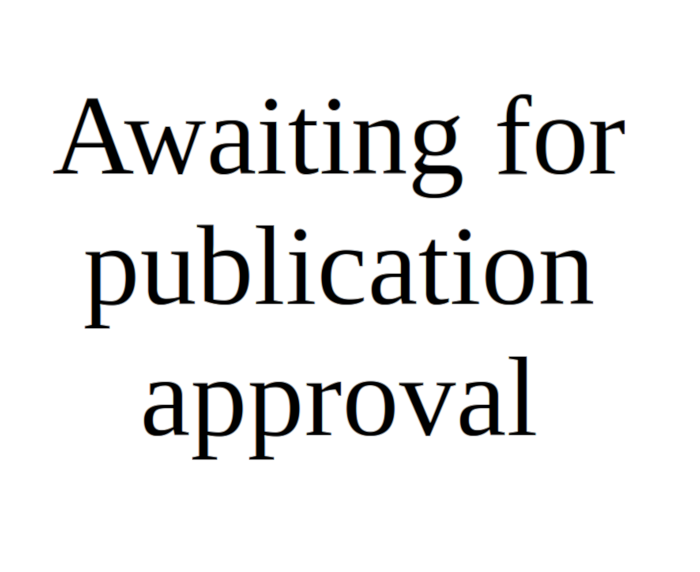}
\label{subfig:rectumSegax1}
}~~~
\centering
\subfloat[][T2w axial slice.]
{
\includegraphics[height=4cm]{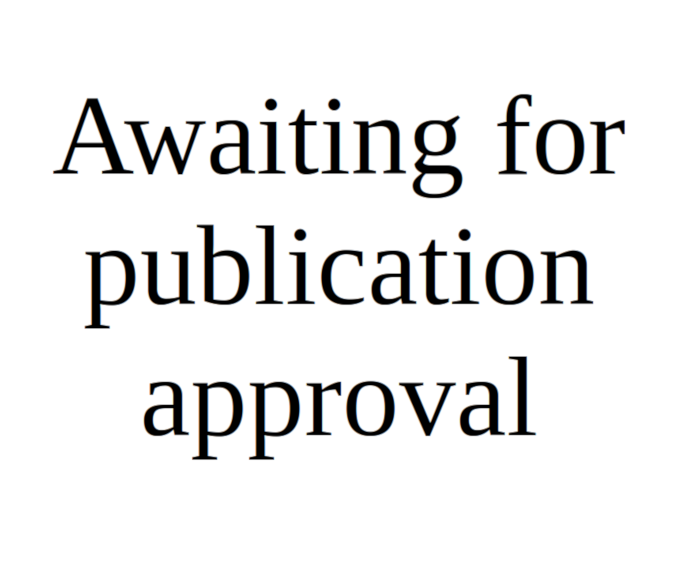}
\label{subfig:rectumSegax2}
}\\
\subfloat[][T2w sagittal slice.]
{
\includegraphics[height=4cm]{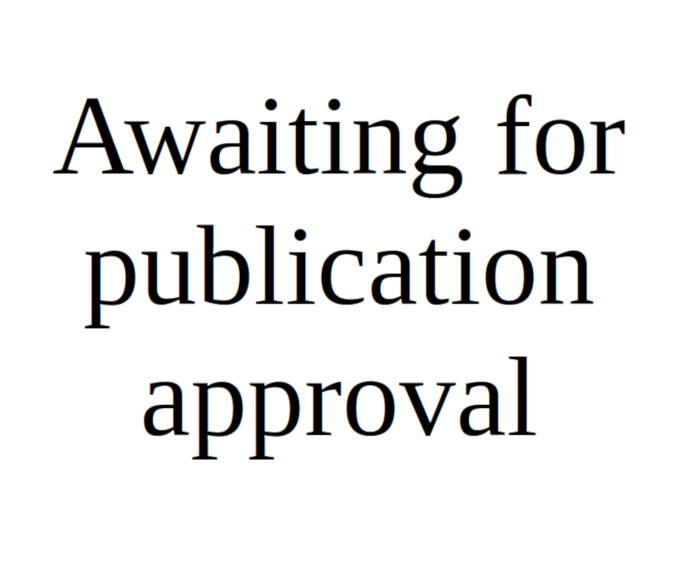}
\label{subfig:rectumSegsag}
}~~~
\centering
\subfloat[][T2w coronal slice.]
{
\includegraphics[height=4cm]{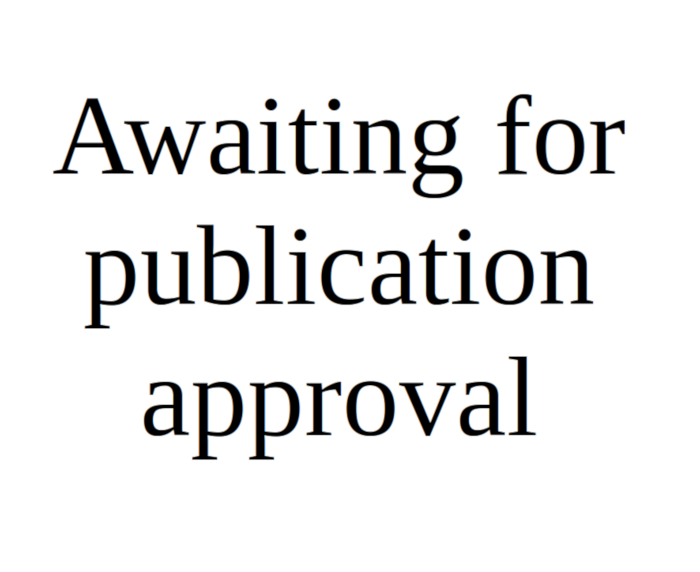}
\label{subfig:rectumSegcor}
}
\caption{Final output of segmentation of rectum in T2w. In green
    \texttt{rectumS} \Siena.}
\label{fig:RectumT2}
\end{figure*}

The segmented rectum (formula \verb!rectum! below) is defined as the
union of the three regions (the green, brown and red areas in
Figure~\ref{fig:RectumSeg1}); the area is expanded, in formula
\verb!rectumS!, to cater for loss of precision that occurs in the
co-registration to the ADC map (see the green area in
Figure~\ref{fig:RectumT2}).

{\small
\begin{verbatim}
Let rectum = hyperT2r |  hypoT2r | intermT2r;
Let rectumS = MDDT(rectum,<5);
\end{verbatim}
}

\begin{figure}
  \centering
  \includegraphics[width=0.47\textwidth]{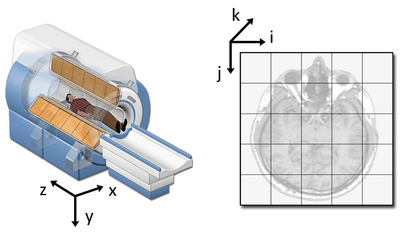}
  \caption{World $(x,y,z)$ and image $(i,j,k)$ coordinate
    systems. Image based on image shared in
    \texttt{https://www.slicer.org/wiki/Coordinate\_systems}}
  \label{fig:coordSys}
\end{figure}

\paragraph{Co-registration}
T2w and DWI images are acquired in the same
acquisition session with the patient. Co-re\-gi\-stra\-tion between the T2w
and the ADC maps makes use of the image header, that in medical images contains the necessary information to translate image
coordinate systems to the scanner (world) coordinate system. More precisely, each voxel has
coordinates $(i,j,k)$ within the image and dimension
$(ps_i,ps_j,ps_k)$. In addition, also the
corresponding position of the voxel in world coordinates is stored in the header. 
Using such information, the
coordinates $(i,j,k)$ of each voxel within the image is mapped to the position
$(x,y,z)$ of the voxel in world coordinates
(Figure~\ref{fig:coordSys}).

\begin{figure*}
\centering
\subfloat[][ADC axial slice.]
{
\includegraphics[height=4cm]{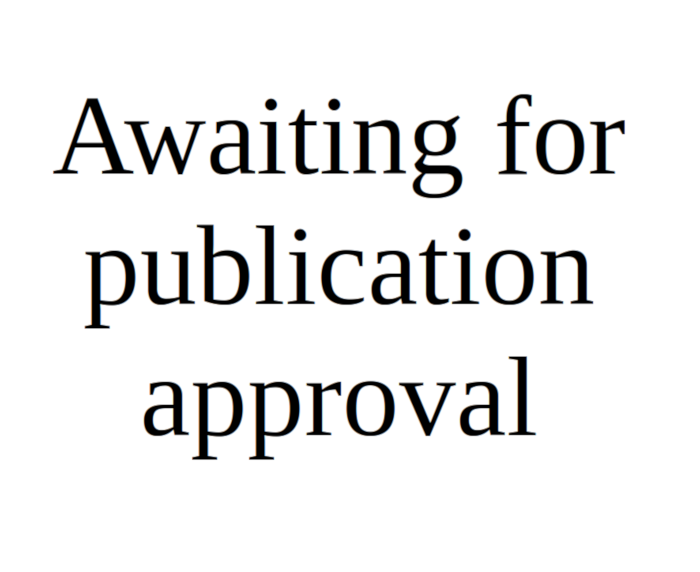}
\label{subfig:rectum2ADCax1}
}~~~
\centering
\subfloat[][ADC axial slice.]
{
\includegraphics[height=4cm]{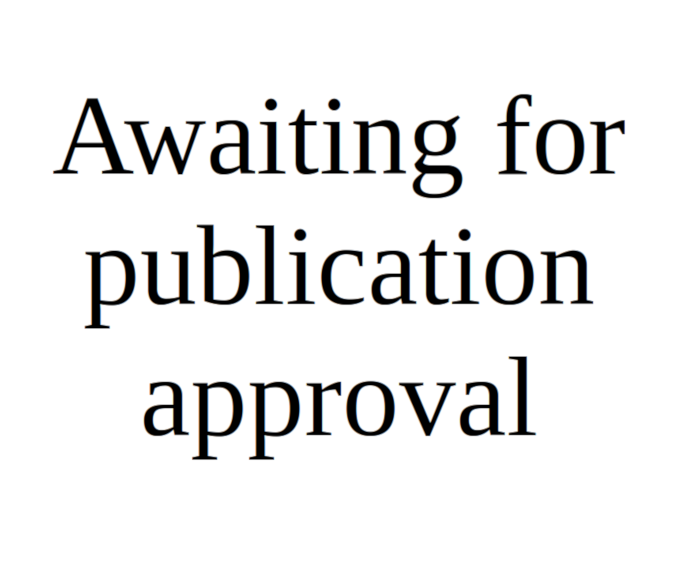}
\label{subfig:rectum2ADCax2}
}\\
\subfloat[][ADC sagittal slice.]
{
\includegraphics[height=4cm]{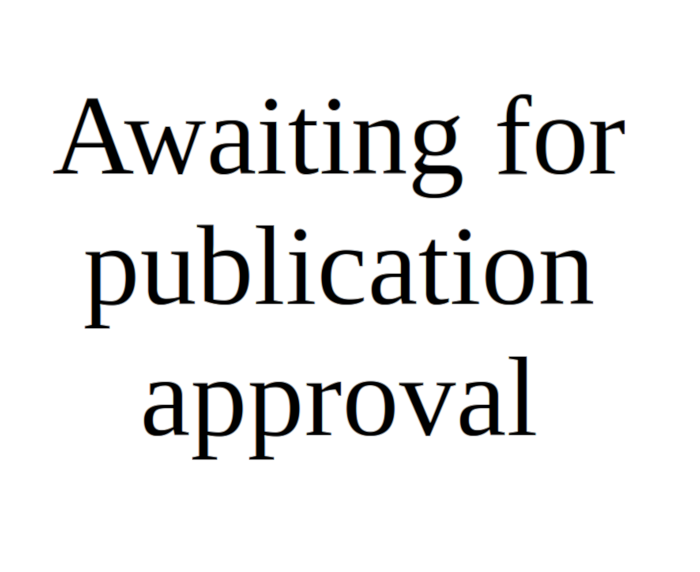}
\label{subfig:rectum2ADCsag}
}~~~
\centering
\subfloat[][ADC coronal slice.]
{
\includegraphics[height=4cm]{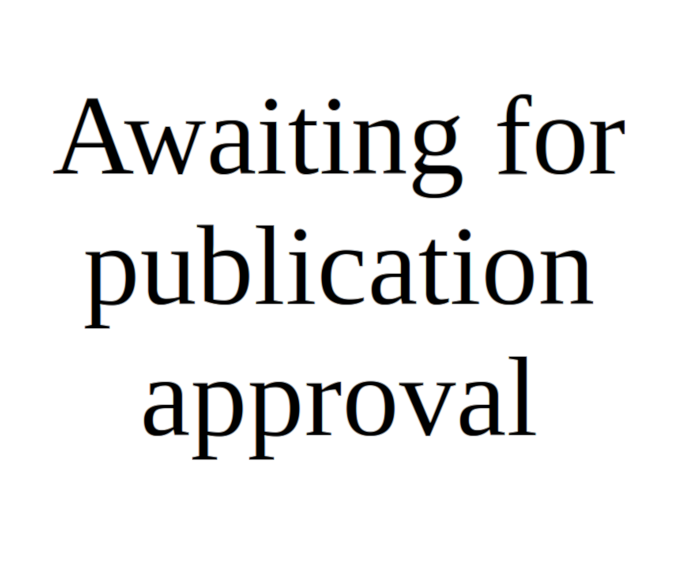}
\label{subfig:rectum2ADCcor}
}
\caption{Co-registration of rectum ROI (green) segmented in T2w to ADC map \Siena.}
\label{fig:ROI_2_ADC}
\end{figure*}

In order to co-register the ROI of the rectum segmented in T2w to the
ADC map, we map the image coordinates $(i,j,k)_{T2}$ of the T2w image
to the world coordinates (x,y,z) and back to the image coordinates
$(i,j,k)_{ADC}$ of the ADC map. In Figure~\ref{fig:ROI_2_ADC}, the
  green area represents voxels on the ADC map that correspond to
  voxels in T2w satisfying \verb!rectumS!.

\subsubsection{Tumour segmentation}

Finally, tumour segmentation is performed on the ADC map. Below, we load the (normalised) ADC
map (\verb!ADC!) and the rectum segmented on T2w and co-registered to
ADC (\verb!ROI!). We define formula \verb!rectumS! selecting voxels
defined in \verb!ROI!. 

{\small
\begin{verbatim}
Model "med:ADC=ADC-norm.nii,
            ROI=ROI_T2-2-ADC.nii";

Let rectumS=[ROI>0];
\end{verbatim}
}

\begin{figure*}
\centering
\subfloat[][Output of threshold operator on one ADC axial slice.]
{
\includegraphics[height=3cm]{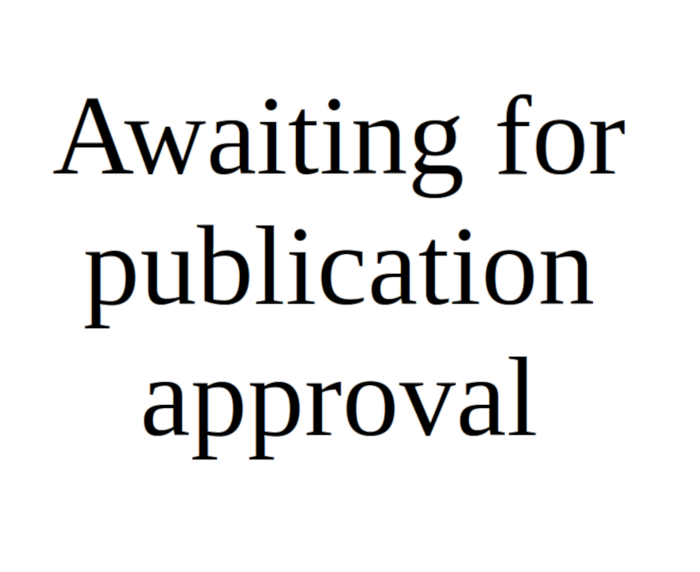}
\label{subfig:tumorThrAx}
}~~~
\centering
\subfloat[][Output of threshold operator on one ADC sagittal slice.]
{
\includegraphics[height=3cm]{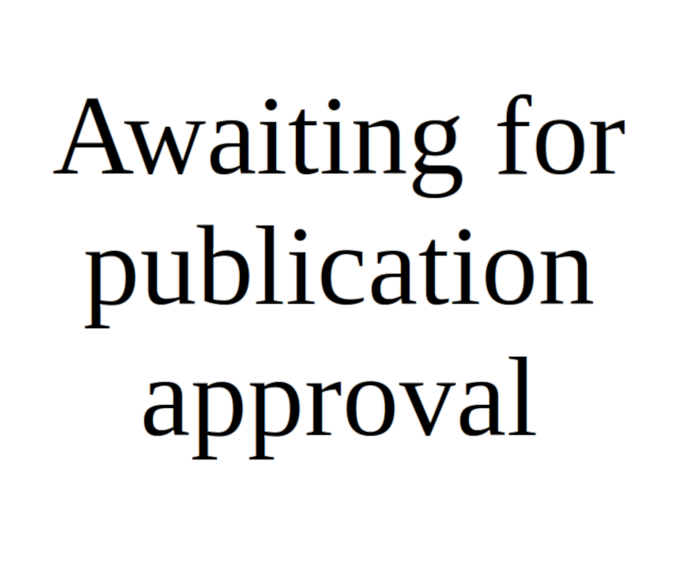}
\label{subfig:tumorThrSag}
}~~~
\centering
\subfloat[][Output of threshold operator on one ADC coronal slice.]
{
\includegraphics[height=3cm]{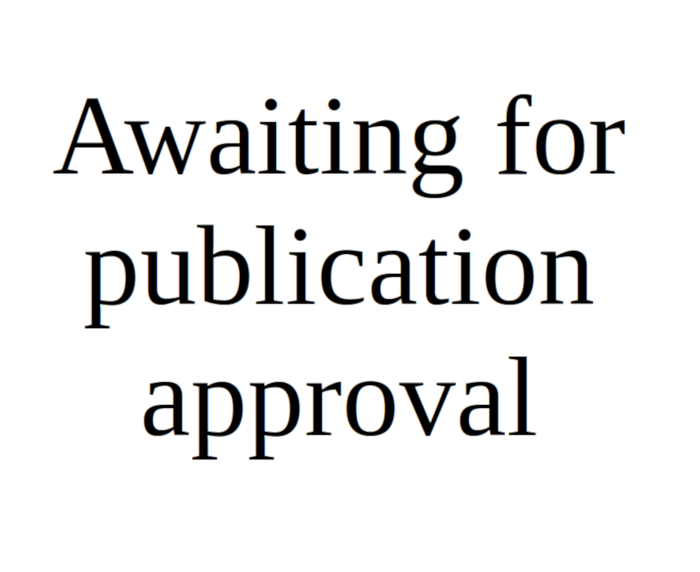}
\label{subfig:tumorThrCor}
}\\
\centering
\subfloat[][Output of \texttt{flt3D} operator on one ADC axial slice.]
{
\includegraphics[height=3cm]{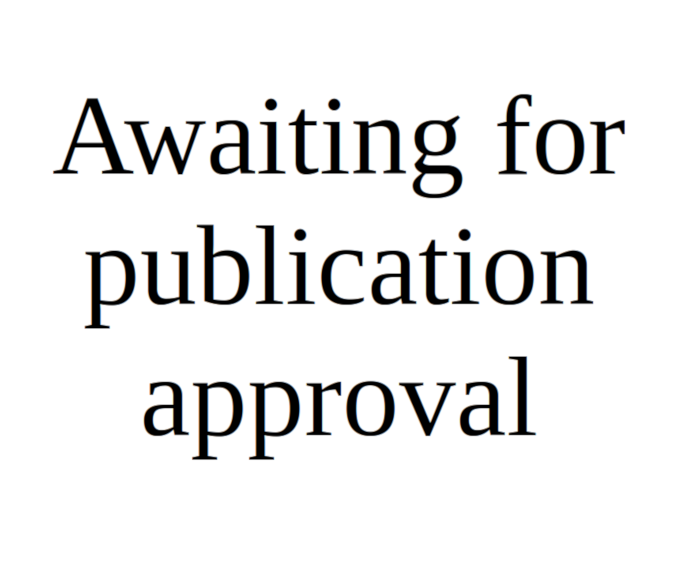}
\label{subfig:tumorFltAx}
}~~~
\centering
\subfloat[][Output of \texttt{flt3D} operator on one ADC sagittal slice.]
{
\includegraphics[height=3cm]{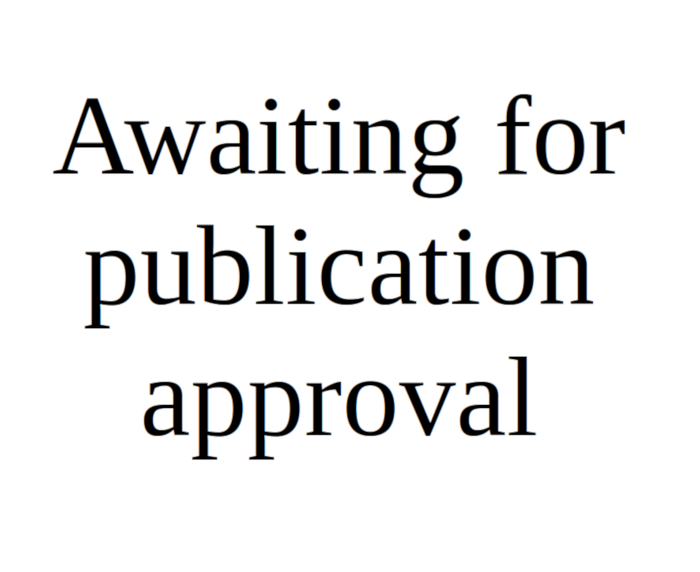}
\label{subfig:tumorFltSag}
}~~~
\centering
\subfloat[][Output of \texttt{flt3D} operator on one ADC coronal slice.]
{
\includegraphics[height=3cm]{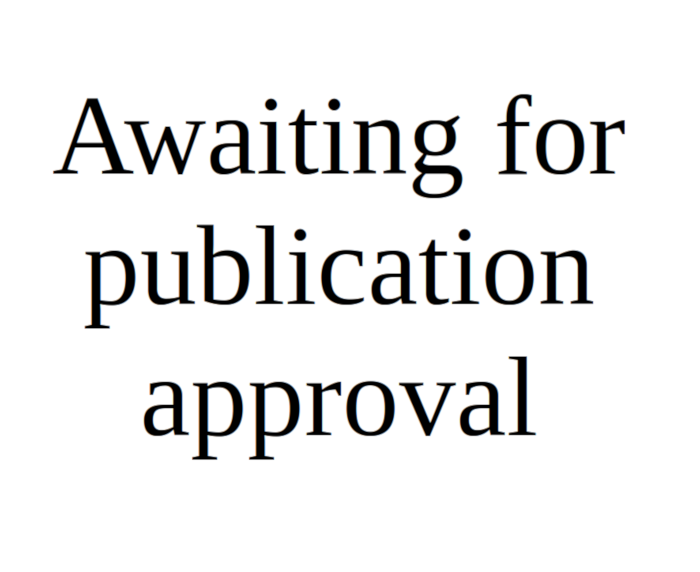}
\label{subfig:tumorFltCor}
}\\
\centering
\subfloat[][Output of \texttt{touch} operator on one ADC axial slice.]
{
\includegraphics[height=3cm]{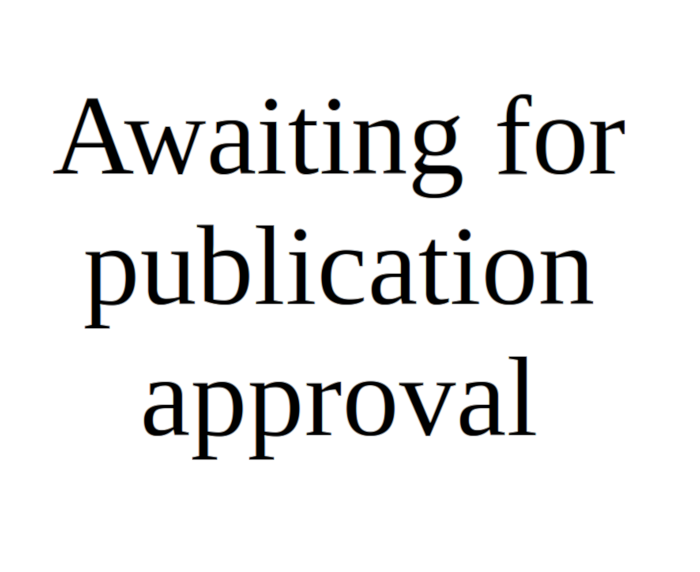}
\label{subfig:tumorReachAx}
}~~~
\centering
\subfloat[][Output of \texttt{touch} operator on one ADC sagittal slice.]
{
\includegraphics[height=3cm]{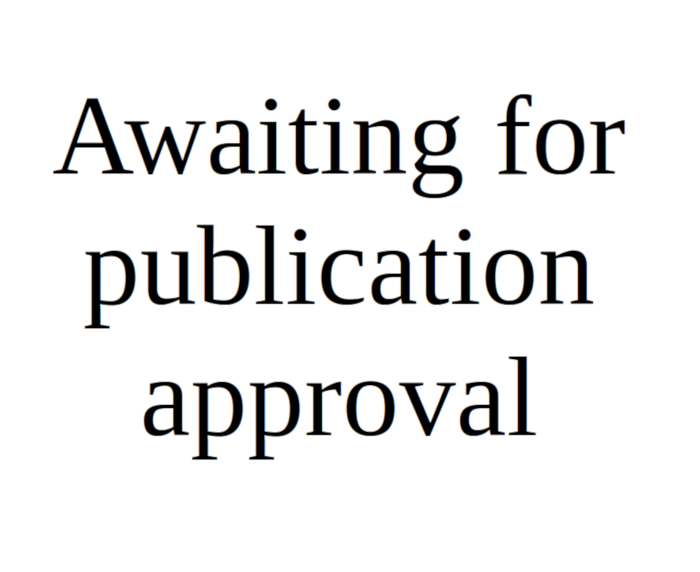}
\label{subfig:tumorReachSag}
}~~~
\centering
\subfloat[][Output of \texttt{touch} operator on one ADC coronal slice.]
{
\includegraphics[height=3cm]{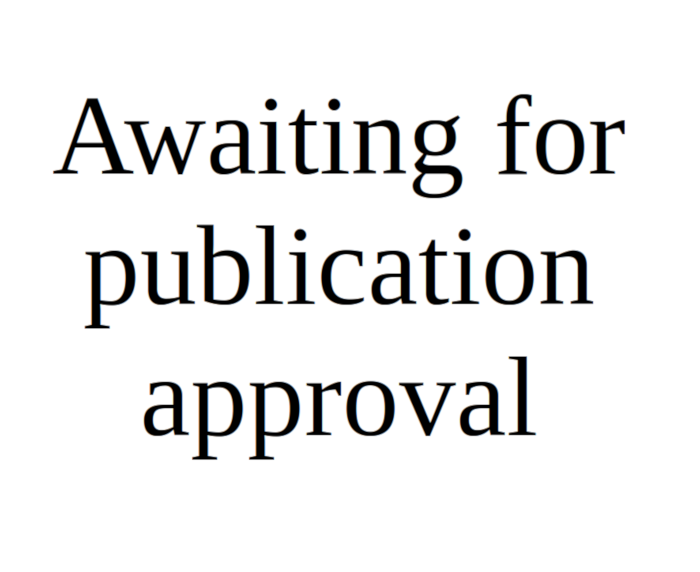}
\label{subfig:tumorReachCor}
}\\
\caption{Output of segmentation of the tumour (green) and the rectum (brown) \Siena.}
\label{fig:rectum-thrfltreach}
\end{figure*}

\noindent We delineate the initial estimate of the tumour based on grey levels
(the green area in Figure~\ref{subfig:tumorThrAx}-\ref{subfig:tumorFltCor}).

{\small
\begin{verbatim}
Let tumor1 = [ADC>0.96] & [ADC<1.56];
Let tumor2 = flt3D(tumor1);
\end{verbatim}
}

Formula \verb!tumour3!, below, constrains the tumour region to an area
that touches the ROI that has been segmented for the rectum
(see the green area in Figure~\ref{subfig:tumorReachAx},\ref{subfig:tumorReachSag},\ref{subfig:tumorReachCor})

{\small
\begin{verbatim}
Let tumor3 = touch(tumor2,rectumS);
\end{verbatim}
}

\begin{figure*}
\centering
\subfloat[][ADC axial slice.]
{
\includegraphics[height=3cm]{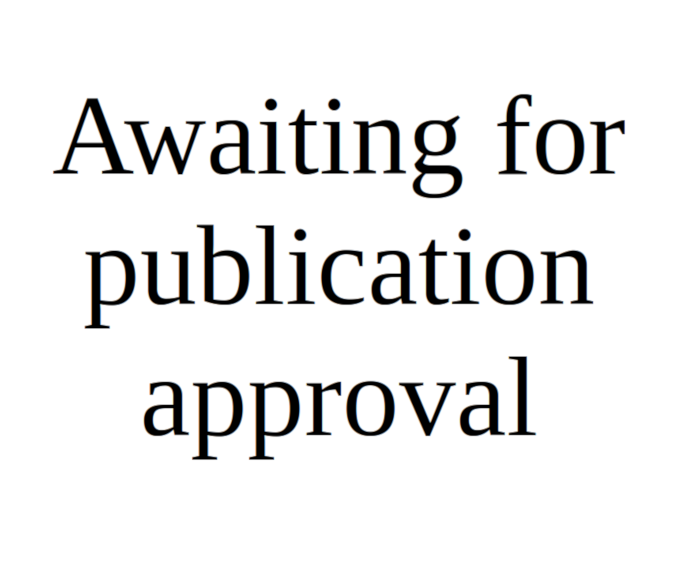}
\label{subfig:tumorStatAx}
}~~~
\centering
\subfloat[][ADC sagittal slice.]
{
\includegraphics[height=3cm]{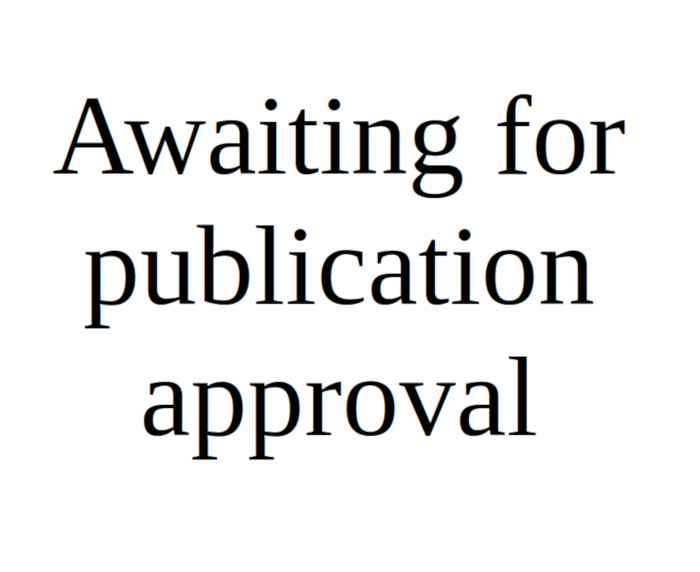}
\label{subfig:tumorStatSag}
}~~~
\centering
\subfloat[][ADC coronal slice.]
{
\includegraphics[height=3cm]{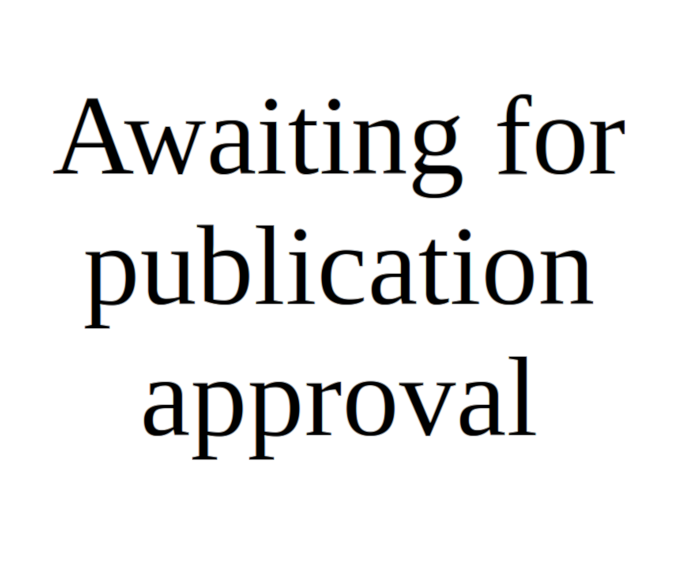}
\label{subfig:tumorStatCor}
}\\
\caption{Output of the \texttt{SCMP} operator (green) and the
    searching space \texttt{tumorSpace} (brown) \Siena.}
\label{fig:rectum-stat}
\end{figure*}

Statistical texture analysis is then used to find regions that are
similar to \verb!tumor3! (cross correlation $> 0.8$). The search is
restricted to areas close to \verb!tumor3!, i.e. the region of radius 20mm
around \verb!tumor3! defined as \verb!tumorSpace!
(Figure~\ref{fig:rectum-stat}).

{\small
\begin{verbatim}
Let tumorSpace = MDDT(tumor3,<20);
Let tumorStat = 
SCMP(ADC,tumorSpace,3,>0.8,0.01,2.7,100)
(ADC,tumor3);
\end{verbatim}
}

\begin{figure*}
\centering
\subfloat[][ADC axial slice.]
{
\includegraphics[height=4cm]{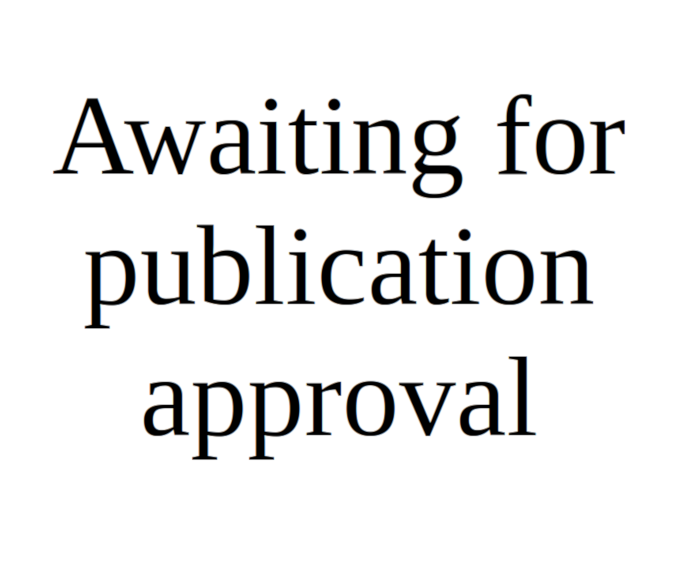}
\label{subfig:tumorFinalAx}
}~~~
\centering
\subfloat[][ADC axial slice.]
{
\includegraphics[height=4cm]{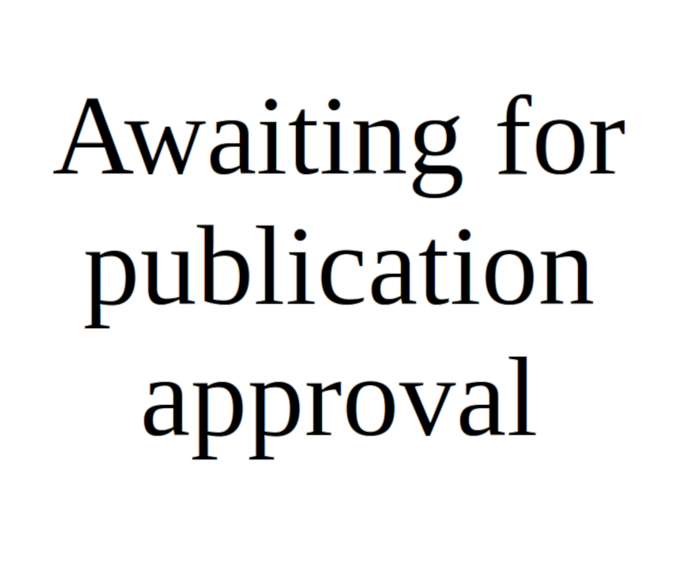}
\label{subfig:tumorFinalAx2}
}\\
\centering
\subfloat[][ADC sagittal slice.]
{
\includegraphics[height=4cm]{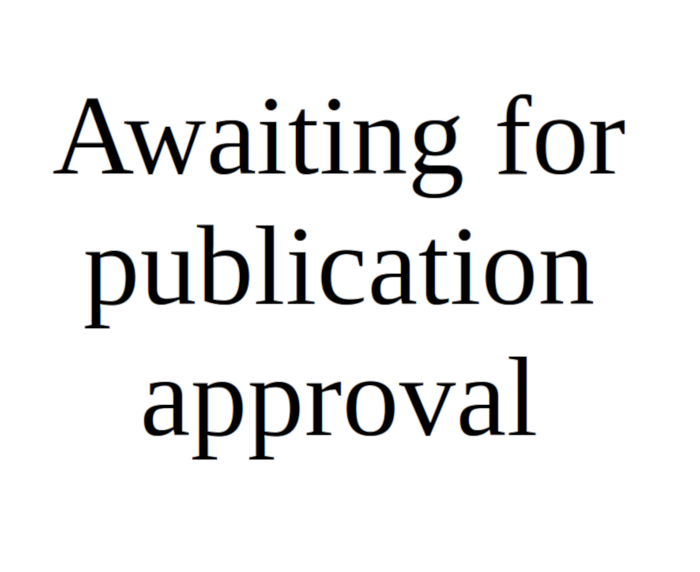}
\label{subfig:tumorFinalSag}
}~~~
\centering
\subfloat[][ADC coronal slice.]
{
\includegraphics[height=4cm]{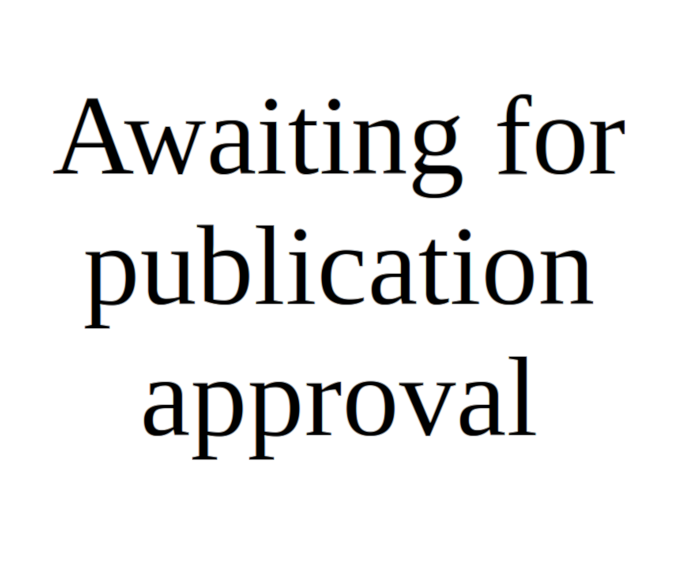}
\label{subfig:tumorFinalCor}
}\\
\caption{Final output of rectal cancer segmentation in green \Siena.}
\label{fig:rectum-final}
\end{figure*}

Finally, the tumour region is the union of \verb!tumor3! and \verb!tumorStat! shown in green in Figure~\ref{fig:rectum-final}.

{\small
\begin{verbatim}
Let tumor = tumor3 | tumorStat;
\end{verbatim}
}

\subsection{Validation}\label{subsec:validation}

The work presented in this section is aimed at providing an illustration of the analysis capabilities of our logic-based methodology, rather than providing complete clinical case studies. For instance, consider the glioblastoma specification, which is rather concise, consisting of a less than 30 lines long logical specification, and a simple preprocessing step. Although such procedure was successfully tested on five images from different sources and
acquired in very different conditions, this is certainly not sufficient to validate our example as a glioblastoma segmentation methodology for future clinical usage. 
Future work aims at improving the method, eliminating corner cases in the formulas as much as possible, making it robust to different acquisition conditions and properly validating it. More generally speaking, clinical experimentation is the next step in our research program.
However, some conclusions can already be drawn from the data we have, both with respect to efficiency and to accuracy of the obtained results. We do so for the glioblastoma case study, as there is enough literature for a comparison with the state of the art. For rectum carcinoma, less data is available, and clinical testing will be essential in understanding the applicability of the procedure.

Analysis time is proportional to the size of the image  (the algorithm is linear). In the glioblastoma example, MR-FLAIR very often have a slice size of $256 \times 256$ pixels, multiplied by $20-30$ slices. As a rough estimate, the execution time for the analysis of a single $1024 \times 1024$-voxels slice---including preprocessing---on a standard laptop (equipped with an Intel CORE i7 CPU, and 8 gigabytes of RAM), currently stays below one minute.
This information, although not being a fully-fledged benchmark, provides a first indication that, efficiency-wise, our approach is in par with the state-of-the art in semi-automatic glioblastoma segmentation procedures (see \cite{Fyllingen2016}). We remark that our procedure makes use of a prototype general-purpose model checker, that could be amenable to further optimisation, e.g. employing specialised,  well-known flood-filling algorithms for images for model checking the {\em surrounded} connective --- in place of the current graph-theoretical method. 


A preliminary assessment of the quality of the obtained results in the case of glioblastoma was
  performed for the patient in Figure~\ref{fig:GBMpatSeg}. The patient
  underwent first surgery and then radiotherapy. We compared our
  results on the post-surgery MR-FLAIR with target volumes delineated on
  the pre-treatment Computed Tomography (CT) by one experienced radiotherapist. In
  particular, we considered the \emph{gross tumour volume} (GTV),
  i.e. what can be seen or imaged, and the \emph{clinical target
    volume} (CTV), which contains the GTV, plus a margin for
  sub-clinical disease spread which therefore cannot be fully imaged
  \cite{Burnet2004}. Usually for glioblastomas the CTV is defined as a 2-2.5 cm
  isotropic expansion of GTV within the brain. In order to quantify
  the effectiveness of our segmentation we computed the \emph{Dice
    coefficient} ($\mathit{DC}$), that we used to measure the morphological
  similarity between the manual segmentation $\mathit{MS}$ and automatic
  segmentation $\mathit{AS}$. The coefficient is defined as
  $DC=\frac{2V(\mathit{MS}\cap \mathit{AS})}{V(\mathit{MS})+V(\mathit{AS})}$, where $V(a)$ is the
  volume of $a$, that is, the number of voxels that belong $a$; $\mathit{DC}$ ranges from 0 to 1, 0
  indicates no overlap and 1 indicates complete overlap.

  The CT volume was co-registered to the FLAIR volume. Then, we considered 
  the region $R$ obtained as the union of the oedema and tumour regions, as found using
  our method. We compared $R$ to the GTV contour, and furthermore we compared $R$, expanded by 2.5cm (as explained above) to the CTV contour. We obtained
  $DC=0.76$ for GTV and $DC=0.81$ for CTV. Although a single case does not have clinical significance, these results are very encouraging, and aligned with
  state-of-the-art methods for automatic and semi-automatic segmentation of glioblastoma \cite{Dupont2016}.


In \cite{Be+17}, a variant of the method we described was assessed on a dataset of 7 patients affected by GBM, that have undergone radiotherapy. The obtained results, and average execution time per patient on the same machine used for the experiments in the current paper, were in line with the state-of-the-art.

Finally, we note that numeric thresholds and other parameters (e.g., the number of nested \mv{N} constructs in some formulas, the number of bins for statistical analysis, etc.) have been chosen by the medical physicist in charge of the analysis, on the basis of expert knowledge on the matter and in some cases by trial-and-error. The values that we used might prove stable in clinical validation (and this is the purpose of the preliminary normalisation of images that we use), but this is not yet to be taken for granted, or even to be expected in more general situations. Instead, \emph{parameter calibration} on a per-image or per-study basis will be an important subject in our future research. Such calibration may be fully automatic (e.g. through \emph{machine learning} techniques), but this is just one possibility. It would also make sense to adopt a semi-automatic approach (which is also frequent in state-of-the-art techniques, see e.g. \cite{Dupont2016,Fyllingen2016,Simi2015,Zhu2012}), involving human interaction with an expert to merely calibrate the parameters, rather than performing a full manual segmentation, in order to save a large part of the time (and costs) required for preparation to radiotherapy or surgery.

\section{Conclusions and future work}
\label{sec:conclusions}

This work provides a first, promising exploration of logical methods for declarative medical
image analysis in the domain of radiotherapy. A declarative approach makes analysis  transparent, reproducible, human-readable, exchangeable, and permits domain experts who are not technicians to understand the specifications. Such advantages  are akin to those obtained in other domains, such as the application of the \emph{Structured Query Language} (SQL) in the field of databases, or the introduction of query languages (XPATH, XSLT, ...) in semi-structured data management. 

Logical properties are used as classifiers for points
of an image; this can be used both for colouring regions that may be
similar to diseased tissues, and therefore being diseased tissue in turn,
and for colouring regions corresponding to organs of the human
body. Envisaged applications range from \emph{contouring} to
\emph{computer-aided diagnosis}. Our logic \SLCSMI is able to predicate on both shortest-path and Euclidean distance at the same time, and \topochecker implements both operators. In MI, shortest-path distances proved useful so far mostly to speed up interactive development; this is implementation-dependent, as the Modified Dijk\-stra transform that we use (see Section~\ref{subsec:distance}) currently performs faster than Maurer distance transform in our tests. We also considered the embedding of specific operators for MI in \SLCSMI such as an operator for texture analysis based on first order statistical methods. Other options and operators could be considered following a similar approach, providing a way to include state-of-the-art analysis techniques that can be conveniently combined using the spatial operators of the logic. 

It is noteworthy that the analysis we designed for glioblastoma segmentation can be used, with mild modifications, also to analyse the whole 3D volume of an image at once. 3D analysis is a relatively new application in medical imaging, leveraging the precision/efficiency trade-off of more classical methods. Furthermore, 3D analysis may be combined with existing applications of \emph{3D printing} in preparation for surgery (see \cite{Rengier2010}), by providing practitioners with models of a patient's body, with the relevant regions, identified by our method,  printed in  different colours. Such aspects constitute a further interesting line of research for future work.

Part of our ongoing work consists in identifying novel logical operators that are useful in medical imaging. So far, we only used operators that classify individual pixels or voxels. However, drawing inspiration from the family of \emph{region calculi} (see \cite{HBSL}), one could also classify regions, taking advantage of ``collective'' observations on sets of voxels that belong to the same area. Some work in this direction is~\cite{CLLM16}, including the definition of operators related to connectedness of regions; further work will be directed to the investigation of properties related to the size of regions, or to their morphological properties. Also, the ``distance-bounded surrounded'' operator defined in~\cite{NBCLM15} could be useful in medical imaging. A limitation of the model checking algorithm in~\cite{NBCLM15} is its quadratic complexity. We have shown that the application of distance transforms yields a {\em linear} algorithm for a weaker variant of the bounded surrounded operator for the case of images (that is, regular grids). 

We recall that \topochecker is a spatio-temporal model checker.  Temporal reasoning could be exploited in future work to consider, for instance, the sequence of acquisitions of a patient in order to reason about the evolution of image features such as tumours, which is very important in radiotherapy applications.

Our experiments show that typical analyses carried out using spatial model checking in medical imaging require careful calibration of numeric parameters (for example, a threshold for the distance between a tumour and the associated oedema, or the size of areas identified by a formula, that are small enough to be considered \emph{noise}, and ought be filtered out). The calibration of such parameters might be performed using machine-learning techniques. In this respect, future work could be focused on application, in the context of our research line, of the methodology used in the development of the logic SpaTeL, aimed at signal analysis (see \cite{Gr+09,GBB14,BBMNS15,Bartocci2016}), that pursues \emph{machine learning} of the logical structure of image features. We emphasise that such a development, if implemented, would be a radical improvement in application of machine learning to medical imaging. It can be framed under the recent research trend on {\em explainable artificial intelligence}, as it would yield a procedure that can explain in terms of a human readable language the methodology that a machine learning algorithm extrapolates from data. Our topological approach to spatial logics would be a key enabling technique for this purpose, as the formulas obtained in the SpaTeL approach are not meant to be intelligible by humans. It is worth noting that machine learning and deep learning methods have also been applied to the detection of tumours in very recent literature \cite{SGBM16,Akkus2017}. On the other hand, our application of machine learning 
could as well be focused simply on the identification of numeric parameters, rather than logic formulas, that may depend on complex features of images. 

The example of glioblastoma that we illustrated in Section~\ref{subsec:GBM} has immediate practical relevance. As we already mentioned, cleanup and clinical validation of the procedure is in progress. The normalisation step that we employ could be improved using state-of-the-art methods (see \cite{Madabhushi2006,lemieux1999fast}, and the references therein).  

Planned future developments also include means for interactive refinement of analysis, based on visual fine-tuning of specific values (e.g. thresholds or distances) that may have a \emph{non-linear} effect on the results of complex queries, with significant impact on methods that require human interaction---e.g. \emph{interactive segmentation} in preparation for surgery, or \emph{contouring} for radiotherapy planning.

\section{Acknowledgments}

The authors wish to thank dr. Marco Di Benedetto for suggesting the application of distance transforms to reduce  the complexity of model checking of formulas with distances, and dr. Valerio Nardone for contributing to the preliminary assessment of our experimental results.


\bibliographystyle{plain}

\bibliography{bibliography}   


\end{document}